\documentclass[12pt,a4paper]{article}

\usepackage{jheppub}

\usepackage[latin1]{inputenc}
\usepackage{amsmath}

\DeclareMathOperator{\artanh}{artanh} 
\DeclareMathOperator{\Erf}{Erf}

\DeclareMathOperator\Erfi{Erfi}

\usepackage{amsfonts}
\usepackage{physics}
\usepackage{amssymb}
\usepackage{faktor}
\usepackage{graphicx}
\usepackage[export]{adjustbox}
\usepackage{tikz}
\usepackage{amsthm}
\numberwithin{equation}{section}
\numberwithin{figure}{section}
\usepackage[english]{babel}
\usepackage{mathtools}
\usepackage{nicefrac}
\usepackage{lscape}
\usepackage{textgreek}
\usepackage{parskip}
\usepackage{enumerate}

\usepackage{tcolorbox}

\usepackage{todonotes}

\usepackage{dsfont}

\usepackage{url}

\usepackage{romannum}

\theoremstyle{definition}

\usepackage{subcaption}
\usepackage{cleveref}
\crefname{figure}{figure}{figures}
\Crefname{figure}{Figure}{Figures}
\hypersetup{
	pdftitle={Topological gravity for arbitrary Dyson index},
	pdfauthor={Torsten Weber}}
\usepackage{upgreek}
\usepackage{array}
\usepackage{bm}

\newcommand{\TW}[1]{\textcolor{black}{#1}}

\author[a]{Torsten Weber}
\author[a]{Marco Lents}
\author[a]{Johannes Dieplinger}
\author[a]{Juan Diego Urbina}
\author[a]{and Klaus Richter}

\affiliation[a]{Institut f\"ur Theoretische Physik, 
Universit\"at Regensburg, \\
Universit\"atsstr. 31, D-93053 Regensburg, Germany}

\emailAdd{torsten.weber@ur.de}
\emailAdd{marco.lents@ur.de}
\emailAdd{johannes.dieplinger@ur.de}
\emailAdd{juan-diego.urbina@ur.de}
\emailAdd{klaus.richter@ur.de}
\title{Topological gravity for arbitrary Dyson index}

\abstract{
We use the well established duality of topological gravity to a double scaled matrix model with the Airy spectral curve to define what we refer to as topological gravity with arbitrary Dyson index $\upbeta$ ($\upbeta$ topological gravity). On the matrix model side this is an interpolation in the Dyson index between the Wigner-Dyson universality classes, on the gravity side it can be thought of as interpolating between orientable and unorientable manifolds in the gravitational path integral, opening up the possibility to study moduli space volumes of manifolds ``in between''. Using the perturbative loop equations we study correlation functions of this theory and prove several structural properties, having clear implications for the generalised moduli space volumes. Additionally we give a geometric interpretation of these properties using the generalisation to arbitrary Dyson index of the recently found Mirzakhani-like recursion for unorientable surfaces. Using these properties, we investigate whether $\upbeta$-topological gravity is quantum chaotic in the sense of the Bohigas-Giannoni-Schmit conjecture. Along the way we answer this question for the symplectic Wigner-Dyson class, not studied in the literature yet, and establish strong evidence for quantum chaos for this version of the theory, and thus for all bosonic varieties of topological gravity. We further argue for quantum chaoticity in the general $\upbeta$ case, based on novel constraints we find to be obeyed by genuinely non-Wigner-Dyson parts of the moduli space volumes. As for the general $\upbeta$ case the universal behaviour expected from a chaotic system is not known fully analytically we give a novel way to approach it, starting with the result of $\upbeta$ topological gravity and compare the results to a numerical evaluation of the universal result.
}

\makeindex
\begin{document}
\maketitle

\section{Introduction}

The duality of Jackiw Teitelboim (JT) gravity with a specific matrix model, first described in \cite{Saad2019}, has attracted a lot of attention in recent years and sparked a plethora of insights into two-dimensional quantum gravity. We give a short recapitulation of the aspects of this duality relevant for the present work in \cref{sec:Back_JT_matrix}\footnote{\TW{We would like to clarify, that by ``duality'' of a matrix model with a gravitational theory we always mean the order-by-order agreement of the topological (perturbative) expansions of the two theories. More details can be found in \cref{sec:Back_JT_matrix}.}}.

The original version of the duality was given for a setting where the gravitational path integral, which defines correlation functions in JT gravity, is restricted to contain only orientable manifolds. There, the dual matrix model belongs to the unitary symmetry class. From the seminal work \cite{Altland1997} the full classification of symmetry classes of random matrices is known to be tenfold. Consequently, the question was asked if the duality of the matrix model to JT gravity in the unitary symmetry class extends to a duality for all the symmetry classes and if so, what the dual gravitational theory is. This was answered affirmatively in \cite{Stanford2019}, where it was found that the direct generalisation of the theory by allowing also unorientable manifolds in the path integral yields theories that are dual to the orthogonal or the symplectic symmetry class which together with the unitary symmetry class form the so-called Wigner-Dyson classes. The realisation of the other seven symmetry classes requires the inclusion of supermanifolds in the gravitational theory. We shall restrict our discussion to the bosonic case on the gravitational side, i.e. to the Wigner-Dyson classes on the matrix model side. 

A particularly interesting aspect of this duality is the ability to study geometric objects appearing in the gravitational theory, specifically moduli space volumes, with matrix model techniques. This relation, i.e. that of the volumes of the moduli spaces of orientable hyperbolic surfaces of genus $g$ with finitely many geodesic boundaries of given lengths, known as Weil-Petersson (WP) volumes, with a unitary matrix model was known prior to the discovery of the JT/matrix model duality \cite{Eynard2007}. The relation for unorientable surfaces was proven by relating the perturbative expansion of the matrix model in the orthogonal symmetry class to a generalisation of Mirzakhani's well known recursion relation for the orientable WP volumes \cite{Mirzakhani2007} to the unorientable case in \cite{Stanford2023}. This is especially interesting due to the moduli space volumes for unorientable hyperbolic surfaces suffering from problems, like being actually divergent and in need of regularisation, not necessary for their tame orientable brothers \cite{Norbury2008,Gendulphe2017}. Though these problems can be overcome and the recursion for the volumes given in \cite{Stanford2023} can be iterated for small genus and numbers of boundaries, working on the matrix model side of the duality and inferring the WP volumes from there, the approach used here, has proven to be the more economic way to determine these objects. 

Additionally to working on the matrix model side, it is useful not to study directly the theory of JT gravity but first a simplification of this theory, known as the Airy model or topological gravity. This theory can be regarded as the low-energy limit of JT gravity: The leading-order contribution to the JT gravity genus $0$ density of states, $\rho^{\text{JT}}_0=\frac{1}{4\pi^2}\sinh\qty(2\pi\sqrt{E})$, corresponds to the Airy density of states $\frac{1}{2\pi}\sqrt{E}\eqqcolon\rho_0^{\text{Airy}}$. Furthermore, in terms of WP volumes, this theory gives the behaviour of the full WP volumes for large boundary lengths in the orientable case \cite{Blommaert2022,Saad2022}. It still does so in the unorientable case where additionally it does not have any divergences, i.e. it produces the leading-order contributions to the non-divergent part of the full WP volumes \cite{Stanford2023}. The study of unorientable topological gravity was started with giving the first-order genuinely unorientable contribution in \cite{Saad2022} and extended to explicit results to higher and structural results to all orders in \cite{Weber2024}. Harnessing these results and the used matrix model techniques, the general structure of the full unorientable WP volumes was found in \cite{Tall2024}, though a proof thereof is a matter of present investigation.

Matrix models, such as the ones dual to JT and topological gravity are thus useful tools to study moduli space volumes efficiently. In order to state the aims of this work it is useful to briefly recall the definition of a matrix model, a more complete review of which can be found e.g. in \cite{Eynard2018}. A matrix model can be defined by a partition function written as an integral over the respective class of matrices one is interested in with respect to a measure determined by the choice of a potential $V$, which for our case of interest can be assumed to be a polynomial. For the Wigner-Dyson case those classes are complex hermitian (unitary), quaternionic hermitian (symplectic) and real symmetric (orthogonal) matrices. These classes of matrices have the property that all their elements are diagonalisable. Consequently, if one restricts to observables dependent only on the spectrum of the matrices one can integrate out the diagonalising matrices\footnote{For historic reasons the respective (compact) groups of matrices determine the names of the symmetry classes.} and write the partition function as an integral over the eigenvalues only. Then, the partition function defining the matrix model reads \cite{Stanford2019}
\begin{align}
\label{eq:Z}
\mathcal{Z}=\mathcal{N}\int_{\mathbb{R}^N}\dd\Lambda\abs{\Delta(\Lambda)}^{\upbeta}e^{-N\frac\upbeta 2 \sum_{i=1}^{N}V(\lambda_i)}.
\end{align}
with $\Lambda=\qty(\lambda_1,\dots,\lambda_N)$, the matrix size $N$ and $\mathcal{N}$ being a normalisation constant not relevant in the following. Furthermore, $\Delta(\Lambda)$ denotes the Vandermonde determinant defined as
\begin{align}
    \Delta(\Lambda)=\prod_{1\leq i< j\leq N}\qty(\lambda_j-\lambda_i).
\end{align}
Interestingly, the choice of matrix ensemble has now boiled down to the value of the so-called \textit{Dyson index} $\upbeta$, where $\upbeta=2$ corresponds to the unitary, $\upbeta=1$ to the orthogonal and $\upbeta=4$ to the symplectic symmetry class. However, at the level of eq.~(\ref{eq:Z}) there is no reason not to choose a general value $\upbeta\in\mathbb{R}_+$ which enables one to interpolate between the Wigner-Dyson classes. This generalisation beyond the Wigner-Dyson classes has been intensively studied in the last decades; an extensive list of references can be found in \cite{Forrester2010}, extending many known results for the Wigner-Dyson classes by an interpolation in terms of the Dyson index to change between the classes. Of most direct importance for our work here is the realisation of the Gaussian matrix model for arbitrary $\upbeta$ as a specific ensemble of tridiagonal matrices in \cite{Dumitriu2002}, enabling the numerical computations shown later. 
\TW{It is interesting to note, that one can further extend the direct generalisation to arbitrary Dyson index of \cref{eq:Z} by choosing functions $f,g:\mathbb{R}_+\to\mathbb{R}_+$ which have the property that $\underset{\upbeta\in\{1,2,4\}}{\forall}:f(\upbeta)=g(\upbeta)=\upbeta$ and by writing
\begin{align}
\mathcal{Z}=\mathcal{N}\int_{\mathbb{R}^N}\dd\Lambda\abs{\Delta(\Lambda)}^{f(\upbeta)}e^{-N\frac{g(\upbeta)}{2} \sum_{i=1}^{N}V(\lambda_i)}.\label{eq:Z[fg]}
\end{align}
This also forms a valid generalisation of the Wigner-Dyson cases depending on two arbitrary functions. As we will see in the main text, at least for the perturbative setting which we are interested in, it is possible to eliminate the dependence on the function $g$ and hence reduce the degree of ambiguity to choosing one function. 
Performing all computations for the choice $f(\upbeta)=\upbeta$ and the mapping by another choice for $f(\upbeta)$ after the desired evaluation, one can thus address this more general case by considering only the direct generalisation we discussed previously.}

\TW{
The way we treat transitions between symmetry classes is not the only available method to do this based on matrix models. The most well-known method uses the so-called Pandey-Mehta model \cite{Pandey1983}, which is a matrix model in its original version describing the transition from the orthogonal to the unitary symmetry class. In this model, the considered matrix $H_{\text{PM}}$ is devised as 
\begin{align}
    H_{\text{PM}}=H^\alpha_{1}+ H_{2}^\alpha,
\end{align}
where the matrices $H^\alpha_{\upbeta}$ are chosen from the Gaussian ensembles for the specified value of $\upbeta$. They depend on a parameter $\alpha$ in such a way that $H_{\text{PM}}$ is from the orthogonal ensemble in the case of $\alpha=0$ and in the unitary ensemble at $\alpha=1$, while being in a crossover
regime in between. The range of this model can be modified to treat the other possible transitions between symmetry classes \cite{Mehta2004}. However, to the best of our knowledge, it is not possible to find a variety of the model which allows for a crossover between all three Wigner-Dyson classes by varying a single parameter. In the classic setting of using random matrix theory to describe universal properties of a quantum chaotic system, this model has a distinct advantage over the transitional procedure we use here. This is due to the possibility to describe a transition from the orthogonal (presence of time-reversal symmetry) to the unitary (no time-reversal symmetry) symmetry class by turning on e.g.\ a magnetic flux in a previously time-reversal invariant system, i.e.\ by adding a time-reversal breaking term to the Hamiltonian. This is essentially what is implemented in the Pandey-Mehta model. Hence it is not surprising that in the universal regime these systems upon breaking time-reversal symmetry agree with its prediction, as shown using semiclassical methods in \cite{Bohigas1995,Turek2003,Saito2006} and seen in explicit examples like in spin chains \cite{Kundu2022}.
}
\TW {
However, in our setting, this way of interpolating between the ensembles is less appealing. This is due to the fact, that for JT/topological gravity the connection with a matrix model does not arise via the agreement of certain universal observables but as a precise perturbative duality. Hence, we can study, in fact define, the gravitational theory dual to the arbitrary Dyson index generalisation of the matrix model by extending this duality. Notably, by this we can treat all Wigner-Dyson classes and transitions between them at once.
}

The aim of the first part of this work now is precisely to apply this thought of interpolating between the symmetry classes by considering a varying $\upbeta$ to the matrix model dual to topological gravity in the Wigner-Dyson classes. Specifically, we will generalise the method used in \cite{Weber2024} for the case of $\upbeta=1$ to arbitrary $\upbeta$ to compute correlation functions of the general $\upbeta$ matrix model perturbatively. Using these, we assume the duality to hold also for the general $\upbeta$ case and \textit{define} general $\upbeta$ Airy WP volumes ($V^{\upbeta}_{g,n}$) using the matrix model to give the, to our knowledge, only way to define moduli space volumes that interpolate between the purely orientable and purely unorientable setting in a sensible way. Doing this, we find and prove the general structure of the $V^{\upbeta}_{g,n}$ based on the transformation of perturbative contributions of matrix model correlation functions under $\upbeta\rightarrow\frac 4\upbeta$. By this, and based on the explicit results we work out, we can firmly establish that the $V^{\upbeta}_{g,n}$ have additional contributions that are vanishing in all the Wigner-Dyson classes, hence showing that moduli space volumes defined in this way are not mere interpolations of the weights of orientable and unorientable manifolds but rather encompass contributions that suggest the interpretation of being neither orientable nor unorientable. Extending the Mirzakhani-like recursion for the unorientable WP volumes of \cite{Stanford2023} to general Dyson index we give a geometric interpretation of the general structure (which we show also to apply in the case of JT gravity) and the non-Wigner-Dyson contributions to the moduli space volumes. 

In the second part of this work we will turn to another important aspect of the duality, the possibility to study the implications of quantum chaos on moduli space volumes. This connection comes about via the famous conjecture due to Bohigas, Giannoni and Schmit (BGS)\cite{Bohigas1984} that quantum chaos for a given quantum system can be classified by checking whether the expectation value of certain ``universal'' observables in this system coincides with that in the Gaussian matrix model of the system's symmetry class in the respective regime of universality. The observable that we shall be interested in mostly is the spectral form factor, specifically its canonical form, being defined as 
\begin{align}
    \kappa_{\upbeta}(t,\beta)\coloneqq\ev{Z(\beta+i t)Z(\beta-it)}_{c,\upbeta},
\end{align}
i.e. the connected correlation function of (thermal) partition functions of complex conjugate complex temperatures in a matrix model with the Dyson index $\upbeta$. The matrix model in question for the present work will be the matrix model dual to topological gravity. Importantly, the canonical SFF, which in the following is always meant when speaking of a SFF unless stated otherwise, is model dependent while the \textit{microcanonical} SFF only depends on the symmetry class, i.e. the Dyson index in the universal regime. However, one can show that the late-time limit of the canonical SFF can indeed be computed from only the microcanonical SFF for the respective symmetry class and the leading-order density of states \cite{Weber2024}. Consequently, agreement of the late-time SFF as computed from topological gravity/JT gravity with the corresponding prediction of universal RMT can be seen as an indication of chaos in the sense of the BGS conjecture. This has been successfully studied so far for the unitary \cite{Saad2022} and orthogonal \cite{Weber2024,Tall2024} symmetry classes and we give more information and details on this in \cref{sec:Chaos_top_grav}. 

The aim of this work regarding this topic is twofold. First, we close the gap in the literature by studying the symplectic symmetry class ($\upbeta=4$) for which we pursue the established route of computing first the prediction of universal RMT for the late time SFF and then comparing to the corresponding result from topological gravity. We find agreement up to $\tau^4$ where we utilize the techniques used in \cite{Weber2024} to successfully show the corresponding statement for the orthogonal case.

Second, we study the case of arbitrary $\upbeta$, where the relevant correlation functions in topological gravity have been worked out in the first part of this work. For this setting, the established way ceases to work since, to our knowledge, the universal RMT result, i.e. the result for the microcanonical SFF for the Gaussian matrix model with arbitrary Dyson index, has not been fully computed analytically in the literature (the highest order results, in the sense of an expansion of the microcanonical SFF for small times, we are aware of are given in \cite{Forrester2001,Forrester2021}\footnote{We thank P.J. Forrester for pointing out these references to us.}). The alternative way we find to approach the question of chaoticity in $\upbeta$ topological gravity is the following: First, we study the constraints imposed on the Airy WP volumes in the unorientable case $\upbeta=1$\footnote{They are the same as those for $\upbeta=4$.} by matching to the universal RMT SFF, extending the study started in \cite{Weber2024}. These constraints can be seen as the imprint of quantum chaos in the WP volumes. Remarkably, those constraints\footnote{With one exception that is, however, expected as explained in the main text.} are obeyed by the genuinely non-Wigner-Dyson part of the WP volumes discussed above which we interpret as a strong sign for the persistence of quantum chaos for arbitrary Dyson index. Building on this, we can now deduce information on the putative universal microcanonical SFF for arbitrary $\upbeta$ by computing it from the $\upbeta$ Airy WP volumes. In this direction we give a novel way to do so by finding the ``uplift'' to the setting of arbitrary Dyson index of the universal RMT results for the Wigner-Dyson classes. As evidence for our results beyond our analytical arguments we compare to a numerical study of the Gaussian microcanonical SFF for arbitrary $\upbeta$ using the expression of this matrix model as an ensemble of tridiagonal matrices in \cite{Dumitriu2002}, mentioned above. We find good agreement in the limits imposed by us not taking into account the non-Wigner-Dyson parts of the WP volumes, establishing the feasibility of our approach to find the universal RMT result for arbitrary $\upbeta$. The completion of this study, i.e. taking into account the non-Wigner-Dyson contributions is subject of present work to be presented elsewhere \cite{Weber2025b}.

The paper is structured as follows. In \cref{sec:Back_JT_matrix} we give necessary background on the relation of JT/topological gravity with matrix models and in \cref{sec:Chaos_top_grav} on the role of quantum chaos in this duality. We then give a compact overview of the main results of this work in \cref{sec:Main_results}. \Cref{sec:TopGrav_arbBeta} deals with the first part of this paper as described above, i.e. the study of topological gravity for arbitrary Dyson index. For this, we first recapitulate the perturbative solution of the loop equations in \cref{sec:Recap_LE} which we then use to compute the perturbative contributions to certain correlation functions in \cref{sec:Top_exp_upbeta_resolvents}. In order to better understand those, we find their general structure in terms of $\upbeta$ in \cref{sec:Gen_Struct} which translates to that of the $\upbeta$ Airy WP volumes we compute in \cref{sec:AiryWP}. To give a better geometric understanding of $\upbeta$ topological gravity we discuss the generalisation of Kontsevich diagrammatics \cite{Kontsevich1992} to this setting and a Mirzakhani-like recursion for JT gravity for arbitrary Dyson index in \cref{sec:Geometric_Arb_beta}. We then turn to the inquiries regarding quantum chaos in arbitrary $\upbeta$ topological gravity in \cref{sec:Chaos_Top_grav_arb}. To address this question we first compute the late time SFF for $\upbeta$ topological gravity in \cref{sec:SFF_loop_eq}. Using this result, we discuss the presence of quantum chaos in the sense of the BGS conjecture for the symplectic symmetry class in \cref{sec:Comparison_RMT_Top_GSE} and for the case of arbitrary $\upbeta$ in \cref{sec:GeneralCase}. We conclude in \cref{sec:Conclusion}. In the appendices we first give collections of some of our results for the perturbative expansion of resolvents (\cref{sec:Coll_Resolvents}) and Airy WP volumes (\cref{sec:WPV_collection}). Then, we give the proofs of several statements needed for establishing the general structure of correlation functions in terms of $\upbeta$ in \cref{sec:Proof_Relation_R_0,sec:Proof_Relation_R} and the proof of the general structure itself in \cref{sec:Proof_GenStruct}. In \cref{sec:split_upbeta} we give technical details on how to easiest bring our results for the correlation functions into the general form we prove, while the proof of a geometrical statement used for the geometric variety of said proof in the main text is given in \cref{sec:App_Decomposition}. In \cref{sec:SFF_derivation} we derive the prediction of universal RMT for the late-time SFF in the symplectic symmetry class. \Cref{sec:det_canc_func} gives further background for the comparison of this result with topological gravity.

\section{Background and main results}\label{sec:Background}
Before going into the main part of the paper, we collect some important background from the literature to fix notation and to make the discussion more self-contained. The reader familiar with this may continue with a summary of our main results in \cref{sec:Main_results}.
\subsection{Relation of matrix models with topological/JT gravity}\label{sec:Back_JT_matrix}
The objects primarily studied in JT gravity are connected correlation function of partition functions. These correlation functions, caused by the possibility to split the gravitational path-integral, by which they are computed, into a sum over contributions of manifolds of different topology, in fact different genus, have a topological expansion of the following form \cite{Stanford2019}
\begin{align}
    \ev{\prod_{i=1}^n Z(\beta_i)}_c=\sum_{g=0,\frac 12,1,\dots} \frac{Z_{g,n}(\beta_1,\dots,\beta_n)}{\qty(e^{S_0})^{2g-n+2}}.
\end{align}
The different contributions to the correlation function at genus $g$ are given by \cite{Saad2019,Stanford2019}
\begin{align}
\label{eq:Zs}
    Z_{g,n}(\beta_1,\dots,\beta_n)=\qty[\prod_{i=i}^n\int_0^{\infty} b_i \dd{b_i} Z^t\qty(b,\beta_i)]V_{g,n}(b_1,\dots,b_n),
\end{align}
with the ``trumpet'' partition function
\begin{align}\label{eq:def_trumpet}
    Z^t\qty(b,\beta)\coloneqq\frac{1}{\sqrt{4\pi \beta}}e^{-\frac{b^2}{4 \beta}},
\end{align}
\begin{figure}
    \centering
    \includegraphics[width=.7\textwidth]{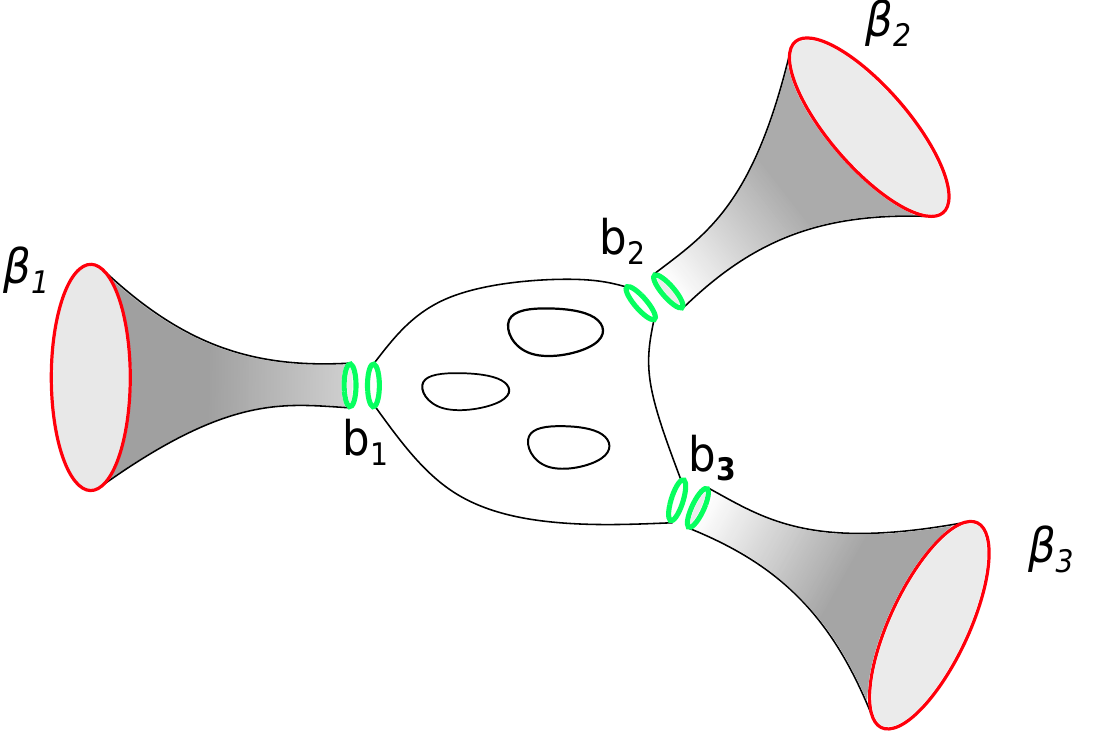}
    \caption{Manifold contributing at genus $g=3$ to the correlation function of three partition functions with (complex) inverse temperatures $\beta_1,\beta_2,\beta_3$. In grey are the ``trumpets'', cut off along geodesic boundaries of lengths $b_1,b_2,b_3$. }
    \label{fig:example_manifold}
\end{figure}
and the $V_{g,n}\qty(b_1,\dots,b_n)$ denoting for JT gravity the Weil-Petersson volumes, i.e. the volumes of the moduli space of hyperbolic two-manifolds of genus $g$ and $n$ geodesic boundaries of lengths $b_1,\dots,b_n$. The expression in eq.~(\ref{eq:Zs}) can be thought of as ``glueing'' a hyperbolic two-manifold of genus $0$ with an asymptotic boundary of renormalised length $\beta_i$ and a geodesic boundary of length $b$ (the ``trumpet'') to a hyperbolic manifold of potentially non-zero genus along a geodesic boundary of the same length while doing so for each partition function in the correlator. This is illustrated in \cref{fig:example_manifold}. In this framework, the different varieties of (bosonic) JT gravity can be constructed by allowing only orientable hyperbolic manifolds which is dual to a matrix model of unitary symmetry class ($\upbeta=2$), or also admitting unorientable ones which will then result in a duality with a matrix model of orthogonal ($\upbeta=1$) or symplectic ($\upbeta=4$) symmetry class \cite{Stanford2019}. In order to state this duality, which is at the heart of this work, we define $n$-point resolvents
\begin{align}\label{def:resolvents}
    R(x_1,\ldots,x_n)\coloneq\prod_{i=1}^n\Tr\frac{1}{x_1-H},
\end{align} 
where in the following the abbreviated notation
\begin{align}
    I = \{x_1,\dots,x_n\},
\end{align}
is often used. For the type of matrix models of interest here, i.e. double-scaled matrix models, there is a topological expansion like that for JT gravity given by 
\begin{align}\label{eq:R_top_exp}
    \expval{R(I)}^{\text{MM}}_c=\sum_{g=0,\frac 12 ,1, \dots} \frac{R^{\text{MM}}_g(I)}{e^{\qty(2g+\abs{I}-2)S_0}},
\end{align}
where the algorithm to actually work out the $R^{MM}_g(I)$ for the matrix models of interest will be the considered in the main text. The correlation functions of resolvents are of course also computable from the correlation functions of partition functions and vice versa, so the statement of the duality amounts to saying that the computation of the correlation functions of choice for a specific matrix model and for JT gravity give the same result. For our purpose, it is best to translate this to a relation of the Weil-Petersson volumes, the prime objects of interest on the gravitational side, to the resolvents, which are the most natural objects to consider on the matrix model side. Doing this, one finds that the duality implies
\begin{align}\label{eq:V[R]}
    V_{g,n}(b_1,\dots,b_n)=\mathcal{L}^{-1}\qty[R^{\text{MM}}_{g}(-z_1^2,\dots,-z_n^2)\prod_{i=1}^n\qty(\frac{-2 z_i}{b_i}),\qty(b_1,\dots,b_n)],
\end{align}
where $\mathcal{L}^{-1}$ denotes the inverse Laplace transformation.
This statement, for the unitary/orientable case, has been shown first in \cite{Eynard2007} considering a matrix model of unitary symmetry class with a leading-order density of states 
\begin{align}
    \rho_0^{\text{JT}}(E)=\frac{1}{4\pi^2}\sinh\qty(2\pi\sqrt{E}),
\end{align}
 that later was shown to be that of JT gravity \cite{Saad2019}. Specifically, the proof worked by showing that the recursion determining the perturbative contributions to the matrix model correlation functions is equivalent to Mirzakhani's well known recursion \cite{Mirzakhani2007} enabling the computation of the moduli space volumes.

For the case of unorientable JT gravity, the claim of duality was first made and motivated in \cite{Stanford2019} and extended in \cite{Stanford2023} by giving a generalisation of Mirzakhani's recursion to the unorientable setting and showing, like in the orientable case, that this is equivalent to the matrix model recursion of the matrix model determined by $\rho_0^{\text{JT}}$ for the orthogonal symmetry class. The unorientable case is considerably more complicated than the orientable case, because the moduli space volumes are divergent objects requiring regularisation. Nevertheless, it is possible to work out explicit results for the whole volumes for low genus and numbers of boundaries as done in \cite{Stanford2023} and extended, with a different but related regularisation in \cite{Tall2024}.

A way to avoid some of the complications of divergent volumes is to consider the regularisation independent parts of the unorientable volumes, which can be shown to be related to the low-energy limit of JT gravity, known as ``topological gravity'' or the ``Airy model'', i.e. a matrix model with the leading-order density of states 
\begin{align}
    \rho^{\text{Airy}}_0(E)=\frac{1}{2\pi}\sqrt{E}.
\end{align}
For this limit, as performed in \cite{Weber2024}, the volumes can be computed to much higher genus and numbers of boundaries which enables a thorough study of their properties, which was vital to the consideration and already showed many of the features found for the full JT case. Consequently, also the generalisation to arbitrary $\upbeta$ performed in \cref{sec:TopGrav_arbBeta} will mainly focus on the Airy model.

\subsection{Chaos in topological/JT gravity}\label{sec:Chaos_top_grav}

An important application of the explicit computations of (Airy) Weil-Petersson volumes is the possibility to show agreement of the topological/JT gravity correlation functions with the predictions of universal RMT for the canonical SFF, this being proof of chaoticity of the respective theory. 

The canonical spectral form factor is defined as 
\begin{align}\label{def:SFF_can}
    \kappa_{\upbeta}(t,\beta)\coloneqq\ev{Z(\beta+i t)Z(\beta-it)}_{c,\upbeta}\simeq\sum_{g=0,\frac 12,1, \dots} \frac{\kappa_{\upbeta}
    ^g(t,\beta)}{\qty(e^{S_0})^{2g}},
\end{align}
where the topological expansion is induced by that of the two-point correlation function of partition functions and we included indices of $\upbeta$ to make explicit that the respective object depends on the choice of ensemble. It was shown in \cite{Weber2024}, that for a matrix model given by the leading-order density of states $\rho_0$ for a Wigner-Dyson choice of $\upbeta$ in the so-called ``$\tau$-scaled'' limit (i.e. $t\rightarrow \infty$, $e^{S_0}\rightarrow \infty$ with $\tau\coloneqq e^{-S_0} t$ fixed)
\begin{align}\label{eq:univRMT_kappa_canonical}
\begin{aligned}
    \kappa^s_\upbeta(\tau,\beta)&\coloneqq\underset{\substack{e^{S_0}, t\rightarrow \infty \\
    \tau\coloneqq e^{-S_0} t \text{ fixed}}}{\lim} e^{-S_0}\kappa_{\upbeta}(t,\beta) \\
    &=\int_{0}^{\infty}\dd E e^{-2\beta E}\rho_0(E)- 
2\int_{0}^{\infty}\dd E e^{-2\beta E}\rho_0(E)\int_{0}^{\infty}\dd{x} \cos(\frac{\tau}{\rho_0(E)}x)\Upsilon^\upbeta\qty(x)\\
    &\eqqcolon\int_{0}^{\infty}\dd E e^{-2\beta E}\rho_0(E)- 
\int_{0}^{\infty}\dd E e^{-2\beta E}\rho_0(E)b^{\upbeta}\left(\frac{\tau}{2\pi \rho_0(E)}\right).
\end{aligned}
\end{align}
Here $\Upsilon^\upbeta$ and thus $b^\upbeta$ can be determined from the Gaussian ensemble of the respective Wigner-Dyson class and the analytical results for them are given in \cite{Mehta2004}.
The cases already considered in the literature are $\upbeta=2$ \cite{Okuyama2021,Saad2022} and $\upbeta=1$ \cite{Weber2024}. For the Airy density of states one finds
\begin{align}
    \kappa^s_1(\tau,\beta)&=2\kappa^s_2(\tau,\beta)-\frac{\tau e^{-8\beta \tau^2}}{8\pi\beta}\Bigg{[}  \Gamma(0, 2\beta \tau^2)(1 - e^{8\beta \tau^2}) + \nonumber\\[5pt]
    & 16 \beta  \tau ^2 \, _2F_2\left(1,1;\frac{3}{2},2;8 \beta  \tau ^2\right)+\pi \Erfi\left(\sqrt{8\beta\tau^2 }  \right)-\label{eq:kappa_1_s}\\[5pt]& \sum_{n=1}^{\infty}\left(\sum_{m=1}^{n}\frac{(-1)^{n+m}(2)^{2m}}{(m)!(n-m)!(n-\frac{m}{2})}\right)\left(2\beta\tau^2\right)^n \Bigg{]},\nonumber\\
    \kappa^s_2(\tau,\beta)&=\frac{1}{4\sqrt{\pi}}\frac{1}{\qty(2\beta)^{3/2}}\Erf\qty(\sqrt{2\beta}\tau)\label{eq:kappa_2_s}.
\end{align}
Perturbative agreement of the topological/JT gravity result for the canonical SFF with this prediction for $\upbeta=2$ has been shown in \cite{Saad2022} with the specific properties of the orientable WP volumes necessary for this to happen worked out and explored in \cite{Blommaert2022,Weber2022}. With perturbative agreement we specifically mean that both sides of \cref{eq:univRMT_kappa_canonical} agree as power series in $\tau$ and $\beta$. This is the natural form one finds (after $\tau$ scaling) for the topological/JT gravity SFF while for the universal RMT answer one can find it by expanding the exact result. Explicitly, for the case of $\upbeta=1$ (\cref{eq:kappa_1_s}) the first terms of this expansion can be found to be \cite{Weber2024}
\begin{align}\label{eq:kappa_1_s_expanded}
    \kappa^s_1(\tau,\beta)&=\frac{\tau}{2\pi \beta}-\frac{\tau^2}{\sqrt{2\pi\beta}}-\frac{\gamma +\log\qty(2\beta \tau^2)+\frac{1}{3}}{\pi}\tau^3+\frac{8\sqrt{2\pi \beta}}{3\pi}\tau^4+\nonumber\\ &+\frac{\beta\qty(4\gamma+4\log\qty(2\beta \tau^2)-\frac{7}{15} )}{\pi }\tau^5-\frac{64\qty(2\pi\beta)^{\frac 32}}{15 \pi^2}\tau^6+\order{\tau^7}.
\end{align}
For the unorientable case, as one could expect, the reasoning is more involved, as well for the side of the WP volumes as the universal RMT side, but agreement was shown for the unorientable Airy model in \cite{Weber2024} and for unorientable JT gravity in \cite{Tall2024}. 

An important complication, occurring already for the Airy model\footnote{Actually only there, in the sense that of course the part of the JT result corresponding to the Airy limit still contains it but the rest doesn't, see \cite{Tall2024} for details.}, is the remaining dependence on $t$ of the coefficients of the expansion in $\tau$ and $\beta$ of the canonical SFF. To better explain this, we rewrite
\begin{align}
    \kappa_\upbeta^s(\tau,\beta)=\lim_{e^{S_0}, t\rightarrow \infty}\sum_{g=0,\frac 12,1, \dots} \frac{\kappa_{\upbeta}
    ^g(t,\beta)\tau^{2g+1}}{\qty(e^{S_0})^{2g+1}\tau^{2g+1}}=\lim_{t\rightarrow\infty}\sum_{g=0,\frac 12,1, \dots} \underbrace{\frac{\kappa_{\upbeta}
    ^g(t,\beta)}{t^{2g+1}}}_{\coloneqq \kappa^{s,g}_\upbeta(t,\beta) }\tau^{2g+1}.
\end{align}
For the orientable case, one finds that sum and limit can be interchanged and $\lim_{t\rightarrow\infty}\kappa^{s,g}_2(t,\beta)$ is independent of $t$. For the unorientable Airy model, it was found in \cite{Weber2024} that one can not interchange sum and limit as starting from $g=\frac 3 2$ all $\kappa^{s,g}_1(t,\beta)$ retain a $t$ dependence that does not vanish with the limit. It was possible to make sense of this by grouping terms with the same $\beta$ dependence and finding that by adding and subtracting certain hypergeometric functions and using their asymptotic expansions for the limit $t\rightarrow\infty$, it was possible to remove the $t$ dependence.
Explicitly, for the contribution to $\tau^3 \beta^0$ one found \cite{Weber2024}
\begin{align}
\label{eq:hyper}
    \frac{1}{\pi} \qty[\frac{-10}{3}+\log\qty(\frac{2t}{\beta})-\frac{\sqrt{2\pi}}{3}\left(t\tau^2\right)^{1/2} +\frac{\sqrt{2\pi}}{30}\left(t \tau^2\right)^{3/2}-\frac{2\left(t\tau^2\right)^2}{45}+\frac{\sqrt{2\pi}}{210}(\tau^2 t)^{5/2}+\dots],
\end{align}
where the dots indicate terms coming from higher orders in the topological expansion. Now, one defines
\begin{align}
\begin{aligned}
f(t,\tau)\coloneqq&\frac{\left(t\tau^2\right)^2}{45\pi}\left(6{}_2F_2\left(2,2;3,\frac{7}{2};-t\tau^2\right) - 4{}_1F_1\left(\frac{3}{2};\frac{7}{2};\frac{-t\tau^2}{2}\right)\right).
\end{aligned}
\end{align}
For this function, it holds that
\begin{align}
    f(t,\tau)&=\frac{2\qty(\tau^2 t)^2}{45\pi}+\underbrace{\frac{\left(t\tau^2\right)^2}{45\pi}\sum_{k=1}^{\infty}a_k (t \tau^2)^k}_{\order{\tau^6}}\\
    &\overset{t \to \infty}{=} \frac{1}{\pi}\left(-\frac{\sqrt{2\pi}}{3}\left(t\tau^2\right)^{1/2} +\log \left(4 t \tau ^2\right)+\gamma-3 \right),
\end{align}
where the first line is the definition of the hypergeometric functions as a power series with coefficients $a_k$, which has an infinite radius of convergence, and the second line is its asymptotic expansion. The trick is to add $f(t,\tau)-f(t,\tau)$, then write out its power series definition for the first occurrence and finally to take the limit of $t\rightarrow\infty$, resulting in 
\begin{align}
    -\frac{1}{\pi}\qty[\frac 1 3 +\log\qty(2\beta \tau^2)-\gamma]+\order{\tau^3},
\end{align}
which is precisely the result at order $\tau^3$ from the expansion of $\kappa^s_1(\tau,\beta)$ obtained from universal RMT, \cref{eq:kappa_1_s_expanded}. It can be argued that by adding and subtracting another hypergeometric function the remaining correction terms will be cancelled, however, this function is determined by contributions of higher genus than those computed in \cite{Weber2024} the relevant of which are given in \cref{eq:hyper}. 

Due to these complications, it is a non-trivial question whether the symplectic ($\upbeta=4$) symmetry class, which has not been studied from this point of view, does show agreement of the (perturbative) topological gravity canonical SFF with its RMT prediction. This is answered affirmatively in \cref{sec:Comparison_RMT_Top_GSE}, thus completing the study of all Wigner-Dyson symmetry classes/``standard'' choices of manifolds in bosonic topological gravity, i.e. showing the presence of quantum chaos, as seen through the lens of the BGS conjecture, for all (bosonic) cases.
\subsection{Main results of this work}\label{sec:Main_results}

After having given the necessary background and notational conventions we give a brief overview of the main results of this work.

\paragraph{Structure of the resolvents for the general $\upbeta$ Airy model.}
Solving the loop equations for the Airy model for arbitrary Dyson index $\upbeta$ in \cref{sec:TopGrav_arbBeta}\TW{, using the direct generalisation i.e. $f(\upbeta)=\upbeta$,} we find the explicit results for the resolvents up to $g=4, n=1$. Furthermore, we show that they have the general structure
\begin{align}
    R^\upbeta_g(I)=\frac{1}{\upbeta^{2g+n-1}}\left(\mathcal{R}_g^0(I)\upbeta^g+(2-\upbeta)^2\sum_{i=1}^g\mathcal{R}_g^i(I)\upbeta^{i-1}((1-\upbeta)(4-\upbeta))^{g-i}\right)
\end{align}
for integer $g$ and
\begin{align}
    R^\upbeta_g(I)=\frac{1}{\upbeta^{2g+n-1}}\left((2-\upbeta)\sum_{i=1}^{g+\frac 12}\mathcal R_g^i(I)\upbeta^{i-1}((1-\upbeta)(4-\upbeta))^{g+\frac 12 -i}\right)
\end{align}
for half-integer $g$, where the $\mathcal R_g^i$ do not depend on $\upbeta$. Actually, we show this structure to be valid for all one-cut matrix models. For topological gravity, we also give the structure of the $\mathcal R_g^i$ to be 
\begin{align}
    \mathcal{R}^0_g(I)=\frac{P^0_{g,n}(I)}{\prod_{j=1}^n\qty(z_j)^{6g+2n-3}},
\end{align}
for $i=0$ with $P^0_{g,n}(I)$ being a polynomial with rational coefficients of combined order $2(n-1)(3g-3+n)$ and 
\begin{align}
    \mathcal{R}^i_g(I)=\frac{P^i_{g,n}(I)}{\prod_{j=1}^n\qty(z_j)^{6g+2n-3}\prod_{j<k}^n\qty(z_j+z_k)^{(2g+2)}},
\end{align}
for the other cases. Again, $P^i_{g,n}(I)$ is a polynomial with rational coefficients, now of the combined order $(n-1)\qty[3\qty(n-2)+g\qty(n+6)]$. 

Along the way we show the following interesting relation of resolvents for general $\upbeta$ one-cut matrix models that are crucial for the proof of the aforementioned results. First,
\begin{align}
    R^\upbeta_0\qty(I)=\frac{1}{\upbeta^{\abs{I}-1}}R^1_0\qty(I),
\end{align}
reducing the computation of contributions at genus $0$ to n-point resolvents to the computation of those for $\upbeta=2$ which is a vast simplification. Second 
\begin{align}
    R^{\frac 4 \upbeta}_g(I)=\qty(-1)^{2g}\qty(\frac \upbeta 2)^{2\qty(g+\abs{I}-1)}R_g^{\upbeta}(I),
\end{align}
being vital for the proof of the general form of the resolvents and a crucial sanity check for their explicit computation.
\paragraph{Structure of the WP volumes for arbitrary $\upbeta$ topological gravity.} 
We build on the duality of the matrix model defined by the Airy spectral curve for the Wigner-Dyson classes with topological gravity in its various bosonic incarnations discussed above by using the arbitrary $\upbeta$ matrix model to define arbitrary $\upbeta$ topological gravity. For the WP volumes of this theory (and also those for JT gravity) we find the general structure
\begin{align}
    V^\upbeta_{g,n}(\vec{b})=\frac{1}{\upbeta^{2g+n-1}}\begin{cases}
        \mathcal{V}_{g,n}^0(\vec{b})\upbeta^g+(2-\upbeta)^2\sum_{i=1}^g\mathcal{V}_{g,n}^i(\vec{b})\upbeta^{i-1}((1-\upbeta)(4-\upbeta))^{g-i} \quad \text{int. } g,\\
        \qty(2-\upbeta)\sum_{i=1}^{g+\frac 12}\mathcal V_{g,n}^i(\vec{b})\upbeta^{i-1}((1-\upbeta)(4-\upbeta))^{g+\frac 12 -i} \quad \text{half int. } g,
    \end{cases}
\end{align}
which for $g>1$ notably includes genuinely non-Wigner-Dyson terms, i.e. non-zero terms that vanish in all the Wigner-Dyson classes.
The contributions at $i=0$ in the Airy case have the form familiar from the orientable Airy WP volumes
\begin{align}
    \mathcal{V}_{g,n}^0(\vec{b})=\sum_{\vec{\alpha}\in \mathbb{N}_0^n}^{\norm{\vec{\alpha}}_1 = 3g-3+n}C^g_{\vec{\alpha}}\prod_{i=1}^{n}b_i^{2\alpha_i},
\end{align}
where $C^g_{\vec{\alpha}}\in\mathbb{Q}_{\geq0}$ and totally symmetric. The other contributions have, for $n=2$, the form
\begin{equation}
\mathcal{V}_{g}^i\left(b_1,b_2\right)=\mathcal V_g^{i,>}(b_1,b_2)\theta(b_1-b_2)+\mathcal V_g^{i,>}\theta(b_2-b_1).
\end{equation}
with
\begin{equation}
\mathcal V_{g}^{i,>}(b_1,b_2)=\sum_{\substack{{\alpha_1,\alpha_2\in\mathbb{N}_0}\\\alpha_1+\alpha_2=6g-2}}C^{g,i}_{\alpha_1,\alpha_2}b_1^{\alpha_1}b_2^{\alpha_2},
\end{equation}
where the $C_{\alpha_1,\alpha_2}\in\mathbb{Q}_{\geq0}$ are not necessarily symmetric under $\alpha_1\leftrightarrow\alpha_2$. This is the form familiar from the unorientable Airy WP volumes from \cite{Weber2024}. For $n>2$ one can make a statement about the general form of the Airy WP volumes but it is more useful to use the general structure of resolvents shown in the previous paragraph instead.

We also give a geometric interpretation of the WP volumes for arbitrary Dyson index by discussing the generalisation of the Kontsevich diagrammatics and the Mirzakhani-like recursion of \cite{Stanford2023} to this setting. Using the recursion we give a geometric proof for the general structure in terms of the Dyson index of the (Airy) WP volumes.

\paragraph{Universal RMT SFF for GSE.}
We find the $\tau$-scaled canonical SFF from the usual combination of universal RMT results for fixed energy and the explicit form of the leading level density. For the symplectic symmetry class we get
\begin{align}
    \kappa^{s}_{4}(\tau,\beta)&=\kappa_{2}^{s}\qty(\frac\tau2, \beta)-\frac{\tau}{8\pi}\chi\left(\tau,\beta\right)
\end{align}
with $\kappa_{2}^{s}\qty(\frac\tau2, \beta)$ defined in \cref{eq:kappa_2_s} and
\begin{align}
\begin{aligned}
    \chi\left(\tau,\beta\right)=&-\frac{1}{4\beta}\left(-\gamma-\log\left(\beta\frac{\tau^{2}}{2}\right)-\sum_{n=1}^{\infty}\frac{\left(-\beta\frac{\tau^{2}}{2}\right)^{n}}{nn!}\right)\\
 & +\sum_{n=0}^{\infty}\frac{1}{4\beta}\frac{\left(-2\beta\tau^{2}\right)^{n}}{n!}\left(-\log2\beta\tau^{2}+\psi\left(n+1\right)\right)\\
 & +\sum_{n=0}^{\infty}\frac{1}{4\beta}\Gamma\left(-\frac{2n+1}{2}\right)\left(2\beta\tau^{2}\right)^{\frac{2n+1}{2}}\\
 &+\sum_{k=0}^{\infty}\frac{1}{8\beta}\left(-\frac{\beta\tau^{2}}{2}\right)^{k}\frac{1}{k!}\frac{2 \, _2F_1\left(1,2 k+1;2 k+2;\frac{1}{2}\right)}{2 k+1},
\end{aligned}
\end{align}
where $\gamma$ denotes the Euler-Mascheroni constant and $_2F_1(a,b;c,z)$ the Gauss Hypergeometric function. Expanding it to the order to which we compute the contributions on the topological gravity side we find
\begin{align}
\begin{aligned}
    \kappa^s_{\TW{4}}(\tau,\beta)=&\frac{\tau}{8\pi\beta}+\sqrt{\frac{2}{\pi}}\frac{\tau^{2}}{16\sqrt{\beta}}-\frac{\tau^{3}\left(3\log\left(\beta\frac{\tau^{2}}{2}\right)+3\gamma+1\right)}{48\pi}-\sqrt{\frac{2}{\pi}}\frac{\sqrt{\beta}\tau^{4}}{12}\\
    &+\frac{\beta\tau^{5}\left(60\log(\beta\frac{\tau^{2}}{2})+60\gamma-7\right)}{960\pi}+\sqrt{\frac{2}{\pi}}\frac{\beta ^{3/2}\tau^6}{15} +\\
    &-\frac{\beta^2\tau^7\qty(1120\log\qty(\beta\frac{\tau^{2}}{2})+1120\gamma-501)}{26880\pi}-\sqrt{\frac{2}{\pi}}\frac{4\beta ^{5/2}\tau^8}{105} +\order{\tau^9}.
\end{aligned} 
\end{align}
Using the mechanism recalled in \cref{sec:Chaos_top_grav} we show this is matched by the result we find from topological gravity with $\upbeta=4$ which up to $g=\frac 72$ is given by
\begin{align}
    \kappa_4^{s,\text{Airy}}(\tau,\beta)=&\lim_{t\rightarrow\infty}\Bigg\{ \frac{\tau}{8\pi\beta}+\sqrt{\frac{2}{\pi}}\qty(\frac{\tau^{2}}{16\sqrt{\beta}}-\frac{\sqrt{\beta}\tau^{4}}{12}+\frac{\beta ^{3/2}\tau^6}{15} -\frac{4\beta ^{5/2}\tau^8}{105}+\dots)+ \nonumber\\
    &+\frac{\tau^3\beta^0}{\pi}\qty[-\frac{5}{24}+\frac{3}{48}\log\qty(\frac{2t}{\beta})+\sqrt{\frac{\pi}{2}}\frac{\sqrt{t}\tau}{48}-\sqrt{\frac{\pi}{2}}\frac{\qty(\sqrt{t}\tau)^3}{1920}-\frac{t^2\tau^4}{5760}-\sqrt{\frac{\pi}{2}}\frac{\qty(\sqrt{t}\tau)^5}{53760}+\dots]\nonumber\\
    &+\frac{\tau^5\beta}{\pi}\qty[\frac{163}{960}-\frac{1}{16}\log\qty(\frac{2t}{\beta})-\sqrt{\frac{\pi}{2}}\frac{17\sqrt{t}\tau}{768}+\sqrt{\frac{\pi}{2}}\frac{3\qty(\sqrt{t}\tau)^3}{5120}+\dots]\nonumber\\
    &+\frac{\tau^7\beta^2}{\pi}\qty[-\frac{8297}{80640}+\frac{1}{24}\log\qty(\frac{2t}{\beta})+\sqrt{\frac{\pi}{2}}\frac{881\sqrt{t}\tau}{61440}+\dots]\nonumber\\
    &+\dots \Bigg\}.
\end{align}
\paragraph{Universal microcanonical SFF for general $\upbeta$.} Building on the agreement of the topological gravity result with the universal RMT prediction in the Wigner-Dyson classes and other arguments, primarily its compliance to the constraints on the general $\upbeta$ Airy WP volumes equivalent to those fulfilled by those for $\upbeta=1$ and $\upbeta=4$, we conjecture the agreement to hold also for the non-Wigner-Dyson choices of the Dyson index. Building on this we derive an important part of the, to our knowledge not yet fully computed, universal microcanonical SFF for general Dyson index. Specifically, we show that its ``Wigner-Dyson part'' i.e. the part arising from the uplift of the Wigner-Dyson contributions to arbitrary $\upbeta$ in the regime of small times is given by
\begin{align}
\kappa^{s,\text{WD}}_\upbeta(\tau,E)&=\rho_0(E)\qty[1-b^{\text{WD}}_\upbeta\qty(\frac{\tau}{2\pi\rho_0(E)})],
\end{align}
with
\begin{align}
\begin{aligned}
    b^{\text{WD}}_\upbeta\qty(x)=1-\frac{2}{\upbeta}x+&\frac{\qty(2-\upbeta)}{2\upbeta^2}x\Bigg[\qty(2-\upbeta+\sqrt{\upbeta})\log\qty(1+\frac{2x}{\sqrt{\upbeta}})
    \\
    & +\qty(2-\upbeta-\sqrt{\upbeta})\log\qty(\abs{1-\frac{2x}{\sqrt{\upbeta}}})\Bigg].
\end{aligned}
\end{align}
To justify this, we compare against numerical evaluations of the microcanonical SFF for an ensemble of general $\upbeta$ Gaussian random matrices.

\section{Topological gravity for arbitrary $\upbeta$}\label{sec:TopGrav_arbBeta}
The object of this section is to study the behaviour of topological gravity for any $\upbeta$, not restricted to the unitary ($\upbeta=2$) or orthogonal ($\upbeta=1$) cases as it was done in the previous works on this subject. In order to perform this computation in the generality needed, we build on the formalism used in \cite{Weber2024} to compute correlation functions of topological gravity for the case of $\upbeta=1$. The idea there, and thus also here, is to use the duality of topological gravity with a double-scaled matrix model with the Airy density of states at genus $0$, i.e. $\rho^{\text{Airy}}_0(E)=\frac{1}{2\pi}\sqrt{E}$. Consequently, the computation of some correlation function in the specified matrix model suffices, to know the respective correlation function in topological gravity.\\
To compute the $n$-point correlation function of e.g. resolvents in a matrix model (which then determine all other correlation functions by suitable integral transforms) it is necessary to provide the \textit{spectral curve}, derived from the genus $g=0$ contribution to the density of states and the symmetry class one wishes to consider. As we explained in the introduction, this symmetry class is fully classified by the exponent $\upbeta$ of the Vandermonde determinant when rewriting the matrix integral as an eigenvalue integral. 
\TW{The generalisation beyond the standard Wigner-Dyson classes is thus straightforward in this representation. It is given by \cref{eq:Z[fg]} including the dependence on the two functions $f$ and $g$. They are defined in the introduction such that they coincide with the identity for the three Wigner-Dyson values of $\upbeta$. Notably, in this representation the dependence on $g$ can be absorbed into the potential by defining $\Tilde{V}(x)\coloneqq\frac{g\qty(\upbeta)}{f\qty(\upbeta)}V(x)$. Our approach to compute correlation functions will use the so-called loop equations, which derive from the eigenvalue integral representation. In fact, their derivation for a variable Dyson index $\upbeta$ in the sense of \cref{eq:Z} has been comprehensively reviewed in \cite{Stanford2019}, i.e.\ we can use the results of \cite{Stanford2019} with replacements $\upbeta\to f(\upbeta)$ and $V\to \Tilde{V}$. The dependence on the potential $V$ is included into the spectral curve in the loop equations formalism enabling the definition of the matrix model just by using this object instead of the potential. Hence, when defining the generalised matrix integral in this way, the only remainder of the generalisation, \cref{eq:Z[fg]}, is the function $f$. Thus, one can perform all computations using $f=\text{id}_{\mathbb{R}_+}$ and then uniquely determine the result for an arbitrary choice of $f$ by mapping $\upbeta\to f(\upbeta)$. Therefore, we will use throughout the rest of the work $f=\text{id}_{\mathbb{R}_+}$, thereby keeping the possibility to introduce another choice for $f(\upbeta)$ afterwards.}

\TW{
Having clarified how to generalise to arbitrary Dyson index, we will proceed with the actual method we will use to compute the perturbative expansion of correlation functions, the loop equations. One important step of \cite{Weber2024}, necessary for this computation, was to take the loop equations as derived in \cite{Stanford2019} and put them into a convenient form by transforming to double-cover coordinates. We very briefly recall the necessary results in the following section, details and the derivation are given in \cite{Weber2024}.}
\subsection{Recap: The perturbative loop equations in double-cover coordinates}\label{sec:Recap_LE}
The perturbative loop equations are a way to compute perturbatively the correlation functions of $n$-point resolvents (cf. \cref{def:resolvents}) of a matrix model with a given leading-order density of states $\rho_0(E)$. By a perturbative computation of the correlation function of these objects we mean computing the coefficients of the topological expansion of the correlations functions given in \cref{eq:R_top_exp}. Note, that in the following we drop the superscript indicating the resolvent to be computed from a matrix model as this is apparent.
Here, we already restricted ourselves to the case of interest, i.e. double-scaled one-cut matrix models, meaning that the support of $\rho_0(E)$ is $[0,\infty)$ and that the size of the matrix $N$ usually appearing in the topological expansion of matrix model correlation functions is replaced by $e^{S_0}$ by the double-scaling procedure \cite{Saad2019,Stanford2019}.

As a first step in this computation, one has to compute the spectral curve $y(x)$ for the matrix model using \cite{Saad2019}
\begin{align}
    \lim_{\epsilon\to 0} y(x\pm i \epsilon)=\mp i \pi \rho_0(x).\label{eq:def_spectralCurve}
\end{align}
Having found the spectral curve in ''normal`` coordinates, one finds that the spectral curve has a cut precisely coinciding with the support of the leading-order density of states. For the structure of the cut that we required here, one can thus simplify the solution of the loop equations by going over to \textit{double-cover} coordinates $z\in \mathbb{P}^1$ defined via $x=-z^2$. In these coordinates, as shown in \cite{Weber2024} based on \cite{Stanford2019}, the loop equations, or rather the recursive prescription to compute the contributions to the resolvent correlation functions arising from them, can be written as
\begin{align}
\begin{aligned}\label{eq:RecursionRInt_z}
    R^\upbeta_g(-z^2,I)=\frac{1}{2\pi i z}\oint_{\qty[-i \infty+\epsilon,i\infty+\epsilon]}\frac{{z'}^2\dd z'}{{z'}^2-z^2}\frac{1}{y(-{z'}^2)}F^\upbeta_g(-z'^2,I),
\end{aligned}
\end{align}
with $\epsilon>0$ and
\begin{align}\label{eq:F_g(z)}
    \begin{aligned}
			F^\upbeta_g(-z^2,I)\coloneqq&\frac{\upbeta-2}{\upbeta}\frac{1}{-2z}\partial_z R^\upbeta_{g-\frac 12}(-z^2,I)+R^\upbeta_{g-1}(-z^2,-z^2,I)\\
					&+\sum'_{I\supseteq J, h} R^\upbeta_h(-z^2,J)R^\upbeta_{g-h}(-z^2,I\backslash J)\\
					&+2\sum_{k=1}^{n}\qty[R^\upbeta_0(-z^2,-z^2_k)+\frac 1 \upbeta \frac{1}{\qty(z_k^2-z^2)^2}]R^\upbeta_{g}(-z^2,I\backslash\qty{-z^2_k}),
			\end{aligned}
\end{align}
where $\sum'$ is a notation for excluding $R_0(-z^2)$ and $R_0(-z^2,-z^2_k)$ from the sum. Excluded from this procedure are some cases where $g=0$ that {require special treatment} (see \cref{sec:Proof_Relation_R_0} for a recursion valid for all cases). In the following, to abbreviate the notation, we denote the dependence on e.g. $-z^2$ by just a dependence on $z$, keeping in mind the true {quadratic dependence}. Computing the resolvents thus boils down to evaluating a contour integral along the closed curve $[\epsilon-i \infty,\epsilon+i \infty]\in\mathbb{P}^1$ that can be evaluated by the residue theorem.

\subsection{Topological expansion of the $\upbeta$ Airy model resolvents}\label{sec:Top_exp_upbeta_resolvents}
Having recalled the procedure to compute resolvents of a matrix model of the type we are interested in we specialize to the case of interest of this work, the Airy model. For this, we first solve \cref{eq:def_spectralCurve} for $\rho^{\text{Airy}}_0(E)=\frac{1}{2\pi}\sqrt{E}$, finding
\begin{align}
\begin{aligned}
    y^{\text{Airy}}(x)&=\frac{\sqrt{-x}}{2},\\
    \implies y^{\text{Airy}}(z)&=\frac 12 z.
\end{aligned}
\end{align}
Having found this, we are nearly ready to compute the topological expansion of the resolvents for the Airy model. However, we have to treat first the special cases of $g=0$.\\
The case of $n=2$ was already considered in \cite{Weber2024,Stanford2019} resulting in 
\begin{align}
    R^\upbeta_0(z_1, z_2)&=\frac{1}{2\upbeta}\frac{1}{z_1 z_2 \qty(z_1+z_2)^2}=\frac{1}{\upbeta}R^1_0(z_1,z_2).\label{eq:R_0_2_beta}
\end{align}
The case of $n=3$ is considered in \cite{Weber2024} and yields
\begin{align}
    \begin{aligned}
        R^\upbeta_0\qty(z_1,z_2,z_3)&=\frac{1}{2\pi i z_1}\oint_{i\mathbb R +\epsilon}\frac{z'^2\dd z'}{{z'}^2-z_1^2}\frac{2}{y(z')}\\
        &\qty[R^\upbeta_0(z',z_2)R^\upbeta_0(z',z_3)+\frac{1}{\upbeta}\qty(\frac{R^\upbeta_0(z',z_3)}{\qty(z_2^2-{z'}^2)^2}+\frac{R^\upbeta_0(z',z_2)}{\qty(z_3^2-{z'}^2)^2})]\\
        &=\frac{1}{\upbeta^2}R^1_0\qty(z_1,z_2,z_3),\label{eq:R_0_3_upbeta}
    \end{aligned}
\end{align}
where the second line follows by using \cref{eq:R_0_2_beta}. From these two cases one can already suspect that 
\begin{align}\label{eq:relation_genus0}
    R^\upbeta_0\qty(I)=\frac{1}{\upbeta^{n-1}}R^1_0\qty(I),
\end{align}
which is a generalisation of the relation of the $\upbeta=1$ with the $\upbeta=2$ resolvents of \cite{Stanford2019}. We prove this statement, actually the slight generalisation to the one-cut case, in \cref{sec:Proof_Relation_R_0}. This relation is very useful insofar as by the algorithm given in \cite{Do2009} one has a very quick way to compute the orientable genus $0$ WP volumes of $n$ boundaries from which one can compute the $R^2_0(I)$ that in turn determine the contributions for $\upbeta=1$ and thus by the above relation that of general $\upbeta$.\\
Having studied the special cases, we can turn to some examples for resolvents of non-zero genus.
\paragraph{$\mathbf{g=\frac 12,n=1}$} For this case we find
\begin{align}
    F^\upbeta_{\frac12}\qty(z)=\frac{\upbeta-2}{\upbeta}\frac{1}{-2z}\partial_z R^\upbeta_0\qty(z)\rightarrow\frac{\upbeta-2}{\upbeta}\frac{1}{-2z}\partial_z y(z)=\frac{2-\upbeta}{\upbeta}\frac{1}{4 z}=\frac{2-\upbeta}{\upbeta}F^1_{\frac12}\qty(z),\label{eq:F_12_1_upbeta}
\end{align}
where the replacement of $R_0(-z^2)$ by the spectral curve is justified, as they differ only by analytic terms, which vanish under the following contour integration. Thus, by \cref{eq:RecursionRInt_z} we find
\begin{align}
    \begin{aligned}
        R^\upbeta_{\frac 12}\qty(z)&=\frac{2-\upbeta}{\upbeta}\frac{1}{2\pi i z}\int_{i\mathbb{R}+\epsilon}\frac{{z'}^2 \dd{z'}}{{z'}^2-{z}^2}\frac{2}{z'}\frac{1}{4z'}
        =\frac{2-\upbeta}{\upbeta}\frac{1}{2 z}\underset{z'=-z}{\Res}\frac{1}{\qty(z'-z)\qty(z'+z)}\\
        &=\frac{\upbeta-2}{\upbeta}\frac{1}{4z^2},
    \end{aligned}
\end{align}
which reproduces the result from \cite{Weber2024} upon setting $\upbeta=1$ and vanishes when setting $\upbeta=2$, coinciding with the expectation for this symmetry class that is dual to orientable manifolds that cannot have genus $\frac 12$ or any non-integer genus, for that matter. 
\paragraph{$\mathbf{g=\frac 12, n=2}$} Computing, again, first the relevant $F$ we find
\begin{align}
\begin{aligned}
    F^\upbeta_\frac 12 (z',z_2)&=\frac{2-\upbeta}{\upbeta}\frac{1}{2z'}\partial_{z'}R^\upbeta_0(z',z_2)+2 R^\upbeta_\frac 12 (z')\qty[R^\upbeta_0(z',z_2)+\frac{1}{\upbeta\qty({z'}^2-z_2^2)^2}]\\
    &=\frac{2-\upbeta}{\upbeta^2}F^1_\frac 12 (z',z_2),
\end{aligned}
\end{align}
which directly implies
\begin{align}
    R^\upbeta_\frac 12 (z_1,z_2)=\frac{2-\upbeta}{\upbeta^2} R^1_\frac 12 (z_1,z_2)=\frac{2-\upbeta}{\upbeta^2}\frac{z_1^4+3 z_2 z_1^3+3 z_2^2 z_1^2+3 z_2^3 z_1+z_2^4}{2 z_1^4 z_2^4 \left(z_1+z_2\right){}^3}.
\end{align}

\paragraph{$\mathbf{g=1,n=1}$} Here, one finds from \cref{eq:F_g(z)}
\begin{align}
\begin{aligned}
    F^\upbeta_{1}(z')&=\frac{2-\upbeta}{\upbeta}\frac{\partial_{z'} R^\upbeta_{\frac{1}{2}}(z')}{2 z'}+R^\upbeta_{0,2}(z',z')+R^\upbeta_{\frac{1}{2}}(z') R^\upbeta_{\frac{1}{2}}(z')\\
    &=\qty[\qty(\frac{2-\upbeta}{\upbeta})^2\qty(\frac{1}{4}+\frac{1}{16})+\frac{1}{8\upbeta}]\frac{1}{z'^4}\\
    &=\qty[\frac{\qty(2-\upbeta)^2}{\upbeta^2}\frac 5 4+\frac{1}{2\upbeta}]\frac{1}{4z'^4}.
\end{aligned}
\end{align}
Thus, one can compute the resolvent as 
\begin{align}
\begin{aligned}
    R_{1}\qty(z)&=\qty[\frac{\qty(2-\upbeta)^2}{\upbeta^2}\frac 5 4+\frac{1}{2\upbeta}]\frac{1}{2\pi i z}\int_{i\mathbb{R}+\epsilon}\frac{{z'}^2 \dd{z'}}{{z'}^2-{z}^2}\frac{2}{z'}\frac{1}{4{z'}^4}\\
    &=-\qty[\frac{\qty(2-\upbeta)^2}{\upbeta^2}\frac 5 2+\frac{1}{\upbeta}]\frac{1}{4z}\underset{z'=z}{\Res}\frac{ \dd{z'}}{\qty(z'-z)\qty(z'+z){z'}^3}\\
    &=-\qty[\frac{\qty(2-\upbeta)^2}{\upbeta^2}\frac 5 2+\frac{1}{\upbeta}]\frac{1}{8z^5}.
\end{aligned}
\end{align}
This is the first occurrence of a resolvent being split into several parts, each associated with either the $\upbeta=2$ or the other two Wigner-Dyson classes. The distinction can be made by noticing that there is a part of the result that is vanishing upon going to the unitary symmetry class. On the gravity side of the duality one can think of this contribution as arising purely from orientable manifolds as those are the sole contributions to the gravitational path integral when choosing the unitary symmetry class for the dual matrix model. Going over to the case of unorientable manifolds in the path integral, i.e. $\upbeta=1$ one can see that the ``orientable'' contribution has doubled with respect to the orientable case which makes sense geometrically as one counts every orientable manifold twice due to the two possibilities to orient it. Thus, one can uniquely define the ``orientable'' contribution to the resolvent from the general $\upbeta$ result as the non-vanishing contribution in the case of $\upbeta=2$. This would suggest that every other contribution can be associated to a purely unorientable sector of the gravitational path integral.

To investigate this further, we compute the resolvents up to $g=4, n=1$ which notably involves the two-boundary resolvents up to $g=\frac 7 2$, some of which can be looked up in \cref{sec:Coll_Resolvents} and all of which are collected in the supplementary material. The aim of this investigation is to find a general structure of the $\upbeta$ dependence of the resolvents.

\subsection{General structure of the $\upbeta$ Airy model resolvents}\label{sec:Gen_Struct}
The search for an underlying structure of correlation functions is often simplified by symmetries of the considered theory. In the present case such a symmetry would be given by the invariance of the contributions to the topological expansion of the multi-point resolvents under certain transformations of $\upbeta$. The guiding example of how such a transformation might look like is given by the well known fact in the study of matrix models that the contributions to the resolvents of the orthogonal and symplectic symmetry class are directly connected by the relation (\cite{Stanford2019})
\begin{align}
    R_g^{1}(I)=(-1)^{2g}2^{2\qty(g+n-1)}R_g^{4}(I).
\end{align}
In fact, this relation is only an example of a more general invariance of the matrix models rooted in the invariance of the integral definition of the correlation functions under $\qty(\upbeta,N)\leftrightarrow\qty(\frac 4 \upbeta, -\frac{N\upbeta}{2})$ (e.g \cite{Eynard2018} and references therein). At the level of resolvents, we show in \cref{sec:Proof_Relation_R} that for one-cut matrix models the relation above generalises to
\begin{align}\label{eq:Rel_beta_4_beta}
    R^{\frac 4 \upbeta}_g(I)=\qty(-1)^{2g}\qty(\frac \upbeta 2)^{2\qty(g+\abs{I}-1)}R_g^{\upbeta}(I).
\end{align}
This relation now allows one to decompose the $\upbeta$ dependence into invariant parts under $\upbeta\rightarrow \frac{4}{\upbeta}$. To make this more precise, it is useful to generically decompose the genus $g$ contribution to the $n$ boundary resolvent as
\begin{align}\label{eq:R_decomp_gen}
    R^{\upbeta}_g(I)=\frac{1}{{\upbeta}^{2g+n-1}}\sum_{i=1}^k\mathcal{P}_{i}(\upbeta)g_i(I)
\end{align}
with the $\mathcal{P}_i$ being polynomials, the $g_i$ denoting the dependence on the $z$-variables that however is of {secondary} importance here, and $k\in \mathbb{N}$. This can be motivated for example by looking at the example of a not-decomposed resolvent with more complicated $\upbeta$ dependence than the ones presented so far at the beginning of \cref{sec:Coll_Resolvents} but can also be seen rather directly by looking at the recursion \cref{eq:RecursionRInt_z} used to compute the resolvents. Considering and comparing more results for different genera and numbers of boundaries one finds that the maximal order of the polynomials being decomposed into the $\mathcal{P}_i$ is given by $2g$, \footnote{This one can prove then by induction, using the explicit form of $F^\upbeta_g(z,I)$ in \cref{eq:F_g(z)}. Of course, it is also a corollary of our proof of the general structure of the resolvents in terms of $\upbeta$ in \cref{sec:Proof_GenStruct}.} i.e. one is looking for a choice of basis for the $2g+1$-dimensional space of polynomials of degree $2g$. Furthermore, we observe that up to inverse powers of $2$, which can be moved to the $g_i$, the coefficients of the polynomials are positive integers (or vanish)\footnote{The emergence of factors of 2 in the denominator of the z-dependent part is expected at least for the Wigner-Dyson cases. This is due to the representation of the resolvents in terms of ribbon graphs naturally producing factors of $2z_k$ in the denominator arising from propagators with sides of the same label.}. While the whole of the resolvent transforms according to \cref{eq:Rel_beta_4_beta} upon $\upbeta \rightarrow \frac 4 \upbeta$, for a generic choice of the $\mathcal{P}_{i}$ this is not true for the individual summands. This property, that the individual summands transform according to \cref{eq:Rel_beta_4_beta}, is now what we require for the sought for invariant decomposition.

Using this definition, one quickly finds that {demanding for} the decomposing polynomials being invariant amounts to 
\begin{align}
    \mathcal{P}_{i}\qty(\frac 4 \upbeta)=\qty(-1)^{2g}\qty(\frac{2}{\upbeta})^{2g}\mathcal{P}_{i}(\upbeta).
\end{align}
To see that this is a quite restrictive property, we consider an arbitrary polynomial with integer coefficients of degree $k$, i.e. given by $\mathcal{P}_{\vec{b}}(\upbeta)\coloneqq\sum_{i=0}^k b_i \upbeta^i$ with $\vec{b}\in\mathbb{Z}^k$. One directly finds
\begin{align}
    \mathcal{P}_{\vec{b}}\qty(\frac{4}{\upbeta})=\qty(-1)^{2g}\qty(\frac 2 \upbeta )^{2g}\sum_{i=2g-k}^{2g}b_{2g-i}\qty(-1)^{2g}2^{2(g-i)}\upbeta^i.
\end{align}
Thus, the polynomial is invariant iff
\begin{align}
    \underset{i\in\qty[0,k]}{\forall} b_i =(-1)^{2g} 2^{2(g-i)}b_{2g-i}.
\end{align}
From this, one can infer two {facts}. First, for the polynomial to be invariant, it has to hold that $k\leq 2g$. This is in correspondence with our observation that the maximal occurring polynomial order is indeed $2g$. Second,
choosing the $b_i$ for $i \in \qty[0,g]$ in the integer genus and $i\in \qty[0,g-\frac 12]$ in the half-integer genus case fixes the other $b_i$. This implies that the space of invariant polynomials is $g+1$ dimensional in the integer genus case and $g+\frac 12$ dimensional in the half-integer genus case. Consequently, finding the required number of invariant basis polynomials fixes a complete basis for the space of invariant polynomials occurring for genus $g$.

Starting with the most obvious option, we first try $\mathcal{P}^m_i(x)\coloneqq x^i$ for which one directly finds
\begin{align}
    \mathcal{P}^m_i\qty(\frac 4 \upbeta)=\qty(\frac{2}{\upbeta})^{2i} \upbeta^i.
\end{align}
This implies that for half-integer genus this can't be invariant and for integer genus it's only invariant if $i=g$. Thus, at most one element of the 
$\mathcal{P}^m_i(x)$ can contribute. A more elaborate choice that has the chance to be invariant would be to choose $\vec{a}\in\mathbb{Z}^m,\vec{n}\in\mathbb{N}^m$ with $m\in \mathbb{N}$ and consider $\prod_{i=1}^m\qty[\qty(a_i-\upbeta)\qty(\frac{4}{a_i}-\upbeta)]^{n_i}$. Here, the common exponent $n_i$ has been chosen as this is the only way the expression can reproduce itself upon transforming $\upbeta$. Adding a monomial dependence leads to the candidate expression
\begin{align}
    \mathcal{P}^c_{\vec{a},\vec{n},k}(\upbeta)\eqqcolon \upbeta^k \prod_{i=1}^m\qty[\qty(a_i-\upbeta)\qty(\frac{4}{a_i}-\upbeta)]^{n_i}.
\end{align}
For this choice of polynomials one finds
\begin{align}
    \mathcal{P}^c_{\vec{a},\vec{n},k}\qty(\frac 4 \upbeta)=\qty(\frac{2}{\upbeta})^{2\qty(k+\sum_{i=1}^m n_i)}\mathcal{P}_{\vec{a},\vec{n},k}(\upbeta),
\end{align}
which is invariant iff $k+\sum_{i=1}^m n_i=g$ for integer genus and never invariant for half-integer genus.
In fact, one can narrow down the choice of the $a_i$ to the cases of $2$ and $1$ since for all other cases\footnote{Except, of course, for $a_i=4$ which however yields the same basis polynomial as 1 and is thus excluded.}  there would appear coefficients that are not integers, in contrast to our assumption on the coefficients to be purely from $\mathbb{N}$. In the case of $a_i=2$, $(a_i-\upbeta)$ can already be made invariant by itself, thus we allow for it to appear without a ``partner''. This results in the polynomials
\begin{align}
    \mathcal{P}^i_{a,b,c}(\upbeta)\coloneqq\upbeta^a \qty(2-\upbeta)^b\qty[\qty(1-\upbeta)\qty(4-\upbeta)]^c.
\end{align}
Here, we find
\begin{align}
    \mathcal{P}^i_{a,b,c}\qty(\frac 4 \upbeta)=\qty(-1)^b\qty(\frac 2 \upbeta)^{2(a+c)+b}\mathcal{P}^i_{a,b,c}(\upbeta),
\end{align}
which can {solve the invariance condition for both the integer and half-integer case}. We further decrease the number of degrees of freedom for this choice of basis by noting that $\qty(1-\upbeta)\qty(4-\upbeta)=\qty(2-\upbeta)^2-\upbeta$ which implies that as it is chosen now, {there is an} overcounting. This can be avoided by choosing $b$ to be the minimal value compatible with the invariance condition. Thus we set $b=1$ for the case of half-integer genus, which in turn implies $a+c=g-\frac 12$ being equivalent to invariance. For integer genus we could set $b=0$, but for reasons that will be clear below we set $b=2$, implying $a+c=g-1$ for the basis entry to be invariant.

Now one can {see}, that for the half-integer genus case this already yields the required $g+\frac 12$ basis polynomials while for integer genus one finds $g$ from this choice which yields the required $g+1$ basis polynomials upon including $\mathcal{P}_g^m$. These basis polynomials are indeed linearly independent as {they all have different degrees}.

Putting everything together, we find
\begin{align}
    R^\upbeta_g(I)=\frac{1}{\upbeta^{2g+n-1}}\left(\mathcal{R}_g^0(I)\upbeta^g+(2-\upbeta)^2\sum_{i=1}^g\mathcal{R}_g^i(I)\upbeta^{i-1}((1-\upbeta)(4-\upbeta))^{g-i}\right)\label{eq:Resolvent_beta_Integer},
\end{align}
for integer $g$ and
\begin{align}
    R^\upbeta_g(I)=\frac{1}{\upbeta^{2g+n-1}}\left((2-\upbeta)\sum_{i=1}^{g+\frac 12}\mathcal R_g^i(I)\upbeta^{i-1}((1-\upbeta)(4-\upbeta))^{g+\frac 12 -i}\right)\label{eq:Resolvent_beta_halfInteger},
\end{align}
for half-integer $g$, where the $\mathcal R_g^i$ do not depend on $\upbeta$. 

Before interpreting this structure, we present a few examples of resolvents being decomposed in the manner we propose to illustrate less abstractly that it works. For the method we use to decompose the expressions, see \cref{sec:split_upbeta}. Starting with $n=1$ we find
\begin{align}
    R_{1/2}^\upbeta(z)=&-\frac{2-\upbeta}{4 z^2\upbeta},\\
    R_{1}^\upbeta(z)=&-\frac{5 (2-\upbeta)^2}{16 \upbeta^2 z^5}-\frac{1}{8 \upbeta z^5},\\
    R_{3/2}^\upbeta(z)=&-\frac{15 (1-\upbeta) (4-\upbeta) (2-\upbeta)}{16 \upbeta^3 z^8}-\frac{2 (2-\upbeta)}{\upbeta^2 z^8},\\
    R_{2}^\upbeta(z)=&-\frac{1105 (1-\upbeta) (4-\upbeta) (2-\upbeta)^2}{256 \upbeta^4 z^{11}}-\frac{3465 (2-\upbeta)^2}{256 \upbeta^3 z^{11}}-\frac{105}{64 \upbeta^2 z^{11}},\\
    R_{5/2}^\upbeta(z)=&-\frac{1695 (1-\upbeta)^2 (2-\upbeta) (4-\upbeta)^2}{64 \upbeta^5 z^{14}}-\frac{9067 (1-\upbeta) (2-\upbeta) (4-\upbeta)}{64 \upbeta^4 z^{14}}\nonumber\\
    &-\frac{160 (2-\upbeta)}{\upbeta^3 z^{14}},
\end{align}
from which we can clearly see that the decomposition works. Going to $n=2$ we present the three lowest genus cases where the decomposition contains more than one term, i.e. $g\in\{1,\frac32,2\}$.

\begin{align}
\begin{aligned}
        R^{\upbeta}_1(z_1,z_2)&=\frac{5 z_1^4+3 z_1^2 z_2^2+5 z_2^4}{8 \upbeta^2 z_1^7 z_2^7}+\frac{(2-\upbeta)^2 }{16 \upbeta^3 z_1^7 z_2^7 (z_1+z_2)^4}\times\left(25 z_1^8+100 z_1^7 z_2\right .\\
    &\left .+165 z_1^6 z_2^2+176 z_1^5 z_2^3+184 z_1^4 z_2^4+176 z_1^3 z_2^5+165 z_1^2 z_2^6+100 z_1 z_2^7+25 z_2^8\right),
    \end{aligned}
\end{align}
\begin{align}
    \begin{aligned}
        R^{\upbeta}_{\frac 3 2}(z_1,z_2)=&\frac{(2-\upbeta )}{16 \upbeta ^3 z_1^{10} z_2^{10} \left(z_1+z_2\right){}^5}\times \left(256 z_1^{12}+1280 z_2 z_1^{11}+2752 z_2^2 z_1^{10}+3590 z_2^3 z_1^9+\right .\\
        &\left .+3710 z_2^4 z_1^8+3739 z_2^5 z_1^7+3750 z_2^6 z_1^6+ (z_1\leftrightarrow z_2)\right)\\
        &+\frac{(1-\upbeta ) (2-\upbeta ) (4-\upbeta ) }{16 \upbeta ^4 z_1^{10} z_2^{10} \left(z_1+z_2\right){}^5}\times \left(120 z_1^{12}+600 z_2 z_1^{11}+1290 z_2^2 z_1^{10}+1700 z_2^3 z_1^9+\right .\\
        &\left .+1810 z_2^4 z_1^8+1865 z_2^5 z_1^7+1866 z_2^6 z_1^6+ (z_1\leftrightarrow z_2)\right),
    \end{aligned}
\end{align}

\begin{align}
\begin{aligned}
        R_2^\upbeta\left(z_1,z_2\right)=&\frac{35 \left(33 z_1^{10}+27 z_1^8 z_2^2+29 z_1^6 z_2^4+29 z_1^4 z_2^6+27 z_1^2 z_2^8+33 z_2^{10}\right)}{64 \upbeta^3 z_1^{13} z_2^{13}}\\
    &+\frac{5 (1-\upbeta) (2-\upbeta)^2 (4-\upbeta)}{256 \upbeta^5 z_1^{13} z_2^{13} (z_1+z_2)^6}\\
    &\times \left(2431 z_1^{16}+14586 z_1^{15} z_2+38454 z_1^{14} z_2^2+61322 z_1^{13} z_2^3\right.\\
    &+72455 z_1^{12} z_2^4+76032 z_1^{11} z_2^5+77730 z_1^{10} z_2^6+78412 z_1^9 z_2^7\\
    &\left .+78756 z_1^8 z_2^8+ (z_1\leftrightarrow z_2) \right)\\
    &+\frac{(2-\upbeta)^2}{256 \upbeta^4 z_1^{13} z_2^{13} (z_1+z_2)^6}\\
    &\times(38115 z_1^{16}+228690 z_1^{15} z_2+602910 z_1^{14} z_2^2+957602 z_1^{13} z_2^3\\
    &+1115707 z_1^{12} z_2^4+1146816 z_1^{11} z_2^5+1157346 z_1^{10} z_2^6+1160588 z_1^9 z_2^7\\
    &+1161156 z_1^8 z_2^8+ (z_1\leftrightarrow z_2)).
    \end{aligned}
\end{align}
Again, as it should be due to our above reasoning, the decomposition works and for these examples its usefulness is apparent when comparing to the non-decomposed resolvent presented in \cref{sec:Coll_Resolvents}. To give another justification for the decomposition to be generic, we show in \cref{sec:Proof_GenStruct} that it is a ``fixed point'' of the loop equations in the sense of it being reproduced for all contributions to the topological expansion of $n$-boundary resolvents if present for the input to the loop equations. As this is the case, as seen from \cref{eq:relation_genus0}, this shows the structure to be general. In fact, we show a more general statement than the presence of the structure in resolvents for topological gravity. Indeed, our proof, like the ones before, applies to the whole set of one-cut matrix models and thus shows our general structure to appear there as well. 

Having thus found the sought general structure, we can study its implications for the interpretation of the general $\upbeta$ resolvents. The first thing to note is that the three Wigner-Dyson values for the Dyson index, given by $\upbeta=1$ for the orthogonal, $\upbeta=2$ for the unitary and $\upbeta=4$ for the symplectic case, are special as seen from the vanishing of several terms of the general $\upbeta$ result upon $\upbeta$ being one of them.

Specifically for the unitary case of $\upbeta=2$ we find that, as expected, all half-integer contributions vanish and for integer genus the only contribution arises from $R^0_g$. The second observation is the reason why we chose the minimal number of occurrences of $(2-\upbeta)$ in the non-monomial basis elements to be two. Had we not, which would have been a possible choice of invariant basis as well, this would not have led to the nice interpretation of the term arising from $R^0_g$ being the only contribution at $\upbeta=2$, i.e. in the case where the resolvents are related to orientable manifolds on the gravity side of the duality.

Having done so, however makes the general structure a direct generalisation of the structure we had observed for the lowest genus resolvents in the preceding section. This specifically means that for integer genus $R^0_g$ (multiplied with a known power of $\upbeta$) can be thought of as being the part of the resolvent stemming, on the side of the gravitational path integral, from the orientable part of moduli space while the other Wigner-Dyson contribution, the $i=g$ term of the sum, arises from the purely unorientable part of moduli space. 

Beginning with the contributions to the two-point correlation function of resolvents\footnote{This is due to the one-point contributions all depending on the one variable $z$ as $\frac{1}{z^{6g-1}}$, leaving no room for distinction between orientable/unorientable from this dependence.} one can (not surprisingly) also see this split from the $z$ dependence of the $R^i_g$. This is due to the terms from $i=0$ having the dependence one is accustomed to from the unitary resolvents, i.e. no terms of the form $(z_k-z_l)^m$ in the denominator, as we prove at the end of \cref{sec:Proof_GenStruct}, while those terms appear in all the other cases, {clearly showing how they necessarily originate from the unorientable part of moduli space}.

For half-integer genus, there is of course no orientable contribution as it vanishes in the unitary case and the $i=g+\frac 12$ summand in \cref{eq:Resolvent_beta_halfInteger} is the only contribution in the other Wigner-Dyson classes making this directly connected to the moduli space of unorientable surfaces. However, the structure is a clear extension of this simple observations as there manifestly are non-Wigner-Dyson terms which shows that a $\upbeta$ matrix model is not just an interpolation between the Wigner-Dyson classes but has genuine ``general $\upbeta$'' contributions not present for the standard ensembles.

Having now completely explored the structure of the contributions to the resolvents, we conclude this section by considering in more detail their dependence on the $z_i$. It is best to start with the part of the result for which the most is known, i.e. the orientable parts $\mathcal{R}_g^0(I)$. As we discussed in \cref{sec:Background}, there is a direct link between the contributions to the topological expansion of resolvents in the Airy model with the Airy WP volumes, given by \cref{eq:V[R]}. In the orientable case, their structure is known to be given by a polynomial in the boundary lengths of combined order\footnote{By this, we mean that for each monomial constituting the polynomial the sum of powers of the individual arguments is given by this number.} $2(3g+n-3)$ (cf. \cref{eq:V_gn^0}). Inverting the relation \cref{eq:V[R]}, we find that this results in the following structure for the orientable part of the resolvent:
\begin{align}\label{eq:mathcalR^0_struct}
    \mathcal{R}^0_g(I)=\frac{P^0_{g,n}(I)}{\prod_{i=1}^n\qty(z_i)^{6g+2n-3}},
\end{align}
with the combined order of the polynomial $P^0_{g,n}$ given by $2(n-1)\qty(3g-3+n)$.

For the ``unorientable'' contributions, i.e. the $\mathcal{R}_g^i(I)$ with $i\geq 1$, the structure of the volumes is more complicated, as it is discussed in the next section. However, we can make some statements about these objects based on the results we found. Indeed, as it was expected already from the study of the $\upbeta=1$ case, in addition to the factors of $z_i$ one had in the orientable case, there are now also powers of sums of the $z_i$ in the denominator. As the individual contributions to the $R_{g}(I)$ have to be symmetric under permutation of the arguments to preserve the symmetry of their combination, in fact all possible linearly independent sums of two arguments for the given number of arguments have to appear with the same power. For the case of $n=2$, for the $\upbeta=1$ model this power was found to be $2g+n$ which we can see from our results to be reproduced by the general $\upbeta$ results for $n=2$ as well as for all other considered values of $n$. Thus, it is reasonable (and in agreement with all our results) to suspect
\begin{align}\label{eq:mathcalR_struct}
    \mathcal{R}^i_g(I)=\frac{P^i_{g,n}(I)}{\prod_{j=1}^n\qty(z_j)^{6g+2n-3}\prod_{j<k}^n\qty(z_j+z_k)^{(2g+2)}},
\end{align}
where $P^i_{g,n}$ is again a polynomial. The combined order of $P^i_{g,n}$ can be motivated by observing that before the decomposition the whole of $R_g(I)$ has the structure given by \cref{eq:mathcalR_struct}. Thus, for the orientable part to obtain the form of \cref{eq:mathcalR^0_struct}, the sums have to cancel out, meaning that for the unorientable part, their presence has to be accounted for by $P^i_{g,n}$ having a combined order that is precisely increased by the combined order of the product of sums. Since there are $\frac{n}{2}\qty(n-1)$ distinct unordered pairs, and thus distinct linearly independent sums of two arguments, to be chosen from the $n$ arguments, one finds
\begin{align}
    \begin{aligned}
        \text{comb. order}(P^i_{g,n})&=2(n-1)\qty(3g-3+n)+\frac n 2 (n-1) \qty(2g+2)\\
        &=\qty(n-1)\qty[3\qty(n-2)+g\qty(n+6)],
    \end{aligned}
\end{align}
which is in agreement with all the resolvents we have computed.

\subsection{The $\upbeta$ Airy Weil-Petersson volumes}\label{sec:AiryWP}
Having computed the resolvent for the general $\upbeta$ matrix model with the Airy spectral curve, we can now compute the objects of interest on the geometric side of the duality. The duality is given by \cref{eq:V[R]}, expressing the Airy WP Volume at genus $g$ of $n$ geodesic boundaries of lengths $b_1,\dots,b_n$, $V^\upbeta_{g(,n)}$, by the contribution to the topological expansion of the $n$-point resolvent at genus $g$. For the Wigner-Dyson classes this statement is actually the way to prove the duality by relating the recursion that is used on the matrix model side to Mirzakhani's recursion  \cite{Mirzakhani2007} for the orientable volumes \cite{Eynard2007,Saad2019} or a Mirazkhani-like recursion for the unorientable volumes \cite{Stanford2023}. For the general $\upbeta$-case, as we pointed out in the introduction, there is no geometric definition of ``intermediate'' (Airy) WP volumes in terms of intersection numbers or a moduli space integral, as far as we know. Thus, the relation can be viewed as the \textit{definition} of the general $\upbeta$ Airy WP volumes and we can study their properties through the matrix model results. Explicitly put, collecting the boundary lengths in $\vec{b}\in\mathbb{R}^n_{\geq 0}$
\begin{align}
    V^\upbeta_{g(,n)}(\vec{b})\coloneqq\mathcal{L}^{-1}\qty[R^{\upbeta}_{g}(z_1,\dots,z_n)\prod_{i=1}^n\qty(\frac{-2 z_i}{b_i}),\vec{b}].
\end{align}
The general structure of the resolvents, \cref{eq:Resolvent_beta_Integer,eq:Resolvent_beta_halfInteger}, of course, induces a structure of the same kind in the volumes i.e.
\begin{align}\label{eq:V_gen_struct}
    V^\upbeta_{g(,n)}(\vec{b})=\frac{1}{\upbeta^{2g+n-1}}\begin{cases}
        \mathcal{V}_{g,n}^0(\vec{b})\upbeta^g+(2-\upbeta)^2\sum_{i=1}^g\mathcal{V}_{g,n}^i(\vec{b})\upbeta^{i-1}((1-\upbeta)(4-\upbeta))^{g-i} \quad \text{int. } g,\\
        \qty(2-\upbeta)\sum_{i=1}^{g+\frac 12}\mathcal V_{g,n}^i(\vec{b})\upbeta^{i-1}((1-\upbeta)(4-\upbeta))^{g+\frac 12 -i} \quad \text{half int. } g.
    \end{cases}
\end{align}
The structure of the constituting $\mathcal{V}_{g,n}^i$, not surprisingly, derives from the $\mathcal{R}^i_g$. Specifically, we thus have to distinguish between the $\mathcal{V}_{g,n}^0$, only occurring for integer genus, and the other $\mathcal{V}_{g,n}^i$. This, as we remarked above, is due to the $\mathcal{R}_{g,n}^0$ being the only contributions surviving in the unitary/orientable case of $\upbeta=2$, {thus demanding} them to have the form expected for the unitary setting, while the other $\mathcal{R}_{g,n}^i$ have the structure one is accustomed to from the generic form found for the case of $\upbeta=1$ in \cite{Weber2024}. Consequently, we find for the ``orientable'' part of the volumes  which can be nicely written for $n$ boundaries \cite{Mirzakhani2007}
\begin{align}\label{eq:V_gn^0}
    \mathcal{V}_{g,n}^0(\vec{b})=\sum_{\vec{\alpha}\in \mathbb{N}_0^n}^{\norm{\vec{\alpha}}_1 = 3g-3+n}C^g_{\vec{\alpha}}\prod_{i=1}^{n}b_i^{2\alpha_i},
\end{align}
with $C^g_{\vec{\alpha}}\in\mathbb{Q}_{\geq 0}$ and totally symmetric. The coefficients are indeed determined by the results for the orientable Airy WP volumes in the literature since the volume should reduce to them in the case of $\upbeta=2$ and thus they differ only by a (known) power of two.

For the ``unorientable'' part, i.e. the rest of the contributions, this is not the case due to the emergence of Heaviside $\theta$-functions of sums of lengths complicating matters. The case easiest to present and also most relevant for the purpose of computing the SFF is that of two boundaries, where one finds \cite{Weber2024}
\begin{equation}
\mathcal{V}_{g}^i\left(b_1,b_2\right)=\mathcal V_g^{i,>}(b_1,b_2)\theta(b_1-b_2)+\mathcal V_g^{i,>}(b_2,b_1)\theta(b_2-b_1).
\end{equation}
with
\begin{equation}
\mathcal V_g^{i,>}(b_1,b_2)=\sum_{\substack{{\alpha_1,\alpha_2\in\mathbb{N}_0}\\\alpha_1+\alpha_2=6g-2}}C^{g,i}_{\alpha_1,\alpha_2}b_1^{\alpha_1}b_2^{\alpha_2},
\end{equation}
where the $C_{\alpha_1,\alpha_2}\in\mathbb{Q}_{\geq0}$ are not necessarily symmetric under $\alpha_1\leftrightarrow\alpha_2$. This seems confusing at first since the resolvents are symmetric under exchanges of the arguments. However, this is taken care of by the $\theta$-functions as explained in \cite{Weber2024}. We give the results for low genera here and refer the reader for a more complete list to appendix \ref{sec:WPV_collection} and the supplementary material.
\begin{align}
    \mathcal V^{1,>}_{1/2}(b_1, b_2)=&b_1\\
\mathcal V^{0}_{1}(b_1, b_2)=&\frac{\left(b_1^2+b_2^2\right)^2}{48}\\
\mathcal V^{1,>}_{1}(b_1, b_2)=&\frac{5 b_1^4+10 b_2^2 b_1^2+8 b_2^3 b_1+b_2^4}{96}\\
\mathcal V^{1,>}_{3/2}(b_1, b_2)=&\frac{30 b_1^7+210 b_2^2 b_1^5+175 b_2^3 b_1^4+210 b_2^4 b_1^3+105 b_2^5 b_1^2+91 b_2^6 b_1+5 b_2^7}{40320}\\
\mathcal V^{2,>}_{3/2}(b_1, b_2)=&\frac{\left(64 b_1^7+448 b_2^2 b_1^5+245 b_2^3 b_1^4+560 b_2^4 b_1^3+147 b_2^5 b_1^2+175 b_2^6 b_1+23 b_2^7\right) }{40320}
\end{align}
As a consistency check for these results one can compare the $\mathcal{V}_{g,n}^0(\vec{b})$ to the results in the literature, which we did for all the cases we computed. For the $\mathcal{V}_{g,n}^{i,>}(\vec{b})$ a consistency check is only possible for $i=g$ in the integer and $i=g+\frac 12$ in the half-integer genus case, where the result for $\upbeta=1$ (with the ``orientable'' part, i.e. $\mathcal{V}_{g,n}^0(\vec{b})$, subtracted for integer genus) should be reproduced, which we checked for all the considered cases. Those contributions for lower $i$ are purely non-Wigner-Dyson and consequently so far only accessible by the method discussed here.

For higher numbers of boundaries the amount of $\theta$-functions, necessary to preserve the permutation symmetry of boundaries, increases. Specifically, now also products of $\theta$-functions of sums of different numbers of boundary lengths appear, as one can see exemplary in the case of $(g,n)=\qty(\frac 12, 3)$ we give in \cref{sec:WPV_collection}. The computation of these volumes is of course possible and for the cases for which we computed the resolvents they can be found in the supplementary material. However, a discussion of their general structure is quite more tedious and less illuminating than the discussion for the case of two boundaries, while the structure at the level of resolvents, which we discuss above, is an immediate generalisation of the discussion for two boundaries. Consequently, we leave the implication of our findings for the $n\geq 3$ resolvents on the corresponding Airy WP volumes for future work.

\subsection{Geometrical construction of arbitrary $\upbeta$ topological gravity/JT gravity}\label{sec:Geometric_Arb_beta}

In order to get an idea, what the geometric interpretation of the arbitrary Dyson index ``moduli space'' volumes is, it is worthwhile to consider their computation in a way inspired by Mirzakhani's recursion for orientable moduli space volumes. Talking about this, amounts to briefly leaving topological gravity and going to JT gravity. Returning is easier than going there, since the topological gravity behaviour is always given by the non-divergent leading-order contribution of the full WP volumes appearing in JT gravity \cite{Saad2022,Stanford2023,Tall2024}. Geometrically, one can understand this by thinking of a surface of some genus $g$ with $n$ boundaries and enlarging its boundary lengths. Doing this, since the area of the surface is constrained by the Gauss-Bonnet theorem, the surface will increasingly look like a connection of thin strips, yielding exactly the ribbon graphs appearing in a diagrammatic discussion of the Airy model in terms of Kontsevich graphs \cite{Kontsevich1992,Blommaert2022}. Fortunately, as for topological gravity, the duality between a specific double-scaled matrix model and JT gravity is well established for the orientable as well as the unorientable case (\cite{Stanford2019,Stanford2023,Tall2024}) and consequently it is reasonable to perform the generalisation to arbitrary Dyson index, already performed above for topological gravity, also for the matrix model dual to JT gravity. In fact, since it is still a one-cut double-scaled matrix model, the recursion for the resolvents \cref{eq:RecursionRInt_z} and the following expression for $F^\upbeta_g$ (\cref{eq:F_g(z)}) remain valid while, of course, one has to insert now the JT gravity spectral curve $y^{\text{JT}}(z)=\frac{\sin\qty(2\pi z)}{4\pi }$. 
For our purpose the most interesting result, regarding the duality in the unorientable case, is the discovery and proof of a Mirzakhani-like recursion for the unorientable moduli space volumes, equivalent to the loop equations, in \cite{Stanford2023}. Plugging now the additional factors needed for the generalisation to arbitrary Dyson index of the loop equations and translating them to the Mirzakhani like relation along the lines of \cite{Stanford2023} one finds
\begin{align}
    b_1 V^{\upbeta,\text{JT}}_g(b_1,B)=&\sum_{k=2}^{\abs{B}} \frac{2}{\upbeta}\int_0^\infty b^\prime\dd{b^\prime}\qty[b_1-\mathsf{T}(b_1\rightarrow b^\prime;b_k)] V^{\upbeta,\text{JT}}_g(b^\prime,B\backslash b_k)\label{eq:Mirz_extrBdry}\\
    &\begin{aligned}\label{eq:Mirz_Cont_DiscGlue}
    &+\frac 12 \int_0^\infty b^\prime\dd{b^\prime}\int_0^\infty b^{\prime\prime}\dd{b^{\prime\prime}} \mathsf{D}(b_1,b^\prime,b^{\prime\prime})\times\\
    &\qty[V^{\upbeta,\text{JT}}_{g-1}(b^\prime,b^{\prime\prime},B)+\underset{\substack{h+h'=g\\ B_1\cup B_2=B}}{\sum^{\prime}} V^{\upbeta,\text{JT}}_{h_1}(b^\prime,B_1)V^{\upbeta,\text{JT}}_{h_2}(b^{\prime\prime},B_2)]
    \end{aligned}
    \\
    &+\frac 12 \frac{(2-\upbeta)}{\upbeta}\int_0^\infty b^\prime\dd{b^\prime}\mathsf{c}\qty(b_1;b^\prime) V^{\upbeta,\text{JT}}_{g-\frac 12}(b^\prime,B),\label{eq:Mirz_CrosscapGlue}
\end{align}
where we denote by $\overset{\prime}{\sum}$ the sum excluding the appearance of $V^{\upbeta,\text{JT}}_0(b^\prime,b^{\prime\prime})$, $V^{\upbeta,\text{JT}}_0(b^\prime)$.
Furthermore, we use the notation of \cite{Stanford2023} for the functions $\mathsf{T},\mathsf{D}$ and $\mathsf{c}$ and refer the interested reader there for their definition. It has to be noted, that the recursion needs as an input the results for $V^{\upbeta,\text{JT}}_{0,3},V^{\upbeta,\text{JT}}_{0,2},V^{\upbeta,\text{JT}}_{\frac 12,1}$ which can be computed (for $\upbeta=1$) from considerations on the JT gravity side \cite{Stanford2023} or also from the loop equations on the matrix model side which as an input still require only the spectral curve/leading-order density of states. Taking the second route for the arbitrary $\upbeta$ case, we first note that 
\begin{align}
    R_0^{\upbeta,\text{JT}}(z_1,z_2,z_3)=\frac{1}{\upbeta^2}R_0^{1,\text{JT}}(z_1,z_2,z_3),
\end{align}
by our reasoning of \cref{eq:R_0_3_upbeta}, which continues to hold due to being at the level before doing the contour integral. For the same reason, here based on \cref{eq:F_12_1_upbeta}, it holds that 
\begin{align}
    R_\frac{1}{2}^{\upbeta,\text{JT}}(z_1)=\frac{2-\upbeta}{\upbeta}R_\frac{1}{2}^{1,\text{JT}}(z_1).\label{eq:JT_crosscap}
\end{align}
Furthermore, it was already shown in \cite{Stanford2019} that for all one-cut double-scaled matrix models it holds that
\begin{align}
    R_0^{\upbeta}(z_1,z_2)=\frac{1}{\upbeta}R_0^{1}(z_1,z_2).
\end{align}
Since the integral transformation from resolvents to WP volumes is linear, these relations translates to the WP volumes and hence all input to the recursion for the arbitrary $\upbeta$ case is known. 

At this point we could iterate the matrix model recursion to find, along the lines of \cite{Tall2024}, results for some of the general $\upbeta$ WP volumes. Since the main focus of this work is, however, on topological gravity and our excursion to the JT setting is rather intended to give a geometrical understanding of the results there, it is worthwhile to focus on the geometrical interpretation of the Mirzakhani-like recursion \cref{eq:Mirz_extrBdry,eq:Mirz_Cont_DiscGlue,eq:Mirz_CrosscapGlue}. 

\begin{figure}[t]
     \centering
     \begin{subfigure}[b]{0.45\textwidth}
         \centering
         \includegraphics[width=\textwidth]{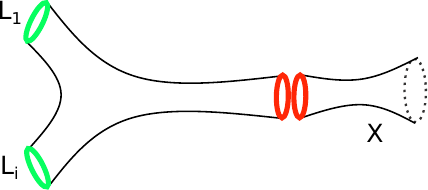}
         \caption{Glueing corresponding to \cref{eq:Mirz_extrBdry}. $X$ has genus $g$ with $n-1$ boundaries.}
         \label{fig:Mirz_extraBdry}
     \end{subfigure}
     \hfill
     \begin{subfigure}[b]{0.45\textwidth}
         \centering
         \includegraphics[width=\textwidth]{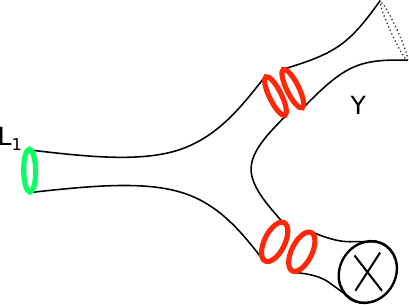}
         \caption{Glueing corresponding to \cref{eq:Mirz_CrosscapGlue}. $Y$ has genus $g-\frac 12$ and $n$ boundaries. The surface attached to the other boundary is a crosscap.}
         \label{fig:Mirz_Crosscap}
     \end{subfigure}
     \\
     \begin{subfigure}[b]{0.45\textwidth}
         \centering
         \includegraphics[width=\textwidth]{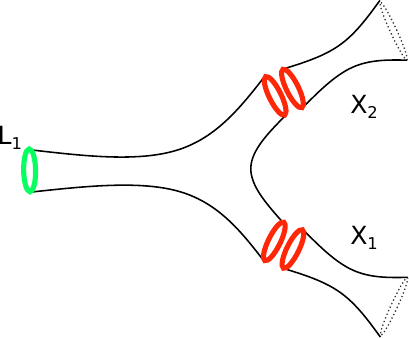}
         \caption{Glueing corresponding to \cref{eq:Mirz_Cont_DiscGlue} 
            (Disconnected part). $X_1$/$X_2$ have genera $h_1$/$h_2$ and $n_1$/$n_2$ boundaries with $h_1+h_2=g$ and $n_1+n_2=n-1$}
         \label{fig:Mirz_Disc}
     \end{subfigure}
    \hfill
         \begin{subfigure}[b]{0.45\textwidth}
         \centering
         \includegraphics[width=\textwidth]{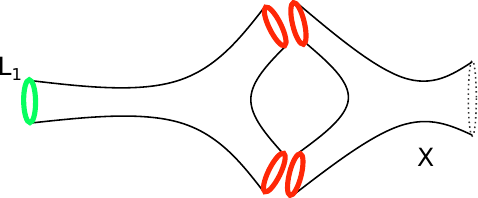}
         \caption{Glueing corresponding to \cref{eq:Mirz_Cont_DiscGlue} (Connected part). $X$ has genus $g-1$ and $n+1$ boundaries.}
         \label{fig:Mirz_Conn}
     \end{subfigure}
        \caption{Depiction of the different ``glueings'' i.e. the separation of a surface of constant negative curvature of genus $g$ and $n$ geodesic boundaries of lengths $L_1,\dots,L_n$ into a 3-holed sphere and another such surface. These separations are the same as for the case of unorientable surfaces in \cite{Stanford2023}. The only difference for our setting, \cref{eq:Mirz_extrBdry,eq:Mirz_Cont_DiscGlue,eq:Mirz_CrosscapGlue}, is, that there is an additional factor of $\frac{1}{\upbeta}$ for the glueing in case a) and an additional factor of $\frac{\qty(2-\upbeta)}{\upbeta}$ for case b).}
        \label{fig:Mirzakhani_Cuttings}
\end{figure}

For this purpose, it is useful to recall that the several integrals in this recursion (of course, also in the ``traditional'' one, found by setting $\upbeta=2$) represent certain ways to attach a 3-holed sphere (a pair of pants) to a surface of a specific genus and number of boundaries to build the desired surface. In \cref{fig:Mirzakhani_Cuttings}, we collect all $4$ ways that are possible where the option a) corresponds to \cref{eq:Mirz_extrBdry}, b) to \cref{eq:Mirz_CrosscapGlue}, d) to the sole volume and c) to the sum over the product of two volumes in \cref{eq:Mirz_Cont_DiscGlue}. Going from the setting of $\upbeta=1$, discussed in \cite{Stanford2023}, to that of arbitrary Dyson index, the only change is in the cases a) and b), where an additional factor of $\frac{1}{\upbeta}$ in case a) and $\frac{\qty(2-\upbeta)}{\upbeta}$ in case b) is introduced. Both of these factors can be understood intuitively. In case b) the factor can be thought of as arising from the attachment of the crosscap which, as seen from \cref{eq:JT_crosscap}, has a dependence on the Dyson index as $\frac{\qty(2-\upbeta)}{\upbeta}$. To understand the factor in the case a) it is useful to recall from \cite{Stanford2023}, that the additional factor 2 in \cref{eq:Mirz_extrBdry} in the $\upbeta=1$ case as compared to the orientable setting occurred due to the possibility, in the unorientable case, to glue the 3-holed sphere with or without a change of orientation for a geodesic going from the boundary of length $L_1$ to that of length $L_i$. Since there is only one possibility in the case of $\upbeta=2$, this factor has to be cancelled, which suggests that the additional factor is $\frac 1 \upbeta$. 

Iterating the $\upbeta$ dependence as determined by the recursion yields precisely the general structure of the WP volumes proven above. This is not surprising, since the arbitrary $\upbeta$ Mirzakhani-like recursion is equivalent to the loop equations used to determine the general structure. In order to get a geometric intuition for the structure, it is however instructive to look at an example, for which we choose the case of $g=\frac 3 2, n=1$. We find the following decomposition, decomposing also the manifolds glued to the 3-holed sphere, to build a surface of genus $\frac{3}{2}$ and one boundary in the Mirzakhani-like recursion
\begin{align}
\begin{aligned}
    V^{\upbeta,\text{JT}}_{\frac 3 2,1} \cong \quad &\includegraphics[width=0.3\linewidth,valign=c]{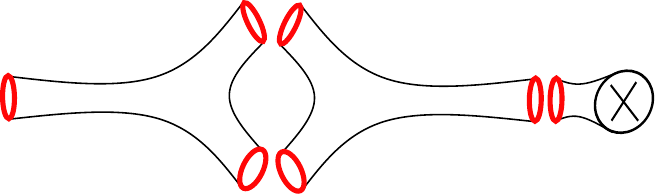}
   +\includegraphics[width=0.3 \linewidth,valign=c]{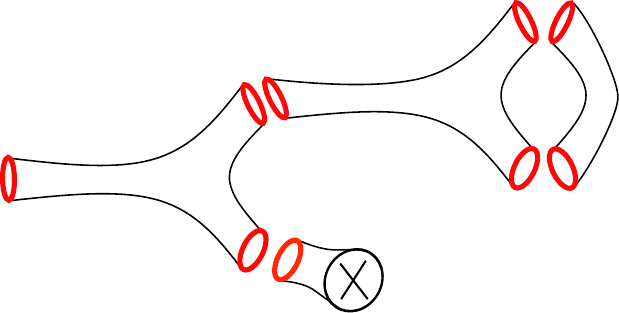}\\
   +&\includegraphics[width=0.3 \linewidth,valign=c]{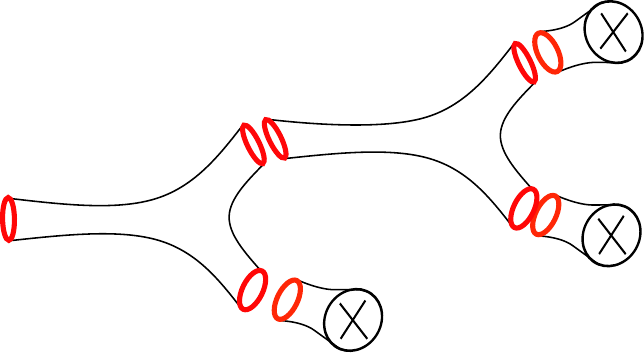}.
\end{aligned}
\end{align}
For the first line we can read off the dependence on $\upbeta$ to be $\frac{2-\upbeta}{\upbeta} \frac{1}{\upbeta}$, while for the second line we find $\qty(\frac{2-\upbeta}{\upbeta})^3$. The terms arising from the first line are thus of the general form already, while the second is not. However, we can use $(2-\upbeta)^2=(1-\upbeta)(4-\upbeta)+\upbeta$ to rewrite it as a sum of a term of the form of the first line and one of the form as in the contribution to the general structure apart from it. From this example it is now apparent that the general structure we have proven, in a sense chosen, above does not correspond to the decomposition of the volumes equivalent to that of the glueings of the Mirzakhani-like recursion. It is thus interesting to note that its distinct advantage, being the decomposition into Wigner-Dyson and non-Wigner-Dyson contributions, is consequently not a property immediately seen from the glueing construction inherent to the Mirzakhani-like recursion. 

However, due to it being the general structure of the matrix model correlation functions, one can use it to derive a structure corresponding to this decomposition. This can be done by transforming the manifestly non-Wigner-Dyson parts of the general structure of the volumes to a structure that reflects the number of inserted crosscaps via $(1-\upbeta)(4-\upbeta)=(2-\upbeta)^2-\upbeta$.
Doing this, one finds for the integer genus case
\begin{align}\label{eq:V_Mirz_int}
    V^\upbeta_{g,n}=\sum_{m=0}^g\frac{1}{\upbeta^{g+n-m-1}}\qty(\frac{2-\upbeta}{\upbeta})^{2m} \Bar{\mathcal{V}}_{g,n}^m,
\end{align}
with
\begin{align}
    \Bar{\mathcal{V}}_{g,n}^m\coloneqq(-1)^{m+g+1}\sum_{i=1}^{g-m+1}(-1)^i\binom{g-i}{m-1}\mathcal{V}^i_{g,n},
\end{align}
for $m>0$ and $\Bar{\mathcal{V}}_{g,n}^0\coloneqq\mathcal{V}_{g,n}^0$.
For the half-integer genus case these considerations yield
\begin{align}\label{eq:V_Mirz_halfint}
    V^\upbeta_{g,n}=\sum_{m=0}^{g-\frac 12}\frac{1}{\upbeta^{g-\frac 12+n-m-1}}\qty(\frac{2-\upbeta}{\upbeta})^{2m+1} \Bar{\mathcal{V}}_{g,n}^m,
\end{align}
with
\begin{align}
    \Bar{\mathcal{V}}_{g,n}^m\coloneqq(-1)^{m+g+\frac 12}\sum_{i=1}^{g+\frac 12-m}(-1)^i\binom{g+\frac 12-i}{m}\mathcal{V}^i_{g,n}.
\end{align}
Of course, one could also go the inverse way of taking this structure to arrive at the one proven above, giving a geometric argument justifying this structure is thus also one for the one proven above. We proceed by giving such a geometric argument. 

Before we go into the geometric proof of \cref{eq:V_Mirz_int,eq:V_Mirz_halfint}, we would like to remark that after this paper was accepted for publication, we became aware of \cite{Borot2013}, which actually provides the equivalent statement for resolvents. However, they have a quite different proof, do not investigate the implications for moduli space volumes and do not provide the structure, split into Wigner-Dyson and non-Wigner-Dyson parts (\cref{eq:Resolvent_beta_Integer,eq:Resolvent_beta_halfInteger}), whose precise form is of highest relevance for most of this work. Consequently, our discussion above is not ``just'' a different proof for a known statement but rather an interesting statement on its own which happens to imply, after the rewriting we performed to find \cref{eq:V_Mirz_int,eq:V_Mirz_halfint}, the statement already given in \cite{Borot2013}.

\begin{figure}[h]
    \centering
    \includegraphics[width=\linewidth]{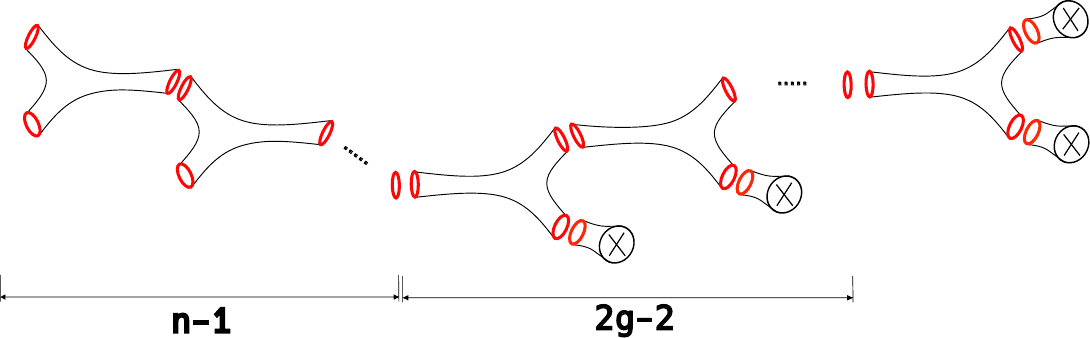}
    \caption{Possible decomposition of a surface of genus $g$ and $n$ geodesic boundaries into 3-holed spheres and crosscaps with the maximal number of crosscaps possible. The numbers below the first and second part of this construction indicate how many 3-holed spheres/3-holed spheres with an attached crosscap are needed.}
    \label{fig:Decomposition}
\end{figure}

For the geometric proof we start by explicitly constructing a surface of genus $g$ and $n$ geodesic boundaries from 3-holed spheres and crosscaps by the general $\upbeta$ Mirzakhani-like construction pictured in \cref{fig:Mirzakhani_Cuttings}. 
We start by constructing the surface with the maximum number of crosscaps, given by $2g$. A particularly nice way to do this is depicted in \cref{fig:Decomposition}, where the surface is decomposed into two parts, one containing the boundaries and one containing the crosscaps. The part containing the boundaries is built from $n-1$ 3-holed spheres, glued in the depicted way as to yield a surface of genus $0$ with $n$ external boundaries and one ``internal'' boundary to glue the genus carrying part. This is in turn constructed from two parts, first $2g-2$ copies of the depicted glueing of a 3-holed sphere and a crosscap such that the number of internal glueing boundaries is conserved, second the glueing of a final 3-holed sphere capped off by two crosscaps. This construction evidently yields a surface with $n$ geodesic boundaries and $2g$ crosscaps, hence genus $g$. It is now an easy task to read off the $\upbeta$ dependence of this surface, which is built from $n-1$ glueings \`a la \cref{fig:Mirz_extraBdry}, $2g-2$ \`a la \cref{fig:Mirz_Crosscap} and two additional crosscaps, hence yielding the total dependence
\begin{align}
    \qty(\frac{1}{\upbeta})^{n-1}\qty(\frac{2-\upbeta}{\upbeta})^{2g-2+2},
\end{align}
which is precisely the dependence obtained in \cref{eq:V_Mirz_int} and \cref{eq:V_Mirz_halfint} upon taking the maximal value for the summation index. It is useful to note here, that the summation index due to it increasing the number of ``crosscap'' factors in the general structures can be interpreted as determining the number of crosscaps in the respective contribution to the volumes as $2m$ in the case of integer genus and $2m+1$ in the case of half-integer genus.

To find the contributions with a lower number of crosscaps geometrically one can substitute two of the crosscaps at a time by a ``hole'', hence keeping the total genus constant. In the case of the final part one can do this by taking out the two crosscaps and glueing the remaining two boundaries directly to one another. In the case of the other parts one takes out two of the constituting blocks and replaces them with two 3-holed spheres as
\begin{align}
    \includegraphics[width=0.3\linewidth,valign=c]{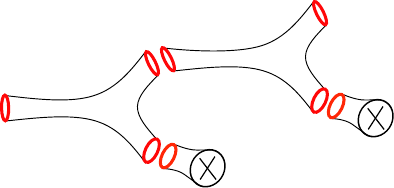}\rightarrow\includegraphics[width=0.3\linewidth,valign=c]{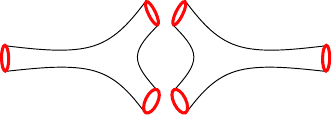}.
\end{align}
To obtain a surface with $k$ crosscaps one obviously has to do a replacement of two crosscaps $\frac 12 \qty(k_{\text{max}}-k)$ times, where $k_{\text{max}}$ denotes the maximal number of crosscaps possible which is given by $2g$. In terms of the $\upbeta$ dependence this replacement cancels out two crosscap factors and adds one $\frac{1}{\upbeta}$ in both cases, hence the dependence on $\upbeta$ of the contribution of the surface with $k$ crosscaps is given by
\begin{align}
    \qty(\frac{1}{\upbeta})^{n-1}\qty(\frac{2-\upbeta}{\upbeta})^{2g-2g+k}\qty(\frac{1}{\upbeta})^{g-\frac 12 k}=\begin{cases}
        \frac{1}{\upbeta^{g-m+n-1}}\qty(\frac{2-\upbeta}{\upbeta})^{2m} &\text{int. }g\\
        \frac{1}{\upbeta^{g-m-\frac 12 +n-1}}\qty(\frac{2-\upbeta}{\upbeta})^{2m+1} &\text{half int. }g
    \end{cases},
\end{align}
where in the last step we put, for reason of comparing with the general result, the number of crosscaps $k$ to be given by $2m$ in the integer genus and $2m+1$ in the half-integer genus case. Comparing this dependence with the general structure put as in \cref{eq:V_Mirz_int} and \cref{eq:V_Mirz_halfint} shows agreement, and consequently we have geometrically shown that all of the terms appearing in the structure arise. To show that this is all the structures that are possible we show in \cref{sec:App_Decomposition} that all decompositions of the surfaces into the components found in the recursion can be reorganised as to give the decomposition we have chosen above and thus can produce no other structure in terms of dependence on $\upbeta$ than the already discussed ones. 

Hence, one can justify the ``Mirzakhani-like'' version of the general structure of the WP volumes geometrically and thus also our other, Wigner-Dyson/non-Wigner-Dyson split, version. In particular this implies, by the reduction of JT gravity to topological gravity/Airy model in the large length limit, the general structure for topological gravity we had shown already above and thus provides a deep geometric reason for it. For JT gravity itself, the computation of the arbitrary $\upbeta$ volumes would extend beyond the scope of this work and is left for future study. A particularly interesting aspect of this study would be to observe the dependence on $\upbeta$ of the divergent parts of the WP volumes which due to their existence for all unorientable incarnations will persist also in the theory for arbitrary Dyson index. This is due to the fact that the divergent moduli space volume of the crosscaps is the ultimate reason for these divergences and it would be interesting to study the interplay of this with the splitting of the volume into contributions with well defined numbers of crosscaps as induced by the general structure in terms of $\upbeta$.

\begin{figure}[h]
    \centering
    \includegraphics[width=0.5\linewidth]{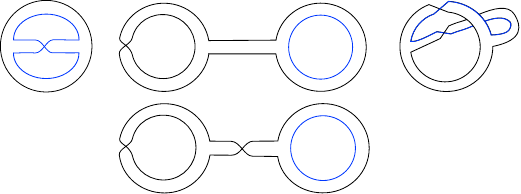}
    \caption{Diagrams contributing to $V_{\frac 12,2}$, i.e. the elements of $\Gamma_{\frac12,2}$. }
    \label{fig:Diagrams_g_12}
\end{figure}

\TW{
Regarding the geometric interpretation of our results in topological gravity, it is worthwhile to consider the generalisation of the aforementioned Kontsevich diagrammatics, recalled e.g. in \cite{Blommaert2022,Saad2022,Weber2024}. This is an equivalent, though quite more tedious, way of computing correlation functions in the matrix model dual to topological gravity. In essence, it is a diagrammatic prescription that allows the computation of the (Laplace transformed) Airy WP volumes of genus $g$ and $n$ boundaries by finding double-line diagrams (ribbons) with three-valent vertices having the same Euler characteristic and number of boundaries. Here, a boundary is defined by giving labels to both edges of all propagators and identifying those that are connected via a vertex. This gives rise to a partition of the set of edges of propagators into a finite number of subsets each of which is denoted as a boundary. In this language, one can introduce unorientable contributions by allowing the propagators to ``twist'', for an example of this cf. the diagrams in \cref{fig:Diagrams_g_12}. Specifically, it holds that\footnote{Note, that the proof of this statement requires Kontsevich's theorem relating intersection numbers with ribbon graphs (cf. \cite{Kontsevich1992}) which is unproven in the unorientable case. Hence, the generalisation of this statement to the unorientable case and beyond it is non-trivial. Regarding the functional dependence of the left and right hand side of the equation, we are however not aware of discrepancies.}
\begin{align}
    \mathcal{L}\qty[V^{\text{Airy}}_{g,n}(L_1,\dots,L_n);\qty(z_1,\dots,z_n)]
    &=\sum_{\gamma\in\Gamma_{g,n}}\frac{2^{2g-2+n}}{\abs{\text{Aut}(\gamma)}}\prod_{k=1}^{6g-6+3n}\frac{1}{z_{l(k)}+z_{r(k)}}\label{eq:V_Graphs} \, .
\end{align}
Here, $\Gamma_{g,n}$ is the set of all ribbon graphs (with labelled edges) of Euler characteristic $2-2g-n$ and $n$ boundaries containing only 3-valent vertices, $\abs{\text{Aut}(\gamma)}$ is the order of the automorphism group of the graph $\gamma$ and $l(k)$ and $r(k)$ denote the labels of the left and right edge of the $k$th propagator.}

\TW{
As the loop equations, the matrix model diagrammatics can be conveniently generalised to the setting of arbitrary Dyson index. This is done, following \cite{Eynard2018}, by defining a $\upbeta$ propagator as
\begin{align}
    \includegraphics[width=0.2 \linewidth,valign=c]{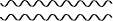}=\includegraphics[width=0.2 \linewidth,valign=c]{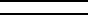}+\frac{\qty(2-\upbeta)}{\upbeta}\includegraphics[width=0.2 \linewidth,valign=c]{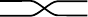},\label{eq:Prop_identity}
\end{align}
which is a result valid for all Wigner-Dyson classes that is readily generalised to the setting of arbitrary Dyson index. Fron another perspective,  this generalised propagator can be perceived as giving an additional factor of $\frac{\qty(2-\upbeta)}{\upbeta}$ for each twisted propagator within a diagram. }

\TW{
As for the definition of the matrix model using the eigenvalue integral \cref{eq:Z[fg]}, there is an ambiguity in generalising the propagator to arbitrary Dyson index here, i.e. the function of the Dyson index multiplying the twisted propagator could be any function $b(\upbeta):\mathbb{R}_+\to \mathbb{R}_+$ coinciding with $\frac{2-\upbeta
}{\upbeta}$ at the Wigner-Dyson values of $\upbeta$, and there could be a function $a(\upbeta):\mathbb{R}_+\to \mathbb{R}_+$ multiplying the untwisted propagator that assumes the value $1$ at those values. Hence, the most general way to write an extension to arbitrary Dyson index of \cref{eq:Prop_identity} would be
\begin{align}
    \includegraphics[width=0.2 \linewidth,valign=c]{Figures/beta_propagator.pdf}=a(\upbeta)\includegraphics[width=0.2 \linewidth,valign=c]{Figures/Propagator.pdf}+b\qty(\upbeta)\includegraphics[width=0.2 \linewidth,valign=c]{Figures/Propagator_twisted.pdf}.\label{eq:Prop_general}
\end{align}
To reduce this to an ambiguity dependent on only one function, we can factor out $a(\upbeta)$ and absorb it into a redefinition of $N$ due to it appearing now in front of every diagram with a power $\#\text{Edges}=6g-6+3n=-3\chi_{g,n}$. The ambiguity remains in the prefactor of the twisted propagator $\frac{b(\upbeta)}{a(\upbeta)}$.
It is convenient to rewrite this function as 
\begin{align}
    \frac{2-h(\upbeta)}{h(\upbeta)},
\end{align}
which exemplifies that one can perform all computations with choosing $h$ as the identity and then reintroduce an arbitrary function $h$ afterwards. This is precisely our procedure of dealing with the ambiguity when using the loop equations to compute correlation functions. One might expect intuitively that the choice of $h=\text{id}_{\mathbb{R}_+}$ coincides with the choice $f=\text{id}_{\mathbb{R}_+}$ in the eigenvalue integral. This is supported  by noting that the twisting of the propagator in a graph is equivalent to inserting a crosscap in the surface dual to the graph. For $f=\text{id}_{\mathbb{R}_+}$ the dependence of this on the Dyson index is given by multiplying the result for $\upbeta=1$ 
by $\frac{2-\upbeta}{\upbeta}$ (\cref{eq:F_12_1_upbeta}), coinciding precisely with the factor of the twisted propagator for $h=\text{id}_{\mathbb{R}_+}$. In this way one can also justify not adding any $\upbeta$ dependence to the prefactor of the non-twisted propagator, since it corresponds to inserting nothing additional in the dual surface. Hence, the two ambiguities are inter-related. To get coinciding results from the diagrammatics with the loop equations one has to choose $h=f$. As above, we continue by setting $h=\text{id}_{\mathbb{R}_+}$, keeping the ambiguity in mind.
}

To illustrate the idea on an example, we consider the case of $g=\frac{1}{2}$ and $n=2$, for which the Kontsevich diagrams for the $\upbeta=1$ case have been studied in \cite{Saad2022}. In general, one can show the graphs relevant for genus $g$ and $n$ boundaries to consist of $4g+2n-4$ vertices, which for the present case motivates us to consider the graphs with two vertices. In \cref{fig:Diagrams_g_12}, we give all the ribbon graphs with two boundaries one can build from two three-valent vertices and twisted/untwisted propagators. 
 
Here we note, that the graph in the first column and the graph in the first line of the second column are those given in \cite{Saad2022}, while the others do not appear there. There is a good reason why not to include them, for the case of $\upbeta=1$, being that the graph in the third column has the same contribution to the Laplace transformed Airy WP volume as that in the first column, while that in the second line and column has the same contribution as that above it. Consequently, they only contribute an additional factor of 2 which is captured by the reasoning of \cite{Saad2022}. For our case, however, it is vital to include these additional diagrams since in the general $\upbeta$ setting their contribution differs by an additional factor of $\frac{\qty(2-\upbeta)}{\upbeta}$ from the respective related graph. However, by this reasoning one can see directly that it holds that
\begin{align}
\begin{aligned}
    \mathcal{L}\qty[V^{\upbeta}_{\frac 12,2}(L_1,L_2);\qty(z_1,z_2)]&=\frac{1}{2}\qty(\frac{2-\upbeta}{\upbeta}+\frac{\qty(2-\upbeta)^2}{\upbeta^2})\mathcal{L}\qty[V^{1}_{\frac 12,2}(L_1,L_2);\qty(z_1,z_2)]\\
    &=\frac{\qty(2-\upbeta)}{\upbeta^2}\mathcal{L}\qty[V^{1}_{\frac 12,2}(L_1,L_2);\qty(z_1,z_2)].
\end{aligned}
\end{align}
This, by the linearity of the Laplace transform implies
\begin{align}
    V^{\upbeta}_{\frac 12,2}(L_1,L_2)=\frac{\qty(2-\upbeta)}{\upbeta^2}V^{1}_{\frac 12,2}(L_1,L_2),
\end{align}
which is precisely what we obtained using the loop equations\footnote{Of course, one would also get the same result counting the diagrams multiplicities and computing the orders of the respective automorphism groups, e.g. by inferring them from the orientable diagrams of the same structure.}. This gives already a good intuition, on what the generalisation to arbitrary Dyson index means geometrically. In fact, the diagrammatics suggest, that the individual contributions are still distinct orientable/unorientable objects while their relative weights differ as determined by the way the vertices are connected.

For now, however, this concludes our discussion of how to compute correlation functions in one-cut double-scaled matrix models with arbitrary Dyson index and what their generic dependence on the Dyson index is. 
The remainder of this work will now focus on leveraging this new knowledge to investigate quantum chaoticity of arbitrary $\upbeta$ topological gravity. 
\section{Quantum chaos in topological gravity for arbitrary $\upbeta$?} \label{sec:Chaos_Top_grav_arb}

As we explained in \cref{sec:Background}, the presence of chaos, according to the BGS conjecture, manifests itself in the spectral properties of a quantum system by rendering the spectral two-point function to be, for small differences of the correlated energies, of a form  that depends only on the spectral density and the system's symmetry class. This form is computed from the Gaussian matrix model for the respective symmetry class and will in the following be referred to as the RMT or ``universal'' prediction. For JT gravity, and thus also topological gravity, it is more convenient to study correlation functions of partition functions. Consequently, the probe for spectral statistics is a two-point function of partition functions, specifically with complex conjugate complex inverse temperatures (cf. \cref{def:SFF_can}), the SFF. This quantity, for large times, is related to the spectral two-point function, in fact its Fourier transform, via \cref{eq:univRMT_kappa_canonical}. Thus, a proof of quantum chaos in a variety of JT gravity/topological gravity, can be performed by showing that its late time SFF matches the predictions of RMT that is known for the Wigner-Dyson cases.

For the non-Wigner-Dyson cases however, this strategy has the problem that the RMT prediction for the spectral two-point function is not fully available and, as far as we know, is quite difficult to compute by the standard approach\footnote{By this, we mean the computation using orthonormal polynomials, given e.g. in \cite{Mehta2004}.}. Our approach provides a novel way to study the spectral two-point function analytically, albeit perturbatively, through the lens of the canonical SFF. In the following section we will begin this study by computing the late time canonical spectral form factor for arbitrary $\upbeta$.

\subsection{The canonical spectral form factor}\label{sec:SFF_loop_eq}
The contribution to the spectral form factor for a given genus $g$ \TW{and a Dyson index $\upbeta$} is 
\begin{equation}
    \kappa_{\upbeta}^g(t,\beta)=\int_0^\infty \dd{b_1}b_1 \int_0^\infty \dd{b_2}b_2 Z^t(\beta_1,b_1) Z^t(\beta_2,b_2) V^{\upbeta}_{g}(b_1,b_2),
\end{equation}
with the trumpet partition function $Z^t(\beta,b)$ given in \cref{eq:def_trumpet}, $\beta_1=\beta+it$ and $\beta_2=\beta_1^*$.
Using this, one can compute the $\kappa_\upbeta^g(t,\beta)$ from the volumes stated above which then naturally inherit their structure regarding the $\upbeta$ dependence. We are interested in the ``universal'' part of the form factor which means the behaviour at large time, i.e. times of order $e^{S_0}$. The large $t$ behaviour\footnote{By this, we mean collecting only terms that after including the factors of $e^{S_0}$ from the genus expansion, expanding square-roots as power series and introducing $\tau$ are not vanishing upon $e^{S_0}\rightarrow \infty$, i.e. all polynomial terms of minimal order $2g+1$. For the non-polynomial terms, more care is needed and the limit is performed as in \cite{Weber2024}, where more details can be found.} of the contributions to the spectral form factor, split up according to the structure in $\upbeta$ found for the resolvents/volumes, is given by \footnote{\TW{To avoid confusion, we remark that in this equation the Dyson index $\upbeta$ and the inverse temperature $\beta$ both appear and should not be confused.}}
\TW{
\allowdisplaybreaks
\begin{align}
\kappa_\upbeta^{0}(t,\beta)\rightarrow&\frac{t}{2 \pi  \beta  \upbeta},\label{eq:kappa_0_general}\\
\kappa_\upbeta^{1/2}(t,\beta)\rightarrow&-\frac{2-\upbeta}{\upbeta^2}\frac{t^2}{\sqrt{2 \pi } \sqrt{\beta }},\\
\kappa_\upbeta^{1}(t,\beta)\rightarrow&\frac{t^{3}(2-\upbeta)^{2}\left(3\log\left(\frac{2t}{\beta}\right)-8\right)}{3\pi\upbeta^{3}}-\frac{2t^{3}}{3\pi\upbeta^{2}},\\
\kappa_\upbeta^{3/2}(t,\beta)\rightarrow&\frac{(1-\upbeta)(2-\upbeta)(4-\upbeta)}{\upbeta^{4}}\left(-\frac{59t^{9/2}}{60\sqrt{2\pi}}+2\sqrt{\frac{2}{\pi}}\sqrt{\beta}t^{4}\right)\nonumber\\
&+\frac{(2-\upbeta)}{3\upbeta^{3}}\left(-\sqrt{\frac{2}{\pi}}t^{9/2}+8\sqrt{\frac{2}{\pi}}\sqrt{\beta}t^{4}\right),\\
\kappa_\upbeta^{2}(t,\beta)\rightarrow&\frac{3 t^6 (4-\upbeta) (2-\upbeta)^2 (1-\upbeta)}{32 \upbeta^5}\nonumber\\
&-\frac{\beta t^{5}(1-\upbeta)(2-\upbeta)^{2}(4-\upbeta)\left(1890\log\left(\frac{2t}{\beta}\right)-3371\right)}{945\pi\upbeta^{5}}\nonumber\\
&-\frac{\beta t^{5}(2-\upbeta)^{2}\left(60\log\left(\frac{2t}{\beta}\right)-151\right)}{15\pi\upbeta^{4}}+\frac{4\beta t^{5}}{5\pi\upbeta^{3}}\label{eq:kappa_g_2},\\
\kappa_\upbeta^{5/2}(t,\beta)\rightarrow&-\frac{t^6 (1-\upbeta)^2 (2-\upbeta) (4-\upbeta)^2 \left(2838528 \beta ^{3/2}+31282 t^{3/2}-966955 \beta  \sqrt{t}\right)}{532224 \sqrt{2 \pi } \upbeta^6}\nonumber\\
&+\frac{t^6 (1-\upbeta) (2-\upbeta) (4-\upbeta) \left(-55824384 \beta ^{3/2}+43966 t^{3/2}+14101571 \beta  \sqrt{t}\right)}{2661120 \sqrt{2 \pi } \upbeta^5}\nonumber\\
&+\frac{t^6 (2-\upbeta) \left(-512 \beta ^{3/2}+2 t^{3/2}+85 \beta  \sqrt{t}\right)}{30 \sqrt{2 \pi } \upbeta^4},\\
\kappa_\upbeta^3(t,\beta)\rightarrow&-\frac{16 \beta ^2 t^7}{21 \pi  \upbeta^4}\nonumber\\
&+\frac{t^7 (4-\upbeta)^2 (2-\upbeta)^2 (1-\upbeta)^2}{145297152 \pi  \upbeta^7}\nonumber\\
&\times \qty(-419374208 \beta ^2+2338048 t^2+290594304 \beta ^2 \log \left(\frac{2 t}{\beta }\right)-22891869 \pi  \beta  t )\nonumber\\
&-\frac{t^7 (4-\upbeta) (2-\upbeta)^2 (1-\upbeta) }{3632428800 \pi  \upbeta^6}\nonumber\\&\times \qty(68734274048 \beta ^2+164049920 t^2-36808611840 \beta ^2 \log \left(\frac{2 t}{\beta }\right)+1378872495 \pi  \beta  t)\nonumber\\
&-\frac{t^7 (2-\upbeta)^2 \left(1151 \beta ^2+2 t^2-480 \beta ^2 \log \left(\frac{2 t}{\beta }\right)\right)}{45 \pi  \upbeta^5},\\
\kappa_\upbeta^{7/2}(t,\beta)\rightarrow&-\frac{t^8 (4-\upbeta)^3 (2-\upbeta) (1-\upbeta)^3}{28229160960 \sqrt{2 \pi } \upbeta^8}\nonumber\\
&\times\left(-120444420096 \beta ^{5/2}-2521705676 \beta  t^{3/2}+61149640 t^{5/2}+47197373065 \beta ^2 \sqrt{t}\right)\nonumber\\
&+\frac{t^8 (4-\upbeta)^2 (2-\upbeta) (1-\upbeta)^2}{2258332876800 \sqrt{2 \pi } \upbeta^7}\nonumber\\
&\times\left(70798377222144 \beta ^{5/2}+435487934100 \beta  t^{3/2}+38265437192 t^{5/2}-22916260990543 \beta ^2 \sqrt{t}\right)\nonumber\\
&+\frac{t^8 (4-\upbeta) (2-\upbeta) (1-\upbeta) }{2258332876800 \sqrt{2 \pi } \upbeta^6}\nonumber\\
&\times\left(146506298425344 \beta ^{5/2}-486701235820 \beta  t^{3/2}+64361951752 t^{5/2}-37542514725263 \beta ^2 \sqrt{t}\right)\nonumber\\
&+\frac{t^8 (2-\upbeta) \left(32768 \beta ^{5/2}-252 \beta  t^{3/2}+8 t^{5/2}-6167 \beta ^2 \sqrt{t}\right)}{840 \sqrt{2 \pi } \upbeta^5}.\label{eq:kappa_7_2_general}
\end{align}
}
One can perform a quick cross-check of these results by comparing to the results for the unitary (e.g. \cite{Saad2022}) and orthogonal symmetry class (\cite{Weber2024}) by plugging the corresponding values of the Dyson index and finding
agreement\footnote{Note, that there was a typo in \cite{Weber2024} in the results for $\kappa_{\frac 5 2}$ and $\kappa_{3}$ which is corrected in the arxiv version.}. This is, of course, expected as the resolvents already agreed. However, at the level of the resolvents, there was no structural difference between the contributions occurring for $\upbeta=1$ and the purely non-Wigner-Dyson ones. For the late time SFF there is a differences as one can see from the contribution at $g=2$. There, a term is found that contributes at $t^6$ for non-Wigner-Dyson $\upbeta$ while vanishing in the Wigner-Dyson classes, where the contribution of largest order is $t^{2g+1}=t^5$. Already from this, one can see that while the comparison to the prediction of universal random matrix theory is involved already in the case of the unorientable ($\upbeta=1$) incarnation of the Airy model as compared to the orientable case, the task will be even more complex for the case of general $\upbeta$. Before going into this discussion however, we will use the result to give evidence for its agreement with the predictions of universal random matrix theory in the remaining, symplectic, symmetry class ($\upbeta=4$) not yet studied in the literature.

\subsection{The case of $\upbeta=4$ (The symplectic class)}\label{sec:Comparison_RMT_Top_GSE}
To compare to the prediction of universal RMT, we first have to compute this for the present case of the Airy model and $\upbeta=4$. We recall from \cref{sec:Background} that the $\tau$-scaled SFF, $\kappa_\upbeta^s(\tau,\beta)$, is given by
\begin{align}
  \kappa_\upbeta^s(\tau,\beta)=\int_{0}^{\infty}\dd E e^{-2\beta E}\rho_0(E)- 
\int_{0}^{\infty}\dd E e^{-2\beta E}\rho_0(E)b_{\upbeta}\left(\frac{\tau}{2\pi \rho_0(E)}\right)
\end{align}
where for GSE, i.e. $\upbeta=4$ one finds \cite{Mehta2004} 
\begin{align}
    b_4\qty(\frac{\tau}{2\pi \rho_0(E)})=\left\{
	\begin{array}{ll}
		1-\frac{\tau}{4\pi\rho_0(E)}+\frac{\tau}{8\pi\rho_0(E)}\log\qty(\abs{1-\frac{\tau}{2\pi\rho_0(E)}}) & \mbox{if } \frac{\tau}{4\pi} \leq \rho_0(E) \\
		0 & \mbox{if } \frac{\tau}{4\pi} \geq \rho_0(E).
	\end{array}.
    \right.
\end{align}
The full calculation of $\kappa_4^s(\tau,\beta)$ can be found in appendix \ref{sec:SFF_derivation}, which results in an exact expression. However, the expressions for the contributions to the SFF we found above from the topological expansion are of the form of an expansion in powers of $\tau$ which requires the RMT result to be expanded in $\tau$ in order to compare. The first orders are:
\begin{align}\label{eq:kappa_Airy_4_RMT}
\begin{aligned}
    \kappa_4^s(\tau,\beta)=&\frac{\tau}{8\pi\beta}+\sqrt{\frac{2}{\pi}}\frac{\tau^{2}}{16\sqrt{\beta}}-\frac{\tau^{3}\left(3\log\left(\beta\frac{\tau^{2}}{2}\right)+3\gamma+1\right)}{48\pi}-\sqrt{\frac{2}{\pi}}\frac{\sqrt{\beta}\tau^{4}}{12}\\
    &+\frac{\beta\tau^{5}\left(60\log(\beta\frac{\tau^{2}}{2})+60\gamma-7\right)}{960\pi}+\sqrt{\frac{2}{\pi}}\frac{\beta ^{3/2}\tau^6}{15} +\\
    &-\frac{\beta^2\tau^7\qty(1120\log\qty(\beta\frac{\tau^{2}}{2})+1120\gamma-501)}{26880\pi}-\sqrt{\frac{2}{\pi}}\frac{4\beta ^{5/2}\tau^8}{105} +\order{\tau^9},
\end{aligned} 
\end{align}
where $\gamma$ denotes the Euler-Mascheroni constant and we went up to the maximal order where we can compare our results from the loop equations.
Having found now the universal RMT result we can compare to the results found from the loop equations. For this, we plug $\upbeta=4$ into the results obtained in section \ref{sec:SFF_loop_eq}. Furthermore, we recall from \cref{sec:Background} the definition
\begin{align}
    \kappa_\upbeta^s(\tau,\beta)=\lim_{t\rightarrow\infty}\sum_{g=0,\frac 12,1, \dots} \underbrace{\frac{\kappa_{\upbeta}
    ^g(t,\beta)}{t^{2g+1}}}_{\coloneqq \kappa^{s,g}_\upbeta(t,\beta) }\tau^{2g+1},
\end{align}
and read off the $\kappa^{s,g}_\upbeta(t,\beta)$ to find 
\begin{align}
\kappa^{s,0}_4(t,\beta)&\rightarrow\frac{1}{8 \pi  \beta },\\
\kappa^{s,\frac 12}_4(t,\beta)&\rightarrow\frac{1}{8 \sqrt{2 \pi } \sqrt{\beta }},\\
\kappa^{s,1}_4(t,\beta)&\rightarrow-\frac{-3\log\left(\frac{2t}{\beta}\right)+10}{48\pi},\\
\kappa^{s,\frac 3 2}_4(t,\beta)&\rightarrow\frac{\sqrt{t}}{48 \sqrt{2 \pi }}-\frac{\sqrt{\beta }}{6 \sqrt{2 \pi }},\\
\kappa^{s,2}_4(t,\beta)&\rightarrow\frac{\beta \left(-60\log\left(\frac{2t}{\beta}\right)+163\right)}{960\pi}\\
\kappa^{s,\frac 5 2}_4(t,\beta)&\rightarrow\frac{1}{15} \sqrt{\frac{2}{\pi }} \beta ^{3/2} -\frac{17 \beta \sqrt{t}}{768 \sqrt{2 \pi }}-\frac{(\sqrt{t})^3}{1920 \sqrt{2 \pi }}\\
\kappa^{s,3}_4(t,\beta)&\rightarrow-\frac{(\sqrt{t})^4}{5760\pi}-\frac{\beta^{2}\left(-3360\log\left(\frac{2t}{\beta}\right)+8297\right)}{80640\pi}\\
\kappa^{s,\frac 7 2}_4(t,\beta)&\rightarrow-\frac{4}{105}\sqrt{\frac{2}{\pi}}\beta^{5/2}+\frac{881\beta^{2}\sqrt{t}}{61440\sqrt{2\pi}}+\frac{3\beta\left(\sqrt{t}\right)^{3}}{5120\sqrt{2\pi}}-\frac{\left(\sqrt{t}\right)^{5}}{53760\sqrt{2\pi}},
\end{align}
where $\rightarrow$ denotes that we only consider terms that are not vanishing upon $t\rightarrow\infty$. Like for the orthogonal case, which we recalled in \cref{sec:Background}, we see that there are remaining instances of $t$. They remain either as powers, if before $\tau$-scaling the term was of higher order than $2g+1$ in $t$, or in the logarithms, where they have remained for reasons apparent momentarily. As for the orthogonal case, we now group terms having the same dependence on $\beta$ in their prefactor. This is motivated by the observation that the contributions at given order in $\tau$ to the universal RMT result have a common dependence of their prefactor on $\beta$ and thus contributions of different such prefactors do not mix. Consequently, we find
\begin{align}\label{eq:kappa_LE_collected}
    \kappa_4^s(\tau,\beta)=&\lim_{t\rightarrow\infty}\Bigg\{ \frac{\tau}{8\pi\beta}+\sqrt{\frac{2}{\pi}}\qty(\frac{\tau^{2}}{16\sqrt{\beta}}-\frac{\sqrt{\beta}\tau^{4}}{12}+\frac{\beta ^{3/2}\tau^6}{15} -\frac{4\beta ^{5/2}\tau^8}{105}+\dots)+ \nonumber\\
    &+\frac{\tau^3\beta^0}{\pi}\qty[-\frac{5}{24}+\frac{3}{48}\log\qty(\frac{2t}{\beta})+\sqrt{\frac{\pi}{2}}\frac{\sqrt{t}\tau}{48}-\sqrt{\frac{\pi}{2}}\frac{\qty(\sqrt{t}\tau)^3}{1920}-\frac{t^2\tau^4}{5760}-\sqrt{\frac{\pi}{2}}\frac{\qty(\sqrt{t}\tau)^5}{53760}+\dots]\nonumber\\
    &+\frac{\tau^5\beta}{\pi}\qty[\frac{163}{960}-\frac{1}{16}\log\qty(\frac{2t}{\beta})-\sqrt{\frac{\pi}{2}}\frac{17\sqrt{t}\tau}{768}+\sqrt{\frac{\pi}{2}}\frac{3\qty(\sqrt{t}\tau)^3}{5120}+\dots]\nonumber\\
    &+\frac{\tau^7\beta^2}{\pi}\qty[-\frac{8297}{80640}+\frac{1}{24}\log\qty(\frac{2t}{\beta})+\sqrt{\frac{\pi}{2}}\frac{881\sqrt{t}\tau}{61440}+\dots]\nonumber\\
    &+\dots \Bigg\},
\end{align}
where the dots indicate contributions that arise at higher order in the topological expansion than we consider here.

This is now the expression that has to coincide with \cref{eq:kappa_Airy_4_RMT} to show that the variety of topological gravity corresponding to $\upbeta=4$ is chaotic. Considering first the terms in the first line, which are independent on $t$, the limit of $t$ is trivial and we see full agreement with the corresponding terms in \cref{eq:kappa_Airy_4_RMT}, as it was the case in  \cite{Weber2024} for $\upbeta=1$. Things become more interesting when considering the terms that retain a dependence on $t$, which one has to consider independently, as suggested by what happens in the $\upbeta=1$ case. Before going into this discussion however, we note that the coefficients of the logarithmic terms here coincide precisely to the coefficients of the logarithmic terms found from universal RMT, giving already a strong indication that full coincidence can be achieved as the computations are totally independent.

We start with the terms of order $\tau^{3}\beta^0$, which can be slightly simplified to
\begin{equation}
\kappa^s_{4}\left(\tau,\beta\right)\supset\frac{\tau^{3}}{48\pi}\qty[-\left(-3\log\left(\frac{2t}{\beta}\right)+10\right)+\sqrt{\frac{\pi}{2}}\sqrt{t}\tau-\sqrt{\frac{\pi}{2}}\frac{(\sqrt{t}\tau)^{3}}{40}-\frac{\left(t\tau^{2}\right)^{2}}{120}-\sqrt{\frac{\pi}{2}}\frac{\left(\tau\sqrt{t}\right)^{5}}{1120}].\label{eq:tau_cubed_collected}
\end{equation}
We see three types of terms: 
\begin{enumerate}
\item The term containing the logarithm and constants,
\item terms of the structure $c\cdot\left(\sqrt{t}\tau\right)^{2n+1}$ for
some\footnote{It may be noteworthy that the denominators of the $c$ are the sequence \href{https://oeis.org/A283433}{OEIS A283433},
which might suggest that there is only one more term of this type
at $g=9/2$.} $c\in\mathbb{R}$ and $n\in\mathbb{N}_{0}$ and
\item terms of the structure\footnote{In our case, we actually only see the term for $n=0$ and would expect
the next one to appear at $g=4$.} $c\cdot\left(t\tau^{2}\right)^{2}\cdot\left(t\tau^{2}\right)^{n}$
for some $c\in\mathbb{R}$ and $n\in\mathbb{N}_{0}$.
\end{enumerate}
We argue in the following that, as for the case of $\upbeta=1$, the terms of the last type form the defining expansion of a certain function, which in the limit $t\rightarrow\infty$, where it can be written as its asymptotic expansion, cancels the terms of the second type and combines with the first term to give the result expected from universal RMT.

Specifically, as by construction we only know finitely many terms of the second and third type, one has to build up the ``cancelling'' function order by order. In \cref{sec:det_canc_func} we give a method how to do so generically, whose results we use in the following. Specifically, it leads us to consider the two functions
\begin{align}
    f_1(t,\tau,\beta)&\coloneqq\frac{1}{240}\left(t\tau^{2}\right)^{2}\,_{2}F_{2}\left(2,2;\frac{5}{2},\frac{7}{2};-\frac{1}{16}t\tau^{2}\right),\\
    f_2(t,\tau,\beta)&\coloneqq\frac{1}{240}\left(t\tau^{2}\right)^{2}\,_{1}F_{1}\left(\frac{3}{2};\frac{5}{2};-\frac{1}{8\sqrt[3]{50}}t\tau^{2}\right).
\end{align}
The asymptotic expansions of these functions\footnote{The notation $f\overset{t\rightarrow\infty}{\longrightarrow}g$ does
not only mean $\lim_{t\rightarrow\infty}\frac{f}{g}=1$, but the stronger (in the case $g\not\rightarrow0$)
$\lim_{t\rightarrow\infty}\left(f-g\right)=0$.} is given by
\begin{align}
    f_1(t,\tau,\beta)&\overset{t\rightarrow\infty}{\longrightarrow}3\log\left(t\tau^{2}\right)+3\gamma-9,\label{eq:2F2_asymptotic}\\
     f_2(t,\tau,\beta)&\overset{t\rightarrow\infty}{\longrightarrow}\sqrt{\frac{\pi}{2}}\sqrt{t}\tau,
\end{align}
and their definitions as a power series with infinite radius of convergence, are given by
\begin{align}
f_1(t,\tau,\beta)&=\left(t\tau^{2}\right)^{2}\left(\frac{1}{240}-\frac{t\tau^{2}}{8400}+\mathcal{O}\left(\tau^{4}\right)\right),\\
f_2(t,\tau,\beta)&=\left(t\tau^{2}\right)^{2}\left(\frac{1}{240}-\frac{t\tau^{2}}{3200\sqrt[3]{100}}+\mathcal{O}\left(\tau^{4}\right)\right).
\end{align}
As done for the case of $\upbeta=1$, we add and subtract the sum of both functions in the bracket of \cref{eq:tau_cubed_collected}. Writing now the added sum as its defining series, we see that the term of the third type we found to the order to which we considered the expansion is cancelled. Now we take the large $t$ limit, where the asymptotic expansion of the subtracted sum can be used. Subtracting \ref{eq:2F2_asymptotic} from the logarithmic term of
\ref{eq:tau_cubed_collected} we find
\begin{equation}
3\log\left(\frac{2t}{\beta}\right)-10-3\log\left(t\tau^{2}\right)-3\gamma+9=-\left(3\log\left(\beta\frac{\tau^{2}}{2}\right)+3\gamma+1\right),
\end{equation}
which is exactly the term obtained from the universal prediction. Subtracting the asymptotic expansion of $f_2(t,\tau,\beta)$ cancels the first term of the second type. 
Thus, by the present manipulation we have shown the agreement of the result from topological gravity and the universal RMT prediction up to $\order{\tau^3}$, in fact to $\order{\tau^4}$.

To go to higher orders, one has to cancel the additional terms of the second type. A function that has a form useful for this would be 
\begin{align}
\left(t\tau^{2}\right)^{2}\left(t\tau^{2}\right)^{n}\,_{1}F_{1}\left(\frac{3}{2};\frac{5}{2};-\frac{3^{\frac 2 3}}{2}t\tau^{2}\right)&\overset{t\rightarrow\infty}{\longrightarrow}\sqrt{\frac{\pi}{2}}\left(\sqrt{t}\tau\right)^{2n+1},\\
&=\qty(t\tau^2)^n\qty[\qty(t\tau^2)^2-\frac{3}{10} 3^{2/3} \qty(t\tau^2)^3+\order{t^4}].
\end{align}
However, for example, to cancel the next occurring term of second type, i.e. that of order $\qty(\sqrt{t}\tau)^3$ one would need to know the term of third type of order $(t\tau^2)^3$ which will appear only at genus $4$ which would thus have to be computed to go an order higher.

Going to higher orders in the expansion of the universal result, one has to consider the terms grouped with the prefactor $\tau^5\beta$, $\tau^7\beta^2$ etc. in \cref{eq:kappa_LE_collected}. As noted above already, the prefactor of the logarithm matches perfectly with those of the universal result, making it highly plausible that one can find cancelling functions like $f_1$ that yield correspondence. However, as we would need higher order terms to fix these, we can't do this with the results computed in this work.

\subsection{Outlook: The general case}\label{sec:GeneralCase}
Coming back to the general case, we would first like to recall and compare recent results from the literature on (Gaussian) general $\upbeta$ matrix models.

\paragraph{The microcanonical SFF}
These discussions mostly focus on the \textit{microcanonical} SFF $\kappa_\upbeta(t,E)$, not to be mistaken for the \textit{canonical} SFF $\kappa_\upbeta(t,\beta)$ studied up to here. The two quantities are related via
\begin{align}
    \kappa_\upbeta(t,E)&=\mathcal{L}^{-1}\qty[\kappa_\upbeta(t,\beta),2\beta,E].
\end{align}
Consequently, one can speak of a topological expansion of $\kappa_\upbeta(t,E)$ as induced by that of the canonical SFF. We are interested in the $\tau$-scaled limit, where the universal prediction for the microcanonical SFF, using \cref{eq:univRMT_kappa_canonical}, is found as
\begin{align}\label{eq:kappa_E_univ}
    \kappa^s_\upbeta(\tau,E)&=\rho_0(E)\qty[1-b_\upbeta\qty(\frac{\tau}{2\pi\rho_0(E)})].
\end{align}
The dependence on time and energy of this object is, up to a prefactor of $\rho_0(E)$, purely via the specific combination found in the argument of the function $b_\upbeta$ and consequently we will use in the following $x\coloneqq \frac{\tau}{2\pi\rho_0(E)}$ to abbreviate the notation. For later use, we define the normalised $\tau$-scaled microcanonical SFF as 
\begin{align}
    \bar{\kappa}_\upbeta(x)\coloneqq\frac{1}{\rho_0(E)}\kappa^s_\upbeta(\tau,E)=1-b_\upbeta(x).\label{eq:normalised_microSFF}
\end{align}As we have seen in the discussion of \cref{sec:Comparison_RMT_Top_GSE} for $\upbeta=4$ and in \cite{Weber2024} for $\upbeta=1$, starting with $\order{\tau^3\beta}$ one has to be careful as taking the universal limit involves finding certain cancelling functions. Consequently, we first restrict our discussion to the terms of lower order, which can be transformed directly to yield
\begin{align}
    \kappa^s_\upbeta(\tau,E)=\frac{\tau}{\pi \upbeta}-\frac{2-\upbeta}{\upbeta^2}\frac{\tau^2}{\pi\sqrt{E}}+\order{\tau^3}.
\end{align}
This can be used to find the expansion of $b_\upbeta$ for small arguments as
\begin{align}\label{eq:b_upbeta_expanded}
    b_\upbeta(x)=1-\frac{2}{\upbeta} x+2\frac{2-\upbeta}{\upbeta^2} x^2+\order{x^3}.
\end{align}
An important thing to note here is, that each of the contributions to $b_\upbeta$ except that at $\order{x^0}$ stems from a contribution to a two-point correlation function at a specific order in the topological expansion. Consequently, each order has to transform according to \cref{eq:Rel_beta_4_beta} under $\upbeta\to \frac{4}{\upbeta}$ and our result does so by construction. Explicitly, writing $b_\upbeta^g$ for the coefficient of $x^{2g}$ which originates from the genus $g$ contribution to the SFF, this requirement can be put as 
\begin{align}\label{eq:rel_inv_b}
    b^g_\upbeta=\qty(-1)^{2g} \qty(\frac{2}{\upbeta})^{2\qty(g+1)}b^g_{\frac{4}{\upbeta}}.
\end{align}
For the genus $0$ contribution, alternatively, one can use the result \cref{eq:relation_genus0}, fixing the general $\upbeta$ dependence by just knowing the $\upbeta=1$ contribution. Interestingly, for $g=\frac{1}{2}$ one can do the same\footnote{Actually for all contributions originating from the Wigner-Dyson part of the volumes, as it is shown later.}. This is due to the dependence on $\upbeta$ being fixed by the general structure for resolvents \cref{eq:Resolvent_beta_halfInteger}, translated to $b_\upbeta$. This leaves only one unknown constant which can be fixed to be $b^{\frac 12}_1$ by realising that the product of $\upbeta$ dependent factors is one for $\upbeta=1$. Consequently, the general structure in terms of $\upbeta$ in combination with the requirement to give the correct limit in the case of $\upbeta=1$ fixes the whole result up to $\order{x^2}$. Before going into the comparison with the literature, we recall, for convenience, the expressions of $b_\upbeta$ for the three Wigner-Dyson classes to be \cite{Mehta2004}
\begin{align}
        b_1\qty(x)&=
	\begin{cases}
		1-2x+x\log\qty(1+2x) & \mbox{if } x \leq 1 \\
		  -1+x\log\qty(\frac{2x+1}{2x-1})  & \mbox{if } x \geq 1
	\end{cases},\label{eq:b^1(x)}\\
        b_2\qty(x)&=
	\begin{cases}
		1-x& \mbox{if } x \leq 1 \\
		0 & \mbox{if } x \geq 1
	\end{cases},\\
        b_4\qty(x)&=
	\begin{cases}
		1-\frac{x}{2}+\frac{x}{4}\log\qty(\abs{1-x}) & \mbox{if } x \leq 2 \\
		0 & \mbox{if } x \geq 2
	\end{cases},
\end{align}
which near $x=0$ can be expanded as
\begin{align}
    b_1(x)&=1-2x+2x^2+\order{x^4},\\
    b_2(x)&=1-x,\\
    b_4(x)&=1-\frac{x}{2}-\frac{x^2}{4}+\order{x^4}.
\end{align}
This, as expected, is reproduced by our result. 

\paragraph{Numerical evaluation of the microcanonical SFF}
For the general $\upbeta$ models, there has been recent (numerical) work on the microcanonical SFF for the arbitrary $\upbeta$ Gaussian matrix model in \cite{Bianchi2024}. For the numerical computation the authors used the implementation of the general $\upbeta$ Gaussian matrix model as an ensemble of certain tridiagonal matrices found in \cite{Dumitriu2002}. Using this implementation, the matrix model result for the microcanonical SFF can be computed as an average of the microcanonical SFFs of individual draws from the ensemble that can be directly evaluated from the eigenvalues of the drawn matrices. As it is customary for this sort of computation they use spectral unfolding, i.e. perform a mapping of the spectrum (details can be found in \cite{Bianchi2024}) such that after the mapping its density of states is constant, meaning $\rho_0=\frac{1}{N_m}$, with $N_m$ denoting the number of rows of the chosen random matrix, prior to the computation of the microcanonical SFF for individual realisations. This is done to ease the comparison with the analytical predictions that exist for the Wigner-Dyson cases (\cref{eq:kappa_E_univ}) but depend on the density of states. By performing the unfolding one can compare to this prediction if it is known and also compute the function $b_\upbeta$ for the cases for which it is not known.

Based on good agreement with the numerical results evaluated in this way, \cite{Bianchi2024} provides an ansatz for the whole of $b_\upbeta$ in the region of $\upbeta\in\qty[1,2]$ as
\begin{align}\label{eq:Ansatz_1}
\begin{aligned}
    b^{A,1}_\upbeta(x)&\coloneqq\begin{cases}
        1-\frac{2}{\upbeta}x+\frac{2-\upbeta}{\upbeta}x\log\qty(1+2x)& \mbox{if }x<1\\
        1-\frac{2}{\upbeta}+\frac{2-\upbeta}{\upbeta}x\log\qty(\frac{2x+1}{2x-1})& \mbox{if } x\geq1
    \end{cases} \\
    &=\frac{2-\upbeta}{\upbeta}b_1(x)+2\frac{\upbeta-1}{\upbeta}b_2(x),
\end{aligned}
\end{align}
and for $x\leq1$ in the region of $\upbeta\in\qty[2,4]$ as
\begin{align}\label{eq:Ansatz_2}
    b^{A,2}_\upbeta(x)=1-\frac{2}{\upbeta}x+\frac{\upbeta-2}{2\upbeta}x\log\qty(\abs{1-x}).
\end{align}
Both ans\"atze coincide (in the range of their validity) with the analytical results for the Wigner-Dyson cases as it is obvious for \cref{eq:Ansatz_1} and can be seen directly upon putting $\upbeta\in\qty{2,4}$ for \cref{eq:Ansatz_2} furthermore, they can be expanded at $x=0$ to yield
\begin{align}
    b^{A,1}_\upbeta(x)&=1-\frac{2}{\upbeta}x+2\frac{2-\upbeta}{\upbeta}x^2+\order{x^3},\\
    b^{A,2}_\upbeta(x)&=1-\frac{2}{\upbeta}x+\frac{2-\upbeta}{2\upbeta}x^2+\order{x^3}.
\end{align}
We first note that the $\order{x}$ term agrees with our result and thus we give an analytical argument for this numerical finding. For the next order our result nearly, but not completely, agrees. In fact, the difference is an additional factor of $\upbeta^{-1}$ in our result compared to the ans\"atze $b^{A,i}_\upbeta$. 

At this point, it is interesting to consider the behaviour of the ansatz, specifically the second order coefficient $b^{A,i,\frac{1}{2}}_\upbeta$, under $\upbeta\to\frac 4\upbeta$. To study this, we choose $\upbeta\in \qty[1,2]$, the range of validity of $b^{A,1}_\upbeta$. Now, under the mapping this is sent to $\frac{4}{\upbeta}\in\qty[2,4]$, i.e. the range of validity of $b^{A,2}_\upbeta$. Consequently, using \cref{eq:rel_inv_b}, we infer that it has to hold that
\begin{align}
    b^{A,1,\frac12}_\upbeta\overset{!}{=}\qty(-1) \qty(\frac{2}{\upbeta})^{3}b^{A,2,\frac 12}_{\frac 4\upbeta}.
\end{align}
However, one finds
\begin{align}
    \qty(-1) \qty(\frac{2}{\upbeta})^{2\qty(\frac 12+1)}b^{A,2,\frac 12}_{\frac 4\upbeta}=-\frac{8}{\upbeta^3}\frac{2-\frac 4 \upbeta}{2\frac{4}{\upbeta}}=-2\frac{\upbeta-2}{\upbeta^3}\neq b^{A,1,\frac12}_\upbeta,
\end{align}
which shows that the $b^{A,i}_\upbeta$ are not compatible with each other (and also not with itself, but that was expected) if each order in the expansion for small $\tau$ of the microcanonical SFF given by the ans\"atze is determined from the contributions to the topological expansion of this object as determined from the loop equations. 

A possible explanation for this would of course be that while for the Wigner-Dyson cases the perturbative expansion does reproduce the microcanonical SFF order-by-order in the expansion for small times, this ceases to be the case for the general $\upbeta$ case, i.e. implying that the general $\upbeta$ model as determined by the loop equations is either not quantum chaotic in the sense of the BGS conjecture or requires non-perturbative contributions that are not accessible by our present approach. To put this to the test, we implemented the numerical evaluation of the microcanonical SFF for the general $\upbeta$ Gaussian matrix model as it was done in \cite{Bianchi2024} and compared it to our result $b_\upbeta$ and the ans\"atze $b^{A,i}_\upbeta$. 
\begin{figure}[h!]
     \centering
     \begin{subfigure}[b]{0.47\textwidth}
         \centering
         \includegraphics[width=\textwidth]{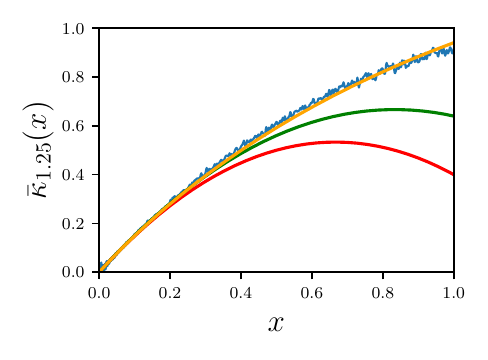}
         \caption{$\upbeta=1.25$}
         \label{fig:upbeta_125}
     \end{subfigure}
     \hfill
     \begin{subfigure}[b]{0.47\textwidth}
         \centering
         \includegraphics[width=\textwidth]{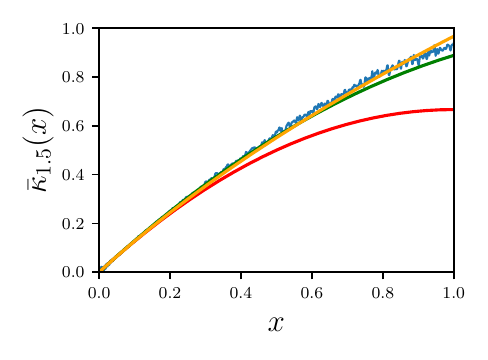}
         \caption{$\upbeta=1.5$}
         \label{fig:upbeta_15}
     \end{subfigure}
     \\
     \begin{subfigure}[b]{0.47\textwidth}
         \centering
         \includegraphics[width=\textwidth]{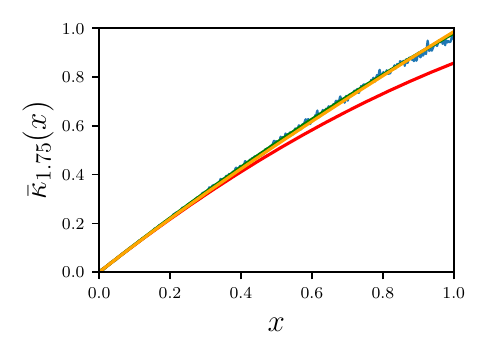}
         \caption{$\upbeta=1.75$}
         \label{fig:upbeta_175}
     \end{subfigure}
    \hfill
         \begin{subfigure}[b]{0.47\textwidth}
         \centering
         \includegraphics[width=\textwidth]{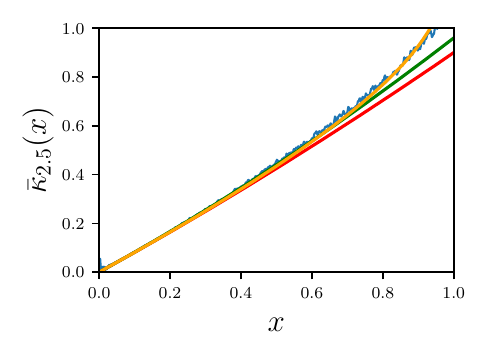}
         \caption{$\upbeta=2.5$}
         \label{fig:upbeta_25}
     \end{subfigure}
     \\
     \begin{subfigure}[b]{0.47\textwidth}
         \centering
         \includegraphics[width=\textwidth]{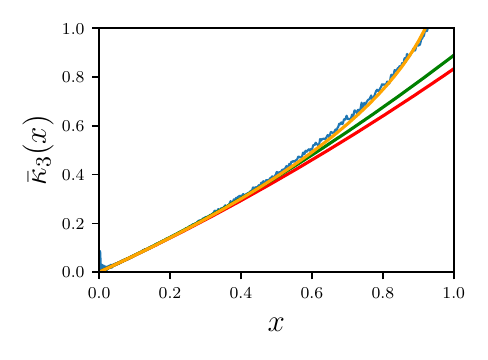}
         \caption{$\upbeta=3$}
         \label{fig:upbeta_3}
     \end{subfigure}
     \hfill
     \begin{subfigure}[b]{0.47\textwidth}
         \centering
         \includegraphics[width=\textwidth]{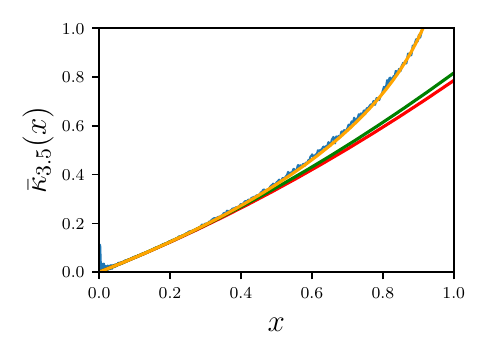}
         \caption{$\upbeta=3.5$}
         \label{fig:upbeta_35}
     \end{subfigure}
        \caption{Comparison of the numerical evaluation of the normalised microcanonical SFF (cf.\cref{eq:normalised_microSFF}) for the general $\upbeta$ Gaussian matrix model (blue line) with our prediction up to second order (green line), the prediction of \cite{Bianchi2024} to all orders (orange line) and up to second order (red line) for different values of $\upbeta\in\qty[1,4]$. We used matrices of size $N_m=200$, averaging over $N_r=8000$ realisations. }
        \label{fig:upbeta_numerics}
\end{figure}
This comparison, for various values of $\upbeta\in\qty[1,4]$ can be found in \cref{fig:upbeta_numerics}, where the numerical result for the (connected part of) the (normalised) microcanonical SFF\footnote{We always plot the normalised SFF in the following, so when referring to the plots we always mean this, leaving stating that it's ``normalised'' implicit, unless indicated otherwise.} is plotted as the blue line and our prediction for it in green\footnote{Numerically, we compute the whole microcanonical SFF and subtract the disconnected part which, for the unfolded spectrum, was evaluated in \cite{Bianchi2024}. Analytically, our correlation functions are defined to be the connected ones from the beginning.}. The whole ans\"atze of \cite{Bianchi2024} are plotted as the orange line and their expansion up to second order, which is what can be compared to our result, is plotted in red. We observe, that the orange curve for all values of $\upbeta$ we consider here nicely follows the numerical line. However, we also see for all considered values of $\upbeta$ that our prediction is following the numerical line with a smaller error than the ansatz appropriate for the respective regime expanded to second order. Notably, we do not have to change from one ansatz to another when switching from $\upbeta\in\qty[1,2]$ to $\upbeta\in\qty[2,4]$. From this we conclude that our result for the microcanonical SFF is a viable approximation to the microcanonical SFF for small times over the whole regime of $\upbeta\in\qty[1,4]$ which, as expected, requires extension to higher orders in order to achieve the accuracy of the full $b^{A,i}_\upbeta$.

Before going into this discussion, we give some further comments pointing at the potential of this study to yield an analytic result for the (universal) microcanonical SFF for the whole range of $\upbeta\in\mathbb{R}_+$. We will do this by studying the extension of the constraints imposed by RMT universality on the (Airy) WP volumes in the Wigner-Dyson classes to arbitrary Dyson index.

\paragraph{Constraints on the $\upbeta$ WP volumes}
First, to recall these, we consider the constraints that the fidelity to the universal RMT predictions imposed on the $\upbeta=1$ Airy WP volumes of integer genus, as it was found in \cite{Weber2024} and those on the $\upbeta=2$ (Airy) WP volumes as found in \cite{Weber2022,Blommaert2022}. For the case of general $\upbeta$ we suspect the volumes to be constrained by analogous relations if the fidelity to universal RMT is preserved.

The constraints discussed in \cite{Weber2022} for $\upbeta=1$ were given by
\begin{align}
    \underset{0\leq l<(g-1)}{\forall} K^{\upbeta=1}_g(l)=0,\label{eq:constraints_GOE}
\end{align}
with
\begin{align}
    K^{\upbeta=1}_g(l)\coloneqq\sum_{\substack{\alpha+\gamma=6g-2\\ \alpha,\gamma  \text{ odd}}}C_{\alpha,\gamma}\frac{\Gamma(1+\frac \alpha 2)\Gamma(1+\frac \gamma 2)}{\pi} (-1)^{\frac{\gamma+1}{2}}
    \sum_{n+m=l}\binom{\frac{\alpha+1}{2}}{n}\binom{\frac{\gamma+1}{2}}{m}(-1)^m,
\end{align}
where the $C_{\alpha,\gamma}$ are the coefficients of the $\upbeta=1$ Airy WP volumes for $b_1>b_2$, i.e. of the $V^{1,>}_g(b_1,b_2)$. Notably, these exactly coincide with the coefficients of the $\mathcal{V}^{g,>}_g$ for the general $\upbeta$ case, thus this part of the volumes fulfils the constraints by the fidelity to universal RMT in the $\upbeta=1$ case. Consequently, this part of the volumes fulfils the constraints for general $\upbeta$.
In the same manner, the constraints imposed by fidelity to the universal result in the unitary, i.e. $\upbeta=2$ case that can be found in \cite{Weber2022} are fulfilled by the $\mathcal{V}^0_{g,2}$, being the only part of the general volumes surviving in this case. Thus, we can infer that the part of the general $\upbeta$ volumes being directly related to the Wigner-Dyson result also in the general $\upbeta$ case fulfils certain constraints. These have the implication of those parts of the volumes not contributing terms to the canonical SFF that are of higher order in $t$ than expected from fidelity to universal RMT, i.e. of no higher order than $e^{S_0}$ after $\tau$-scaling. However, fidelity in the Wigner-Dyson cases does not have any implications on non-Wigner-Dyson terms. Still, we can check the constraints also for the non-Wigner-Dyson terms. The specific constraints we can consider are those for $g=2$ and $g=3$ which are given by
\begin{align}
   K^{\upbeta=1}_2(0)\propto 21 C_{1,9}-7 C_{3,7}+5 C_{5,5}-7 C_{7,3}+21 C_{9,1},
\end{align}
and
\begin{align}
    K^{\upbeta=1}_3(0)&\propto 715 C_{1,15}-143 C_{3,13}+55 C_{5,11}-35 C_{7,9}+35 C_{9,7}-55 C_{11,5}+143 C_{13,3}-715 C_{15,1},\\
    K^{\upbeta=1}_3(1)&\propto 1001 C_{1,15}-143 C_{3,13}+33 C_{5,11}-7 C_{7,9}-7 C_{9,7}+33 C_{11,5}-143 C_{13,3}+1001 C_{15,1}.
\end{align}
Checking now the one constraint for $g=2$ with the only non-Wigner-Dyson contribution in this case, $\mathcal{V}_{2}^{1,>}$ we find that it is not fulfilled. Interestingly, when checking the constraints for $g=3$ with the two non-Wigner-Dyson terms there, i.e. $\mathcal{V}_{3}^{1,>}$ and $\mathcal{V}_{3}^{2,>}$ we find that $K^{\upbeta=1}_3(0)$ is fulfilled while $K^{\upbeta=1}_3(1)$ is not. We recall from \cite{Weber2024} that the contributions arising from the volumes that had to vanish for the $\upbeta=1$ case and thus gave rise to the constraints were an additional logarithmic term \footnote{Actually with additional subleading corrections which a posteriori are however not of relevance for the present discussion since these constraints are fulfilled.} contributing at higher order than $e^{S_0}$ for the constraints $K^{\upbeta=1}_g(l)$ for which $3g-l$ was odd while in the case where it was even the contribution was polynomial in $t$, although without dependence on $\pi$, which is distinct from all other terms contributing to the large time limit of the canonical SFF. Consequently, since $3\times2-0=6$ and $3\times3-1=8$ are even and $3\times3-0=9$ is odd we infer that the additional logarithmic terms cancel while the polynomial term seems to survive. Presently we don't have a good explanation for the non-vanishing of the polynomial type of terms, though it is reasonable to assume that it is involved in the cancellation of terms of remaining $t$ dependence of lower genus non-Wigner-Dyson terms. For example, for the term arising from $g=2$, there is a non-Wigner-Dyson term arising at $g=\frac32$ that, assuming that the terms directly deriving from the Wigner-Dyson contributions still cancel among themselves, can only be taken care of by this term (with additional higher order terms, as expected). The cancellation of the logarithmic term however can be understood as a good sign for fidelity of the canonical SFF to the predictions of universal RMT since its presence would give a term that even by the mechanism of cancelling terms of larger than expected order applied for the unorientable Wigner-Dyson cases could not be cancelled, hence indicating disagreement with the universal RMT prediction that, regardless of the specific form of the microcanonical SFF, is always $\order{e^{S_0}}$.

Adding to this discussion we note that these constraints on the unorientable volumes, as it was already pointed out in \cite{Weber2024}, are only a subset of the full set of constraints fulfilled by the Airy WP volumes to achieve fidelity to universal RMT in the Wigner-Dyson cases. Going beyond that subset one has to consider all the contributions to the Airy WP volumes, not only the contributions where both powers of the lengths are odd as it was done in \cite{Weber2024} in order to focus on the logarithmic terms. To assess whether such constraints even exist, it is useful to study the contribution of the term $b_1^\alpha b_2^\gamma \theta\qty(b_1-b_2)$, in the WP volume to the two-point correlation function of partition functions, denoted as $I(\alpha,\gamma)$. This can, for even $\alpha$ be evaluated as\footnote{For details see appendix D of \cite{Weber2024}}
\begin{align}
    I(\alpha,\gamma)&=\frac{2^{\alpha +\gamma+1} \beta_1^{\frac{1}{2} (\alpha +\gamma+3)} \beta_2^{\frac{1}{2} (\alpha +\gamma+1)}  \Gamma \left(\frac{1}{2} (\alpha +\gamma+4)\right) }{\pi  (\gamma+2)\left(\beta_1+\beta_2\right){}^{\left(\frac{\alpha +\gamma}{2}+1\right)}}\sum_{k=0}^{\frac{\alpha}{2}} \frac{\frac{\alpha}{2}! \left(\frac{\beta_1}{\beta_2}\right)^k \Gamma \left(\frac{\gamma}{2}+2\right)}{\left(\frac{\alpha}{2}-k\right)! \Gamma \left(k+\frac{\gamma}{2}+2\right)}, 
\end{align}
from which one can directly see that if one computes the contribution to the canonical SFF from this, i.e. $\beta_1=\beta+i t, \beta_2=\beta_1^\star$ the leading-order contribution is proportional to
\begin{align}
    \qty(\beta_1\beta_2)^{\frac{\alpha+\gamma+1}{2}}\beta_1=\qty(\beta^2+t^2)^{\frac{6g-2+1}{2}}\qty(\beta+i t).
\end{align}
For the full contribution to the canonical SFF one has to take into account also the complementary term, i.e. $b_1^\gamma b_2^\alpha\theta(b_2-b_1)$, which can be computed directly by exchanging $\beta_1\leftrightarrow\beta_2$ in $I(\alpha,\gamma)$. Consequently, the full contribution to the canonical SFF from this part of the WP volume is proportional to
\begin{align}
    \qty(\beta^2+t^2)^{\frac{6g-1}{2}}\propto t^{6g-1},
\end{align}
which for $g\geq 1$ is of higher order than the maximal order compatible with universal RMT, i.e. $2g+1$. For odd $\alpha$ the same statements follows from the expression for $I(\alpha,\gamma)$ for this case, given in \cite{Weber2024}. The vanishing of the naive leading-order can also be explained by the requirement on the canonical SFF to be a real quantity. Employing this reasoning one is led to conjecture that there generically is a contribution to every odd power $k\in \qty[2g+2,6g-1]$ of $t$ for the integer genus and to every even power $k\in \qty[2g+2,6g-1]$ for the half-integer case. To achieve agreement with universal RMT those contributions have to vanish. To probe this, we evaluated the canonical SFF with an arbitrary choice of coefficients for the contributions to the WP volume for $\upbeta=1$ or $\upbeta=4$, i.e.
\begin{align}
    V_g^{>}(b_1,b_2)=\sum_{\substack{{\alpha_1,\alpha_2\in\mathbb{N}_0}\\\alpha_1+\alpha_2=6g-2}}C_{\alpha_1,\alpha_2}b_1^{\alpha_1}b_2^{\alpha_2},
\end{align}
and found exactly the structure we conjectured. The constraints arising from this for $g\in\qty[1,\frac 5 2]$ are given by
\begin{align}
    g=1:\quad 0=&2 C_{0,4}+C_{1,3}-C_{3,1}-2 C_{4,0},\\
    g=\frac 3 2: \quad 0=&7 C_{0,7}+5 C_{1,6}+3 C_{2,5}+C_{3,4}-C_{4,3}-3 C_{5,2}-5 C_{6,1}-7 C_{7,0},\\
    &0=35 C_{0,7}+35 C_{1,6}+25 C_{2,5}+9 C_{3,4}\nonumber\\
    &-9 C_{4,3}-25 C_{5,2}-35 C_{6,1}-35 C_{7,0},\\
    g=2: \quad 0=&5 C_{0,10}+4 C_{1,9}+3 C_{2,8}+2 C_{3,7}+C_{4,6}\nonumber\\
    &-C_{6,4}-2 C_{7,3}-3 C_{8,2}-4 C_{9,1}-5 C_{10,0},\\
    \quad 0=&55 C_{0,10}+50 C_{1,9}+41 C_{2,8}+29 C_{3,7}+15 C_{4,6}\nonumber\\
    &-15 C_{6,4}-29 C_{7,3}-41 C_{8,2}-50 C_{9,1}-55 C_{10,0},\\
    \quad 0=&495 C_{0,10}+420 C_{1,9}+441 C_{2,8}+394 C_{3,7}+235 C_{4,6}\nonumber\\
    &-235 C_{6,4}-394 C_{7,3}-441 C_{8,2}-420 C_{9,1}-495 C_{10,0},
\end{align}
\begin{align}
    g=\frac 5 2: \quad 0=&13 C_{0,13}+11 C_{1,12}+9 C_{2,11}+7 C_{3,10}+5 C_{4,9}+3 C_{5,8}+C_{6,7}\nonumber\\&-C_{7,6}-3 C_{8,5}-5 C_{9,4}-7 C_{10,3}-9 C_{11,2}-11 C_{12,1}-13 C_{13,0},\\
    \quad 0=&3003 C_{0,13}+2717 C_{1,12}+2343 C_{2,11}+1897 C_{3,10}+1395 C_{4,9}\nonumber\\&+853 C_{5,8}+287 C_{6,7}-287 C_{7,6}-853 C_{8,5}-1395 C_{9,4}\nonumber\\&-1897 C_{10,3}-2343 C_{11,2}-2717 C_{12,1}-3003 C_{13,0},\\
    \quad 0=&63063 C_{0,13}+59345 C_{1,12}+54747 C_{2,11}+47509 C_{3,10}+37023 C_{4,9}\nonumber\\&+23577 C_{5,8}+8099 C_{6,7}-8099 C_{7,6}-23577 C_{8,5}-37023 C_{9,4}\nonumber\\&-47509 C_{10,3}-54747 C_{11,2}-59345 C_{12,1}-63063 C_{13,0},\\
    \quad 0=&105105 C_{0,13}+145431 C_{1,12}+100485 C_{2,11}+80619 C_{3,10}+79065 C_{4,9}\nonumber\\&+64863 C_{5,8}+25613 C_{6,7}-25613 C_{7,6}-64863 C_{8,5}-79065 C_{9,4}\nonumber\\&-80619 C_{10,3}-100485 C_{11,2}-145431 C_{12,1}-105105 C_{13,0}.
\end{align}
These constraints have to be fulfilled by \textit{all} the contributions to the general $\upbeta$ WP volume for a given genus, i.e. all the $\mathcal{V}_g^{i,>}$ with $i>0$\footnote{The cancellations for the $\mathcal{V}_g^{0}$ have been explored above.}. Remarkably, this is also the case for all the non-Wigner-Dyson contributions to the Airy WP volumes we computed\footnote{In \cite{Weber2025c} this is discussed up to $g=\frac7 2$, finding the arising constraints to be fulfilled by all contributions to the $\upbeta$ Airy WP volumes, likewise.}. This gives a strong sign for the agreement of arbitrary $\upbeta$ topological gravity with universal RMT.
\paragraph{Partial resummation} Building on this, we give the first step how to obtain the arbitrary $\upbeta$ universal microcanonical SFF from our results for the canonical SFF. For this first step, we will harness mainly the understanding of the canonical SFF as arising for the Wigner-Dyson classes, where it suffices to restrict to the cases of $\upbeta\in\qty{1,2}$. For these cases, the canonical SFF is obviously determined solely from the Wigner-Dyson part of the Airy WP volumes. As we remarked already above, this part of the general $\upbeta$ Airy WP volumes is fully determined by knowing the volumes for the unitary and orthogonal symmetry class. Specifically, distinguishing the cases of integer and half-integer genus and using the general structure of the Airy WP volumes \cref{eq:V_gen_struct}, we see that the Wigner-Dyson part of the volume is given by
\begin{align}\label{eq:V_WD_integer}
    V^{\upbeta,\text{WD}}_{g,n}(\vec{b})=\frac{2^{g+n-1}}{\upbeta^{g+n-1}}V^2_{g,n}(\vec{b})+\frac{(2-\upbeta)^2}{\upbeta^{g+n}}\qty(V^1_{g,n}(\vec{b})-2^{g+\abs{I}-1}V^{2}_{g,n}(\vec{b})),
\end{align}
for integer genus and
\begin{align}\label{eq:V_WD_halfInteger}
    V^{\upbeta,\text{WD}}_{g,n}(\vec{b})=\frac{(2-\upbeta)}{\upbeta^{g+n-\frac 12}}V^1_{g,n}(\vec{b}),
\end{align}
for half-integer genus. For the case of $n=2$, these volumes give rise to the canonical SFF which matches the behaviour of universal RMT in the Wigner-Dyson cases. This is equivalent to saying that the microcanonical SFF arising from these volumes is that of universal RMT for the Wigner-Dyson classes. This connection is what we will use to compute the ``Wigner-Dyson part'' of the universal microcanonical SFF for arbitrary $\upbeta$.

For the present work we shall restrict our study in this direction to the regime before the point of non-analyticity present in all Wigner-Dyson classes, i.e. to $x=\frac{\tau}{2\pi\rho_0(E)}\leq 1$, leaving the study of the full object for \cite{Weber2025b}\footnote{Note that for all except the symplectic symmetry class this is the point where $b_\upbeta(x)$ changes domain in its piecewise definition while for $\upbeta=4$ at this point the function diverges.}. For this regime we know from the study for the unitary symmetry class that only the genus $0$ contribution from the orientable part of the volumes is relevant, while we know from \cref{sec:Comparison_RMT_Top_GSE} and \cite{Weber2024} that all of the unorientable part is relevant. The orientable contribution to the Wigner-Dyson part of the microcanonical SFF or, directly related, the contribution to the Wigner-Dyson part of the arbitrary $\upbeta$ $b(x)$, which we denote as $b_\upbeta^\text{WD}$, we have already included in \cref{eq:b_upbeta_expanded}. The unorientable contributions we can find by noticing that for $\upbeta=1$ the Airy WP volumes lead to $b_1(x)$ and consequently, since every term of its expansion is thus linked to one and only one term in the topological expansion, i.e. one Airy WP volume, we can uplift the result to arbitrary Dyson index by making use of our knowledge of the generalisation of the Airy WP volumes to this setting (\cref{eq:V_WD_integer} or \cref{eq:V_WD_halfInteger}) and the linearity of the Laplace transform. Explicitly, we find
\begin{align}
    b_{\upbeta}^{\text{WD}}(x)=1-\frac{2}{\upbeta}x+\qty(2-\upbeta)^2\sum_{g\in \mathbb{N}_+}\frac{1}{\upbeta^{g+2}}b^1_{g}(x)
    +\qty(2-\upbeta)\sum_{g\in \frac{\mathbb{N}_+}{2}}\frac{1}{\upbeta^{g+\frac 3 2}}b^1_{g}(x),
\end{align}
where $b_{1,g}(x)$ denotes the term in the series expansion of $b_{1}(x)$ originating from the genus $g$ WP volume and $\frac{\mathbb{N}_+}{2}$ the set of positive half-integers. 
To evaluate this, we expand $b^1(x)$, given in \cref{eq:b^1(x)}, to find
\begin{align}
    b^1_g(x)=\frac{\qty(-1)^{2g+1}2^{2g}}{2g}x^{2g+1}.
\end{align}
Using this, one can write
\begin{align}
\begin{aligned}\label{eq:b_upbeta_WD}
    b_{\upbeta}^{\text{WD}}(x)=1-\frac{2}{\upbeta}x+&\frac{\qty(2-\upbeta)^2}{2\upbeta^2}x\sum_{g=1}^\infty\frac{\qty(-1)^{g+1}}{g}\qty(-\frac{4x^2}{\upbeta})^g\\
    & +\frac{\qty(2-\upbeta)}{\sqrt{\upbeta}^3}x\sum_{k=0}^\infty\frac{1}{2k+1}\qty(\frac{2x}{\sqrt{\upbeta}})^{2k+1}\\
    =1-\frac{2}{\upbeta}x+&\frac{\qty(2-\upbeta)^2}{2\upbeta^2}x\log\qty(1-\frac{4x^2}{\upbeta})
    +\frac{\qty(2-\upbeta)}{\sqrt{\upbeta}^3}x\artanh\qty(\frac{2x}{\sqrt{\upbeta}})\\
    =1-\frac{2}{\upbeta}x+&\frac{\qty(2-\upbeta)}{2\upbeta^2}x\Bigg[\qty(2-\upbeta+\sqrt{\upbeta})\log\qty(1+\frac{2x}{\sqrt{\upbeta}})
    \\
    & +\qty(2-\upbeta-\sqrt{\upbeta})\log\qty(\abs{1-\frac{2x}{\sqrt{\upbeta}}})\Bigg],
\end{aligned}
\end{align}
where, in principle, we have to restrict ourselves to the region of convergence of the two series, i.e. $\abs{x}\leq\frac 12 \sqrt{\upbeta}$. However, one can continue the result analytically beyond this region using the logarithm, as we have already rewritten the result in the last line. As a first sanity check one can put $\upbeta=1$ which yields the expected result.
Secondly, plugging $\upbeta=4$ the GSE result is reproduced up to $x=1$, which is the regime in which we are interested. This is of course expected, but nevertheless one can note here that the combination of the GOE result with the dependence on $\upbeta$ for the individual terms we have explored here leads to this result directly and no additional considerations are necessary. 

\begin{figure}[h]
\centering
\begin{subfigure}[b]{0.49\textwidth}
    \includegraphics[width=\linewidth]{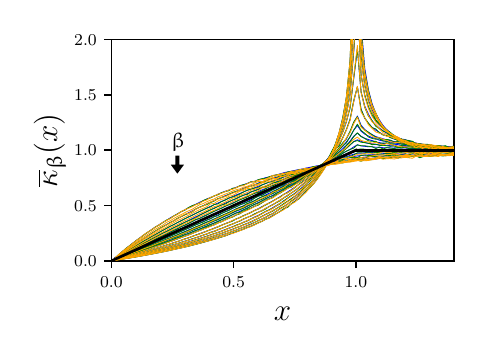}
    \caption{}
    \label{fig:AllSFFs}
\end{subfigure}
\hfill
\begin{subfigure}[b]{0.49\textwidth}
    \includegraphics[width=\linewidth]{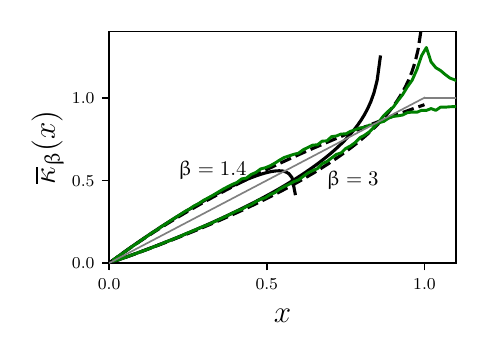}
    \caption{}
    \label{fig:ComparisonSFFs}
\end{subfigure}
\caption{In \textbf{a)}: The connected part of the (normalised) microcanonical SFF for values of the Dyson index between $\upbeta=1$ and $\upbeta=6$, computed from the tridiagonal matrix ensemble of \cite{Dumitriu2002}. The black line is the analytical result for $\upbeta=2$, the colour of the other lines corresponds to the size of the matrices $N_m$, averaged over $N_r$ realisations. Blue corresponds to $N_m=200$, $N_r=5000$, green to $N_m=400$ $N_r=1000$ and orange to $N_m=1000$, $N_r=200$. The plotted microcanonical SFFs are naturally ordered in the regime before intersecting the $\upbeta=2$ curve, where a larger SFF corresponds to a smaller value of $\upbeta$. As one can see clearly for all $\upbeta>2$, a divergence like that for the GSE case appears at $x=1$ while for all $\upbeta<2$ the transition to the plateau is smooth.\\
In \textbf{b)}: Comparison of the connected part of the normalised microcanonical SFF for the values of $\upbeta=1.4$ and $\upbeta=3$ from the ensemble as in \textbf{a)} with $N_m=1000$, $N_r=200$ (green lines), with the various predictions, the black solid line ours (\cref{eq:b_upbeta_WD}), the black broken one that from \cite{Bianchi2024} (\cref{eq:Ansatz_1,eq:Ansatz_2})}.
\end{figure}

Having found this result, it remains to compare it to a numerical evaluation of the microcanonical SFF for the Gaussian matrix model for arbitrary $\upbeta$ and the predictions of \cite{Bianchi2024}. We present the results for the (connected part of the) microcanonical SFFs, computed in the way outlined above, for various choices of the matrix size $N_m$ and numbers of realisations averaged over ($N_r$) in \cref{fig:AllSFFs}. It is interesting to note here that we observe a divergence at $x=1$ for all results with $\upbeta>2$, while for all other values we observe a transition to the plateau without divergences. This feature is not reproduced by our result (\cref{eq:b_upbeta_WD}) which for all cases except $\upbeta\in\qty{1,2}$ diverges logarithmically at $x=\frac 12 \sqrt{\upbeta}$. This can be seen directly in \cref{fig:ComparisonSFFs}, where the numerical results for the values of $\upbeta=1.4$ and $\upbeta=3$ are depicted in green while our prediction is put as the black solid line. These lines diverge at the expected points, indicating that at this point latest, the effects from the non-Wigner-Dyson parts of the Airy WP volumes are crucial. Consequently, we can only expect good results from our prediction for all values of $\upbeta>1$ if we restrict to a range of values of $x$ smaller than $x=\frac 12$. Looking at the numerical curve we can, however, see very good agreement of our result with the numerical data before the onset of the divergence. In fact, the agreement in this range is better than that with the predictions of \cite{Bianchi2024}, put as the black broken line, which however give a good approximation to the result over the whole range of $x$ considered. 

\begin{figure}[h]
\begin{subfigure}[b]{0.49\textwidth}
    \centering
    \includegraphics[width=\linewidth]{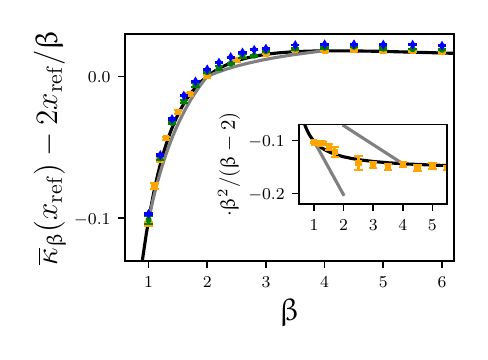}
    \caption{$x_{\text{ref}}=0.25$}
    \label{fig:SFF_analyics_0.25}
\end{subfigure}
\hfill
\begin{subfigure}[b]{0.49\textwidth}
    \centering
    \includegraphics[width=\linewidth]{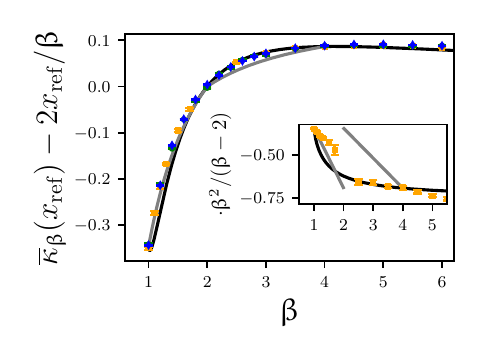}
    \caption{$x_{\text{ref}}=0.5$}
    \label{fig:SFF_analyics_0.4}
\end{subfigure}
\caption{Comparison of the numerical results at a specific point $x_{\text{ref}}$ for the (normalised) microcanonical SFF subtracting the ``ramp'' with our analytical result \cref{eq:b_upbeta_WD} and the ansatz of \cite{Bianchi2024}. In the main plot, the numerical results are put as the coloured dots with error bars, where the colours correspond to the same choices of $N_m$ and $N_r$ as in \cref{fig:AllSFFs}. Furthermore, our prediction is put as the solid black line, while that of \cite{Bianchi2024} is put as the solid grey line. In the inset, all three results are scaled by $\frac{\upbeta^2}{\qty(\upbeta-2)}$, where we only put the best converged numerical results, i.e. that for $N_m=1000$, and retain their and the predictions' depiction from the main plot.}
\label{fig:SFF_analytics_comparison}
\end{figure}

To better compare the two predictions with the numerical results, we do so in \cref{fig:SFF_analytics_comparison} for fixed values of $x$ while varying $\upbeta$ from $\upbeta=1$ to $\upbeta=6$. Here, we chose two examples for the reference values $x_{\text{ref}}$ which exemplify the behaviour we observe generically and decided to present not the full microcanonical SFF but to subtract the arbitrary $\upbeta$ generalisation of the ``ramp'' which is identical in our prediction and in that of \cite{Bianchi2024}, as discussed above. We plot, for the respective $x_{\text{ref}}$, as dots with error bars the numerical results for the choices of the matrix and ensemble sizes $N_m$ and $N_r$ where the colours represent the choice of these in the same way as in \cref{fig:AllSFFs}. Furthermore, we put our analytical prediction as the black and that of \cite{Bianchi2024} as the grey solid line. In the inset, we plot the numerical result with $N_m=1000$ and the analytical predictions, all scaled by $\frac{\upbeta^2}{\upbeta-2}$, which is a scaling suggested by our analytical result. Considering first \cref{fig:SFF_analyics_0.25}, where we set $x_{\text{ref}}=0.25$, we note that over the whole considered range of $\upbeta$ our result agrees very well, nearly perfectly within the error bars, with the numerical result, while the prediction of \cite{Bianchi2024} shows clear deviations. This becomes even more apparent when looking at the inset. In \cref{fig:SFF_analyics_0.4}, where we put $x_{\text{ref}}=0.5$, we can make the same observation, i.e. our predictions fitting nearly perfectly to the numerical results, for $\upbeta\geq 2$. Going to smaller values of $\upbeta$, we see that the numerics is increasingly well described by the prediction of \cite{Bianchi2024}. This, we attribute to the increasing impact of the divergence at $x=\frac 12 \sqrt{\upbeta}$ which for this range of $\upbeta$ is progressively near to $x_{\text{ref}}$. From the inset, we can observe the described behaviour even better. 

Consequently, we find in the range where it is applicable, i.e. for values of $x$ not influenced by the divergence, better agreement of our analytical prediction (\cref{eq:b_upbeta_WD}) with the numerical results as compared to that of the predictions of \cite{Bianchi2024} (cf. \cref{eq:Ansatz_1,eq:Ansatz_2}). This gives rise to the reasonable expectation, that the extension of our prediction for $b_\upbeta(x)$ by the non-Wigner-Dyson contributions will lead to an even better approximation of the numerical result and possibly a full analytical result for the microcanonical SFF. What one can already say for certain about this extension is, that it has to encompass the cancellation of the divergence present in \cref{eq:b_upbeta_WD} and replacing it by one at $x=1$. This could potentially result in the final result of computing the universal microcanonical SFF from $\upbeta$ topological gravity being actually of close similarity to that of \cite{Bianchi2024}. As a final comment, let us briefly remark that the series expansion of \cref{eq:b_upbeta_WD} fully agrees with the Wigner-Dyson parts of the analytic results for the series expansion of the microcanonical SFF for small $x$ in \cite{Forrester2001,Forrester2021} (In fact, with some additional work that is presented in \cite{Weber2025c}, one can see that these results can be reproduced up to $\order{x^8}$ by our results for the topological expansion of the $\upbeta$ topological gravity SFF (\cref{eq:kappa_0_general} to \cref{eq:kappa_7_2_general}). 

In conclusion, we can summarise that we find strong indicators pointing to the presence of quantum chaos, in the guise of universal RMT behaviour, in topological gravity for arbitrary Dyson index. First, the persistence of the constraints fulfilled by the Airy WP volumes for $\upbeta=1$ (and $\upbeta=4$) in their non-Wigner-Dyson contributions indicates the possibility of extension of the mechanism of cancellations necessary to find universal behaviour in the GOE variety of topological gravity to the arbitrary $\upbeta$ case. Secondly, building on this, we could produce an analytical result for an important part of the universal microcanonical SFF, based solely on the general structure of the Airy WP volumes we found in \cref{sec:AiryWP}. The good agreement of this result, which notably is an expression valid for all values of $\upbeta$ without necessitating the inclusion of several cases, with a numerical evaluation of the full microcanonical SFF for arbitrary $\upbeta$ Gaussian matrix models in its regime of validity gives another strong sign for the presence of universal RMT behaviour. Furthermore, it opens up the interesting possibility of a new way to study the whole universal microcanonical SFF for arbitrary Dyson index analytically, a program for which the considerations presented here may represent an important first step.

\section{Conclusion}\label{sec:Conclusion}
In this paper we have successfully implemented the extension to arbitrary Dyson index of the duality between topological gravity and the Airy matrix model in order to define the general $\upbeta$ Airy WP volumes that determine the perturbative expansions of the gravitational correlation functions. The study of the general structure of these volumes in terms of $\upbeta$ revealed that they are not mere interpolations between the orientable and unorientable Airy WP volumes but rather entail contributions that are genuinely non-Wigner-Dyson. From the geometric side, this observation is challenging insofar as, to our knowledge, the distinction between orientable and unorientable is a dichotomy for surfaces. Consequently, geometrically the naive idea of interpolating between orientable and unorientable by scaling the weights of their contributions in the gravitational path integral would be tempting. Our study now shows that the persistence of the duality to this setting implies that this is not the correct way of defining moduli space volumes in between orientability and unorientability. 

We gain further insight into this behaviour by our study of a generalisation of the Mirzakhani-like recursion for the unorientable WP volumes to the setting of arbitrary Dyson index. There, we first find an alternative, albeit equivalent, structure of the volumes in terms of $\upbeta$. In this alternative structure there is a one-to-one correspondence between the $\upbeta$ dependence of a specific term in the volume and a particular decomposition of the respective surface into 3-holed spheres (pairs of pants) and crosscaps, i.e. a particular contribution to the volume for $\upbeta=1$ (or equivalently $\upbeta=4$). From this point of view, the WP volumes are indeed a superposition of orientable and unorientable contributions, resolving the tension of the general structure of the volumes with the existence of only orientable and unorientable surfaces.  However, the weight of a specific surface depends on its properties beyond just orientability/unorientability, like e.g. the number of crosscaps it contains. This structure can thus be understood as the individual prefactors, which are generically non vanishing for the Wigner-Dyson classes (though, of course, all unorientable contributions drop in the unitary class), containing the Wigner-Dyson as well as the non-Wigner-Dyson behaviour. Our original structure can be computed from this by essentially splitting up the contributions into their Wigner-Dyson and non-Wigner-Dyson contributions and combining those that now have the same dependence on the Dyson index.

This decomposition is preferential in the considerations of the second part of our work where we investigated the question whether, in the sense of the BGS conjecture applied for the SFF, arbitrary $\upbeta$ topological gravity is quantum chaotic. While this goes along the same lines as for GOE ($\upbeta=1$) in the case of $\upbeta=4$, for general Dyson index there is the immediate problem that there is no full analytic result for the microcanonical SFF in this setting. In the absence of this, we study the question of chaoticity of the theory by investigating the constraints imposed on the Wigner-Dyson part of the Airy WP volumes by the fidelity to the universal predictions of random matrix theory in these cases, actually extending the discussion of \cite{Weber2024}. Interestingly, we observe that almost all of the constraints are also fulfilled by the non-Wigner-Dyson part of the Airy WP volumes for arbitrary Dyson index. This we interpret as a strong sign for the fidelity of $\upbeta$ topological gravity to a universal result, yet to be determined analytically. We extend this discussion by comparing our (perturbative) results for the microcanonical SFF in the universal regime for times before the plateau to the results of a numerical study of the Gaussian matrix model with arbitrary Dyson index and an ansatz for the SFF proposed in the literature \cite{Bianchi2024}. The comparison indicates good agreement of our results with the numerics in the studied regime which is also present in the proposed ansatz that is however at tension with certain symmetries of the perturbative expansion of the SFF.
Motivated by this agreement and to better compare with the ansatz, we compute an important contribution to the general $\upbeta$ universal microcanonical SFF, namely the contribution of the Wigner-Dyson part of the theory, by putting together the general structure in terms of $\upbeta$, proven in the first part, and the known universal result for $\upbeta=1$. Comparison of this result with the numerics shows improved agreement. This is especially pronounced when scaling the SFF in a particular $\upbeta$ dependent way which also highlights certain deviations of the proposed ansatz from the numerical result. However, our result is not in full agreement with the numerical result due to the $\upbeta$ dependent position of the divergence of the microcanonical SFF, which from the numerics is to be expected at a time independent of $\upbeta$. This we believe to be corrected upon including the contributions of non-Wigner-Dyson origin which is under present scrutiny \cite{Weber2025b}. 

Alternatively, it would be interesting to find the microcanonical SFF of the Gaussian matrix model and arbitrary Dyson index by different means. One way to do this would, of course, be to revisit the derivation of the ``traditional'' results given e.g. in \cite{Mehta2004} and find a way to generalise the discussion to arbitrary $\upbeta$. Apart from this, it is possible to consider any derivation of the Wigner-Dyson results and examine its extendability. One example of such a method could be the application of nonlinear $\sigma$-model techniques (cf. e.g. \cite{Efetov1996}) to compute the matrix integral for arbitrary Dyson index, potentially by rewriting the matrix model of \cite{Dumitriu2002} as a superintegral. Another method, more inspired by geometry, would be to extend the computation of correlation functions of branes in \cite{Saad2019,Blommaert2021}, yielding the unitary version of the universal two-point function of level densities, first to the unorientable setting and then to that of arbitrary Dyson index. Along this line of thinking, it would be very interesting to find an extension of the universe field theory (providing also access to the non-perturbative regime of the theory) introduced in \cite{Post2022} to the unorientable/arbitrary Dyson index setting which would also enable an analytical approach.

Beyond this interesting direction of research, there are several other things that would be worth considering. First of all, there is the obvious question for the actual computation of the WP volumes in JT gravity with arbitrary Dyson index. Of course, we have already studied their general structure in the main text but it would nevertheless be interesting to study the full results, in order to assess the correlation of their expected form as discussed in \cite{Tall2024} with the dependence on the Dyson index. Furthermore, this computation would enable the study of the interplay of the dependence of the logarithmic divergences of the WP volumes, regulated for example by the $\epsilon$-description of \cite{Stanford2023} or by using the matrix model dual to the minimal string as in \cite{Tall2024}, with the Dyson index. One reason why this is interesting is rooted in the observation, made in \cref{sec:Geometric_Arb_beta}, that there is a way to write the arbitrary $\upbeta$ WP volumes in a form from which they manifestly are a sum of contributions arising from decompositions of the surface with different numbers of crosscaps. Thus, one can speak of certain parts of a volume as corresponding to a sector of moduli space with a well-defined number of crosscaps. On the other hand, since the volumes' divergences are purely a result of the divergence of the moduli space volume of the crosscap one is led to associate the decomposition of the volume into parts of different degrees of divergence with a decomposition into parts of different numbers of crosscaps. Performing the computation to find the arbitrary $\upbeta$ WP volumes for some examples, easiest along the lines of \cite{Tall2024}, would shed light on the interplay of this intuition with the geometrically motivated one in terms of the Dyson index and would be an important step in the study of the moduli space volumes of unorientable surfaces. 

Furthermore, it would be interesting to go beyond the bosonic variations of JT/topological gravity between all of which we can now tune by varying the Dyson index $\upbeta$ and go to the variety of the theory including supermanifolds in the gravitational path integral. For these varieties of JT gravity the duality to a double-scaled matrix model, now with the super JT spectral curve $y^{\text{SJT}}(z)=-\frac{\sqrt{2}}{z}\cos\qty(2\pi z)$, persists as shown in \cite{Stanford2019}. Taking also these varieties of JT gravity into account the dual matrix models exhaust the full ten-fold classification of random matrices \cite{Altland1997}. As explained in \cite{Stanford2019}, the considerations of the additional ensembles require the inclusion of the additional parameter $\upalpha$ into the definition of the matrix models via a partition function (cf. \cref{eq:Z} in the bosonic setting). For these models, as for the bosonic varieties, the loop equations enabling the computation of the perturbative expansion of matrix model correlation functions can be written down not only for the values of $(\upalpha,\upbeta)$ corresponding to actual ensembles in the Altland/Zirnbauer classification but also for intermediate values. This opens up the possibility to perform similar considerations as in this work for these matrix models and study their behaviour for arbitrary values of the two parameters $\upalpha$ and $\upbeta$.

\section*{Acknowledgements}
We thank F. Haneder, J. Tall, P.J. Forrester and M. Rozali for valuable discussions and acknowledge financial support from the Deutsche Forschungsgemeinschaft (German Research Foundation) through Ri681/15-1 (project number 456449460) within the Reinhart-Koselleck Programme. M.L. acknowledges financial support from the Hanns Seidel Foundation.
\appendix
\section{Collection of $\upbeta$ resolvents}\label{sec:Coll_Resolvents}
Here we collect a list of the lower genus and boundary resolvents for arbitrary $\upbeta$ we computed, referring for a complete list to the supplementary material. Before we start the list, for which we use the invariant decomposition we have motivated in the main text, we present here an example for the non-decomposed form of a resolvent
\begin{align}
\begin{aligned}
    R_1^\upbeta(z_1,z_2)=&\frac{1}{16 \upbeta ^3 z_1^7 z_2^7 \left(z_1+z_2\right){}^4}\Big [5 (\upbeta  (5 \upbeta -18)+20) (z_1^8+z_2^8)\\
    &+20 (\upbeta  (5 \upbeta -18)+20) \qty(z_2 z_1^7+z_1 z_2^7)
    +33 (\upbeta  (5 \upbeta -18)+20) \qty(z_2^2 z_1^6+z_1^2 z_2^6)\\
    &+16 (\upbeta  (11 \upbeta -40)+44) \qty(z_2^3 z_1^5+z_1^3 z_2^5)
    +8 (\upbeta  (23 \upbeta -85)+92) z_2^4 z_1^4\Big],
\end{aligned}
\end{align}
whose $\upbeta$ dependence is rather spurious compared to the decomposed version presented below. 

In order to make the resolvents' presentation  more compact, we define $f(z_1,z_2;a_0,\dots,a_n):=a_0z_1^{2n}+a_1z_1^{2n-1}z_2+\dots a_{n-1}z_1^{n+1}z_2^{n-1}+a_nz_1^nz_2^n+a_{n-1}z_1^{n-1}z_2^{n+1}\dots$
\paragraph{$\mathbf{n=1}$}
\begin{align}
R_{0}(z_1)=&\frac{z_{1}}{2}\\
R_{1/2}^\upbeta(z)=&-\frac{2-\upbeta}{4 z^2\upbeta}\\
    R_{1}^\upbeta(z)=&-\frac{5 (2-\upbeta)^2}{16 \upbeta^2 z^5}-\frac{1}{8 \upbeta z^5}\\
    R_{3/2}^\upbeta(z)=&-\frac{15 (1-\upbeta) (4-\upbeta) (2-\upbeta)}{16 \upbeta^3 z^8}-\frac{2 (2-\upbeta)}{\upbeta^2 z^8}\\
    R_{2}^\upbeta(z)=&-\frac{1105 (1-\upbeta) (4-\upbeta) (2-\upbeta)^2}{256 \upbeta^4 z^{11}}-\frac{3465 (2-\upbeta)^2}{256 \upbeta^3 z^{11}}-\frac{105}{64 \upbeta^2 z^{11}}\\
    R_{5/2}^\upbeta(z)=&-\frac{1695 (1-\upbeta)^2 (2-\upbeta) (4-\upbeta)^2}{64 \upbeta^5 z^{14}}-\frac{9067 (1-\upbeta) (2-\upbeta) (4-\upbeta)}{64 \upbeta^4 z^{14}}\nonumber\\
    &-\frac{160 (2-\upbeta)}{\upbeta^3 z^{14}}\\
    R_{3}^\upbeta(z)=&-\frac{414125 (1-\upbeta)^2 (4-\upbeta)^2 (2-\upbeta)^2}{2048 \upbeta^6 z^{17}}\nonumber\\
    &-\frac{696205 (1-\upbeta) (4-\upbeta) (2-\upbeta)^2}{512 \upbeta^5 z^{17}}\nonumber\\
    &-\frac{4239235 (2-\upbeta)^2}{2048 \upbeta^4 z^{17}}-\frac{25025}{256 \upbeta^3 z^{17}}\\
    R_{7/2}^\upbeta(z)=&-\frac{59025 (1-\upbeta)^3 (2-\upbeta) (4-\upbeta)^3}{32 \upbeta^7 z^{20}}\nonumber\\
    &-\frac{8709175 (1-\upbeta)^2 (2-\upbeta) (4-\upbeta)^2}{512 \upbeta^6 z^{20}}\nonumber\\
    &-\frac{23421111 (1-\upbeta) (2-\upbeta) (4-\upbeta)}{512 \upbeta^5 z^{20}}-\frac{35840 (2-\upbeta)}{\upbeta^4 z^{20}}\\
    R_{4}^\upbeta(z)=&-\frac{1282031525 (1-\upbeta)^3 (2-\upbeta)^2 (4-\upbeta)^3}{65536 \upbeta^8 z^{23}}\nonumber\\
    &-\frac{13859296175 (1-\upbeta)^2 (2-\upbeta)^2 (4-\upbeta)^2}{65536 \upbeta^7 z^{23}}\nonumber\\
    &-\frac{45213403895 (1-\upbeta) (2-\upbeta)^2 (4-\upbeta)}{65536 \upbeta^6 z^{23}}\nonumber\\
    &-\frac{44972612685 (2-\upbeta)^2}{65536 \upbeta^5 z^{23}}-\frac{56581525}{4096 \upbeta^4 z^{23}}
    \end{align}
    \paragraph{$\mathbf{n=2}$}
    \begin{align}
    R_{0}(z_1, z_2)=&\frac{1}{2 \upbeta z_{1} z_{2} \left(z_{1} + z_{2}\right)^{2}}\\
    R_{1/2}(z_1, z_2)=&- \frac{\left(\upbeta - 2\right) \left(z_{1}^{4} + 3 z_{1}^{3} z_{2} + 3 z_{1}^{2} z_{2}^{2} + 3 z_{1} z_{2}^{3} + z_{2}^{4}\right)}{2 \upbeta^{2} z_{1}^{4} z_{2}^{4} \left(z_{1} + z_{2}\right)^{3}}\\
    R_{1}(z_1, z_2)=&(2-\upbeta)^2 \frac{f\left(z_1,z_2;25,100,165,176,184\right)}{16 \upbeta^3 z_1^7 z_2^7 \left(z_1+z_2\right){}^4}+\frac{5 z_1^4+3 z_2^2 z_1^2+5 z_2^4}{8 \upbeta^2 z_1^7 z_2^7}\\
    R_{3/2}(z_1, z_2)=&\frac{(1-\upbeta)(2-\upbeta)(4-\upbeta)f\left(z_1,z_2;120,600,1290,1700,1810,1865,1866\right)}{16\upbeta^{4}z_{1}^{10}z_{2}^{10}\left(z_{1}+z_{2}\right){}^{5}}\nonumber\\
    &+\frac{(2-\upbeta)f\left(z_1,z_2;256,1280,2752,3590,3710,3739,3750\right)}{16\upbeta^{3}z_{1}^{10}z_{2}^{10}\left(z_{1}+z_{2}\right){}^{5}}
\end{align}
\section{Collection of $\upbeta$ Airy WP volumes}\label{sec:WPV_collection}
Here we collect some of the results for the Airy WP volumes at low genus and number of boundaries. For a complete list we refer to the supplementary material.

For the two-boundary case, making use of the symmetry of the Airy WP volumes under $b_1\leftrightarrow b_2$, we can focus on $V^>_{g(,2)}(b_1,b_2)$, defined by $V_g(b_1,b_2)\eqqcolon V^>_g(b_1,b_2)\theta(b_1-b_2)+V^>_g(b_2,b_1)\theta(b_2-b_1)$, which is a more convenient way of writing the volumes than writing out the individual $\mathcal{V}_{g,2}^i(b_1,b_2)$
\begin{align}
V_{1/2}(b_1)=&\frac{2 - \upbeta}{2 b_{1} \upbeta},\\
V_{1}(b_1)=&\frac{5 b_1^2 (2-\upbeta)^2}{48 \upbeta^2}+\frac{b_1^2}{24 \upbeta}\\
V_{3/2}(b_1)=&\frac{b_1^5 (1-\upbeta) (2-\upbeta) (4-\upbeta)}{384 \upbeta^3}+\frac{b_1^5 (2-\upbeta)}{180 \upbeta^2},\\
V_{0}(b_1, b_2)=&\frac{2 \delta{\left(- b_{1} + b_{2} \right)}}{b_{2} \upbeta},\\
V^>_{1/2}(b_1, b_2)=&\frac{b_1 (2-\upbeta)}{\upbeta^2},\\
V^>_{1}(b_1, b_2)=&\frac{\left(5 b_1^4+10 b_2^2 b_1^2+8 b_2^3 b_1+b_2^4\right) (2-\upbeta)^2}{96 \upbeta^3}+\frac{\left(b_1^2+b_2^2\right)^2}{48 \upbeta^2},\\
V^>_{3/2}(b_1, b_2)=&(4-\upbeta) (2-\upbeta) (1-\upbeta)\nonumber\\
&\times\frac{\left(30 b_1^7+210 b_2^2 b_1^5+175 b_2^3 b_1^4+210 b_2^4 b_1^3+105 b_2^5 b_1^2+91 b_2^6 b_1+5 b_2^7\right) }{40320 \upbeta^4}\nonumber\\
&(2-\upbeta)\nonumber\\
&\times\frac{\left(64 b_1^7+448 b_2^2 b_1^5+245 b_2^3 b_1^4+560 b_2^4 b_1^3+147 b_2^5 b_1^2+175 b_2^6 b_1+23 b_2^7\right) }{40320 \upbeta^3},
\end{align}
\begin{align}
V^>_{2}(b_1, b_2)&=\frac{(1-\upbeta) (2-\upbeta)^2 (4-\upbeta)}{46448640 \upbeta^5}(221 b_1^{10}+3315 b_2^2 b_1^8+2880 b_2^3 b_1^7+9282 b_2^4 b_1^6+\nonumber\\
&+6048 b_2^5 b_1^5+10290 b_2^6 b_1^4+2880 b_2^7 b_1^3+2235 b_2^8 b_1^2+768 b_2^9 b_1+53 b_2^{10})\nonumber\\
&+ \frac{(2-\upbeta)^2}{232243200 \upbeta^4}(3465 b_1^{10}+51975 b_2^2 b_1^8+30720 b_2^3 b_1^7+162330 b_2^4 b_1^6+\nonumber\\
&+64512 b_2^5 b_1^5+158970 b_2^6 b_1^4+46080 b_2^7 b_1^3+32535 b_2^8 b_1^2+10240 b_2^9 b_1+1465 b_2^{10})+\nonumber\\
&+\frac{b_1^{10}+15 b_2^2 b_1^8+58 b_2^4 b_1^6+58 b_2^6 b_1^4+15 b_2^8 b_1^2+b_2^{10}}{552960 \upbeta^3},\\
V_{0}(b_1, b_2, b_3)=&\frac{4}{\upbeta^{2}},\\
\mathcal{V}^1_{\frac{1}{2}}(b_1, b_2, b_3)=& \frac{b_1^3}{6}\Big[-\theta \left(-b_1-b_2+b_3\right)-\theta \left(-b_1+b_2+b_3\right)+\nonumber\\
&+\theta \left(b_2-b_1\right) \left(\theta \left(b_1-b_2+b_3\right)+\theta \left(-b_1+b_2+b_3\right)-2\right)+4\Big]\nonumber\\
&+\frac{1}{2} b_1^2 \Big[\left(b_2-b_3\right) \theta \left(-b_1-b_2+b_3\right)-\left(b_2+b_3\right) \theta \left(-b_1+b_2+b_3\right)+\nonumber\\
&+\theta \left(b_2-b_1\right) \left(b_2 \left(\theta \left(b_1-b_2+b_3\right)+\theta \left(-b_1+b_2+b_3\right)-2\right)\right)+\nonumber\\
&+b_3 \left(\theta \left(-b_1+b_2+b_3\right)-\theta \left(b_1-b_2+b_3\right)\right)\Big]+\nonumber\\
&+\frac{1}{2} b_1 \Bigg\{b_2^2 \Big[-\theta \left(-b_1-b_2+b_3\right)-\theta \left(-b_1+b_2+b_3\right)+\nonumber\\
&+\theta \left(b_2-b_1\right) \left(\theta \left(b_1-b_2+b_3\right)+\theta \left(-b_1+b_2+b_3\right)-2\right)+2\Big]+\nonumber\\
&+2 b_3 b_2 \Big[\theta \left(-b_1-b_2+b_3\right)-\theta \left(-b_1+b_2+b_3\right)+\nonumber\\
&+\theta \left(b_2-b_1\right) \left(\theta \left(-b_1+b_2+b_3\right)-\theta \left(b_1-b_2+b_3\right)\right)\Big]+\\
&+b_3^2 \Big[-\theta \left(-b_1-b_2+b_3\right)-\theta \left(-b_1+b_2+b_3\right)+\nonumber\\
&+\theta \left(b_2-b_1\right) \left(\theta \left(b_1-b_2+b_3\right)+\theta \left(-b_1+b_2+b_3\right)-2\right)+2\Big]\Bigg\}\nonumber\\
&+\frac{b_3 b_2^2}{2} \Big[\theta \left(b_2-b_1,b_1-b_2+b_3\right)+\theta \left(-b_1-b_2+b_3\right)+\nonumber\\
&-\theta \left(b_1-b_2\right) \theta \left(-b_1+b_2+b_3\right)\Big]+\frac{b_3^3}{6} \Big[\theta \left(b_2-b_1,b_1-b_2+b_3\right)+\nonumber\\
&+\theta \left(-b_1-b_2+b_3\right)-\theta \left(b_1-b_2\right) \theta \left(-b_1+b_2+b_3\right)+2\Big]+\nonumber\\
&+\frac{b_2^3}{6} \Big[-\theta \left(-b_1-b_2+b_3\right)+\theta \left(-b_1+b_2+b_3\right)+\nonumber\\
&-\theta \left(b_2-b_1\right) \left(\theta \left(b_1-b_2+b_3\right)+\theta \left(-b_1+b_2+b_3\right)-2\right)+2\Big]+\nonumber\\
&-\frac{b_3^2 b_2}{2} \Big[\theta \left(-b_1-b_2+b_3\right)-\theta \left(-b_1+b_2+b_3\right)+\nonumber\\
&+\theta \left(b_2-b_1\right) \left(\theta \left(b_1-b_2+b_3\right)+\theta \left(-b_1+b_2+b_3\right)-2\right)\Big]\nonumber.
\end{align}

\section{Proof of the relation of resolvents of the $\upbeta$ ensembles for $g=0$}\label{sec:Proof_Relation_R_0}
In this appendix we prove the relation \cref{eq:relation_genus0} of the genus 0 resolvents for arbitrary $\upbeta$, claimed in the main text. For convenience we recall that this relation was given by
\begin{align}\label{eq:claim_relation}
     R^\upbeta_0\qty(I)=\frac{1}{\upbeta^{n-1}}R^1_0\qty(I).
\end{align}
For the proof we will go along the lines of the derivation of the perturbative loop equations in \cite{Stanford2019}. The starting point there are the loop equations for resolvents derived for a $\upbeta$ matrix model determined by a potential $V(x)$. Inserting the perturbative expansion of the resolvents into these it is shown that for $x$ near the cut one finds
\begin{align}\label{eq:R_0_sum_F}
    2y(x)R^\upbeta_0(x,I)+F^\upbeta_0(x,I)=\qty(\text{analytic in }x),
\end{align}
which then by a dispersion relation argument yields the relations recalled in \cref{sec:Recap_LE}. It is important to note that while for the discussion in the main text it is sufficient to focus on the case of double-scaled matrix models, i.e. the support of $\rho_0$ being $\mathbb{R}_+$, the dispersion relation argument is possible in the general one-cut case as well. From \cref{eq:R_0_sum_F} we can already see that for a proof of our claim it suffices to show
\begin{align}\label{eq:lemma_proof_relation}
    F^\upbeta_0\qty(x,I)=\frac{1}{\upbeta^{n}}F^1_0\qty(x,I),
\end{align}
since a multiplicative factor carries through the dispersion relation computation for the double-scaled as well as the one-cut case. 

However, as we remarked in the main text there, the expressions for $F^\upbeta_g(x,I)$ for the case $g=0$ are different from the general. The first task we have to perform though is to derive this expressions. 

Starting with the loop equations given by \cite{Stanford2019}\footnote{Note, that the dependence on $\upbeta$ is tacitly given here by the average only and thus there is no superscript $\upbeta$ put on the observables.}
\begin{equation}\label{eq:LoopEq}
			\begin{aligned}
			-N\ev{ P(x,I)}_c=&\left(1-\frac 2 \upbeta\right)\partial_x\ev{ R(x,I)}_c+\ev{ R(x,x,I)}_c+\sum_{J\supseteq I}\ev{ R(x,J)}_c \ev{R(x,I\backslash J)}_c\\
			& - N V^{\prime}(x)\ev{R(x,I)}_c+\frac 2 \upbeta \sum_{k=2}^{n}\partial_{x_k}\left[\frac{\ev{ R(x,I\backslash\left\{x_k\right\})}_c-\ev{ R(I)}_c}{x-x_k}\right],
			\end{aligned}
\end{equation}
we insert the perturbative expansion of the correlation functions of resolvents and collect the rhs in one sum. This yields
\begin{align}
\begin{aligned}
    (\text{analytic})=&\sum_{g=0}^\infty\frac{1}{N^{2g+\abs{I}-2}}\bigg[ \qty(1-\frac{2}{\upbeta})\partial_x R^\upbeta_{g-\frac 12}(x,I)+R^\upbeta_{g-1}(x,x,I)+\\
    +&\sum_{\substack{J\subseteq I\\
    h+h'=g}} R^\upbeta_h(x,J)R^\upbeta_{h'}(x,I\backslash J)
    -V'(x)R^\upbeta_g(x,I)+\frac{2}{\upbeta} \sum_{k=1}^n\partial_{x_k}\frac{R^\upbeta_g(x,I\backslash x_k)-R^\upbeta_g(I)}{x-x_k} \bigg],
\end{aligned}
\end{align}
where by analytic we mean analytic in $x$ near the cut. From this one finds one equation for every order in $N^{-1}$, labelled by $g$. The relevant case here is $g=0$, plugging this we find
\begin{align}
\begin{aligned}
    \qty(\text{analytic})=\sum_{\substack{J\subseteq I}} R^\upbeta_0(x,J)R^\upbeta_{0}(x,I\backslash J) -V'(x)R^\upbeta_0(x,I)+\frac{2}{\upbeta} \sum_{k=1}^n\partial_{x_k}\frac{R^\upbeta_0(x,I\backslash x_k)-R^\upbeta_0(I)}{x-x_k}.
\end{aligned}
\end{align}
Analogously to \cite{Stanford2019}, we pull the terms containing $R_0(x)$ out of the sum and, assuming the $x_k$ are away from the cut, move all terms from the sum over derivatives to the lhs that are analytic near the cut. Thus we find
\begin{align}
    \qty(\text{analytic})=\underbrace{\qty[2R^\upbeta_0(x)-V'(x)]}_{2y(x)}R^\upbeta_0(x,I)+\sum_{\substack{J\subseteq I\\J\neq\emptyset\\J\neq I}} R^\upbeta_0(x,J)R^\upbeta_{0}(x,I\backslash J) +\frac{2}{\upbeta} \sum_{k=1}^n\frac{R^\upbeta_0(x,I\backslash x_k)}{\qty(x-x_k)^2},
\end{align}
where we used the relation of the spectral curve $y(x)$ with the potential defining the matrix model. Thus we see that $F^\upbeta_g(x,I)$ for $g=0$ is given by\footnote{Note, that this yields the results for $F^\upbeta_0(x,I)$ in the special cases $\abs{I}=1,2$ given in the main text and agrees with the general result in the other cases, where it's applicable.}
\begin{align}
    F^\upbeta_0(x,I)=\sum_{\substack{J\subseteq I\\J\neq\emptyset\\J\neq I}} R^\upbeta_0(x,J)R^\upbeta_{0}(x,I\backslash J) +\frac{2}{\upbeta} \sum_{k=1}^n\frac{R^\upbeta_0(x,I\backslash x_k)}{\qty(x-x_k)^2}.
\end{align}
Having found this we conclude our proof by showing \cref{eq:lemma_proof_relation} by induction. The base clause is given by the case $I=\qty{x_1}$ for which one finds
\begin{align}
    F^\upbeta_0(x,I)=\frac{2}{\upbeta} \frac{R_0(x)}{\qty(x-x_1)^2},
\end{align}
which evidently fulfils the statement we would like to proof. Now we assume \cref{eq:lemma_proof_relation} for all lengths of $I$ ,$k\in \mathbb N$ smaller than a given $n\in \mathbb{N}$. This implies \cref{eq:claim_relation} for all $k\leq n $\footnote{Note, that the $I$ appearing on \cref{eq:claim_relation} has one element less than the $I$ appearing in \cref{eq:lemma_proof_relation}}. For $\abs{I}=n$ we find 
\begin{align}
\begin{aligned}
    F^\upbeta_0(x,I)&=\sum_{\substack{J\subseteq I\\J\neq\emptyset\\J\neq I}} R^\upbeta_0(x,J)R^\upbeta_{0}(x,I\backslash J) +\frac{2}{\upbeta} \sum_{k=1}^n\frac{R^\upbeta_g(x,I\backslash x_k)}{\qty(x-x_k)^2}\\
    &=\sum_{\substack{J\subseteq I\\J\neq\emptyset\\J\neq I}} \frac{R^1_0(x,J)R^1_{0}(x,I\backslash J)}{\upbeta^{\abs{J}+\abs{I\backslash J}}} +\frac{2}{\upbeta} \sum_{k=1}^n\frac{R^1_g(x,I\backslash x_k)}{\upbeta^{\abs{I}-1}\qty(x-x_k)^2}\\
    &=\frac{1}{\upbeta^n} F^1_0(x,I),
\end{aligned}
\end{align}
which completes the induction step and thus our proof.

\section{Relation of $R^\upbeta$ with $R^{\frac 4 \upbeta}$}\label{sec:Proof_Relation_R}
It is a well known fact in the study of matrix models that the resolvents of the orthogonal and symplectic symmetry class are directly connected by the relation \cite{Stanford2019}
\begin{align}
    R_g^{1}(I)=(-1)^{2g}2^{2\qty(g+n-1)}R_g^{4}(I).
\end{align}
This relation follows directly from the loop equations and thus serves as a good sanity check for our general expressions for the resolvents in dependence on $\upbeta$ given in \cref{eq:Resolvent_beta_Integer} and \cref{eq:Resolvent_beta_halfInteger}. In fact, this relation is only an example of a more general invariance of the matrix model given by the invariance of the integral definition of the correlation functions under $\qty(\upbeta,N)\leftrightarrow\qty(\frac 4 \upbeta, -\frac{N\upbeta}{2})$ (e.g \cite{Eynard2018} and references therein). At the level of resolvents this leads us to conjecture
\begin{align}\label{eq:Rel_beta_4_beta_app}
    R^{\frac 4 \upbeta}_g(I)=\qty(-1)^{2g}\qty(\frac \upbeta 2)^{2\qty(g+\abs{I}-1)}R_g^{\upbeta}(I),
\end{align}
which we prove in the following by showing the relation to arise from the loop equations.

Let the $R^{\upbeta}_g(I)$ be the solutions to the loop equations for $\upbeta$. First, for $g=0$ by \cref{eq:relation_genus0} one has 
\begin{align}
\begin{aligned}
    R^{\frac 4 \upbeta}_g(I)&=\frac{\upbeta^{\abs{I}-1}}{2^{2\qty(\abs{I}-1)}}R^{1}_g(I)=\qty(\frac \upbeta 2)^{2\qty(\abs{I}-1)}\frac{1}{\upbeta^{2\qty(\abs{I}-1)}}R^{1}_g(I)\\
    &=\qty(\frac \upbeta 2)^{2\qty(\abs{I}-1)}R^{\upbeta}_g(I),
\end{aligned}
\end{align}
where we used \cref{eq:relation_genus0} to get to the last line. This shows our claim for the case of $g=0$. For $g>0$ it is easiest to use the expression of the respective contribution to the $n$-boundary resolvent as a contour integral, \cref{eq:RecursionRInt_z}, as arising from the loop equations. For convenience we recall this expression to be 
\begin{align}
\begin{aligned}\label{eq:R_F_invariance}
    R^\upbeta_g(-z^2,I)=\frac{1}{2\pi i z}\oint_{\qty[-i \infty+\epsilon,i\infty+\epsilon]}\frac{{z'}^2\dd z'}{{z'}^2-z^2}\frac{1}{y(-{z'}^2)}F^\upbeta_g(-z'^2,I),
\end{aligned}
\end{align}
with $\epsilon>0$ and
\begin{align}\label{eq:F_split_invariance}
    \begin{aligned}
			F^\upbeta_g(-z^2,I)\coloneqq&\underbrace{\qty(1-\frac 2 \upbeta)\frac{1}{-2z}\partial_z R^\upbeta_{g-\frac 12}(-z^2,I)}_{\Romannum{1}}\\ 
                        &+\underbrace{R^\upbeta_{g-1}(-z^2,-z^2,I)}_{\Romannum{2}}\\
					&+\underbrace{\sum'_{I\supseteq J, h} R^\upbeta_h(-z^2,J)R^\upbeta_{g-h}(-z^2,I\backslash J)}_{\Romannum{3}}\\
					&+2\underbrace{\sum_{k=1}^{n}\qty[R^\upbeta_0(-z^2,-z^2_k)+\frac 1 \upbeta \frac{1}{\qty(z_k^2-z^2)^2}]R^\upbeta_{g}(-z^2,I\backslash\qty{-z^2_k})}_{\Romannum{4}},
			\end{aligned}
\end{align}
where $\sum'$ is a notation for excluding $R_0(z)$ and $R_0(z,z_k)$ from the sum. For the sake of brevity we will use $z$ in the arguments of the contributions to the resolvents in the following.
Due to the recursive nature of this way of computing the topological expansion we use the cases of $g=0$, which we have shown above, as base clauses and perform the proof by induction. Assuming thus that for all resolvents necessary to compute $R^\upbeta_g(z,I)$, with $\abs{I}\eqqcolon n$, our claim holds, it remains to show that this implies our claim for $R^\upbeta_g(z,I)$. To do this, due to \cref{eq:R_F_invariance}, it suffices to show \footnote{Note that $R^\upbeta_g(z,I)$ has $n+1$ arguments.}
\begin{align}
    F_g^\upbeta(z,I)=\qty(-1)^{2g}\qty(\frac 2 \upbeta)^{2\qty(g+n)}F_g^{\frac 4 \upbeta}(z,I),
\end{align}
which we shall prove by considering each line of \cref{eq:F_split_invariance} separately and plugging our base assumption that the relation between $\upbeta$ and $\frac 4 \upbeta$ holds.

\begin{align}
\begin{aligned}
    \Romannum{1}&=\qty(1-\frac 2 \upbeta)(-1)^{2g-1}\qty(\frac 2 \upbeta)^{2g+2n-1}\frac{1}{-2z}\partial_z R^{\frac 4 \upbeta}_{g-\frac 12}(z,I)\\
    &=(-1)^{2g}\qty(\frac 2 \upbeta)^{2\qty(g+n)}\qty(1-\frac{\upbeta}{2})\frac{1}{-2z}\partial_z R^{\frac 4 \upbeta}_{g-\frac 12}(z,I)\\
    &=(-1)^{2g}\qty(\frac 2 \upbeta)^{2\qty(g+n)} \Romannum{1}(\upbeta\leftrightarrow \frac 4 \upbeta).
\end{aligned}
\end{align}

\begin{align}
    \begin{aligned}
        \Romannum{2}&=(-1)^{2g-2}\qty(\frac{2}{\upbeta})^{2\qty(g-1+n+2-1)}R^{\frac 4 \upbeta}_{g-1}(z,z,I)=(-1)^{2g}\qty(\frac{2}{\upbeta})^{2\qty(g+n)}R^{\frac 4 \upbeta}_{g-1}(z,z,I)\\
        &=(-1)^{2g}\qty(\frac 2 \upbeta)^{2\qty(g+n)} \Romannum{2}(\upbeta\leftrightarrow \frac 4 \upbeta).
    \end{aligned}
\end{align}
\begin{align}
    \begin{aligned}
        \Romannum{3}&=\sum'_{I\supseteq J, h}\qty(-1)^{2h+2g-2h}\qty(\frac{2}{\upbeta})^{2(h+\abs{J}+g-h+n-\abs{J})} R^{\frac 4 \upbeta}_h(z,J)R^{\frac 4 \upbeta}_{g-h}(z,I\backslash J)\\
        &=\qty(-1)^{2g}\qty(\frac{2}{\upbeta})^{2(g+n)}\sum'_{I\supseteq J, h} R^{\frac 4 \upbeta}_h(z,J)R^{\frac 4 \upbeta}_{g-h}(z,I\backslash J)\\
        &=(-1)^{2g}\qty(\frac 2 \upbeta)^{2\qty(g+n)} \Romannum{3}(\upbeta\leftrightarrow \frac 4 \upbeta).
    \end{aligned}
\end{align}

\begin{align}
    \begin{aligned}
        \Romannum{4}&=\sum_{k=1}^{n}\qty[\qty(\frac 2 \upbeta)^{2} R^{\frac 4 \upbeta}_0(z,z_k)+\frac 1 \upbeta \frac{1}{\qty(z_k^2-z^2)^2}]\qty(-1)^{2g}\qty(\frac 2 \upbeta)^{2\qty(g+n-1)}R^{\frac 4\upbeta}_{g}(z,I\backslash\qty{z_k})\\
        &=\qty(-1)^{2g}\qty(\frac{2}{\upbeta})^{2\qty(g+n)}\sum_{k=1}^{n}\qty[ R^{\frac 4 \upbeta}_0(z,z_k)+\frac{\upbeta}{4} \frac{1}{\qty(z_k^2-z^2)^2}]R^{\frac 4\upbeta}_{g}(z,I\backslash\qty{z_k})\\
        &=(-1)^{2g}\qty(\frac 2 \upbeta)^{2\qty(g+n)} \Romannum{4}(\upbeta\leftrightarrow \frac 4 \upbeta).
    \end{aligned}
\end{align}
\qed

Another way of performing the proof would of course be to use the general form of the resolvents proven in \cref{sec:Proof_GenStruct} and to show by an explicit computation that these obey the relation \cref{eq:Rel_beta_4_beta_app}.

As a final comment, we note that our proof generalises directly to the general one-cut case since it is purely based on the behaviour of $F_g^\upbeta$ under the transformation of $\upbeta$ and this does not change upon going to the general one-cut case. The only modification occurs in the necessity to modify \cref{eq:R_F_invariance}, which however doesn't affect our argument.

\section{Proof of the general structure}\label{sec:Proof_GenStruct}
Here we prove the statement made in the main text that the general form of the resolvents, in terms of their $\upbeta$ dependence, is given by \cref{eq:Resolvent_beta_Integer,eq:Resolvent_beta_halfInteger}, i.e.
\begin{align}\label{eq:Resolvent_integer}
    R^\upbeta_g(I)=\frac{1}{\upbeta^{2g+n-1}}\left(\mathcal R_g^0(I)\upbeta^g+(2-\upbeta)^2\sum_{i=1}^g \mathcal R_g^i(I)\upbeta^{i-1}((1-\upbeta)(4-\upbeta))^{g-i}\right),
\end{align}
for integer $g$ and
\begin{align}\label{eq:Resolvent_halfInteger}
    R^\upbeta_g(I)=\frac{1}{\upbeta^{2g+n-1}}\left((2-\upbeta)\sum_{i=1}^{g+\frac 12}\mathcal R_g^i(I)\upbeta^{i-1}((1-\upbeta)(4-\upbeta))^{g+\frac 12 -i}\right),
\end{align}
for half-integer $g$. 

We will perform the proof by means of induction, where the base clause, due to recursive nature of the computation of resolvents by means of the loop equations, is given already by the form of $R^\upbeta_{0}(z_1,z_2)$ but can of course be thought of to be provided by the set of examples computed in the main text that are in correspondence with the claimed structure. To be more precise, we use the expression of the $n+1$ boundary and genus $g$ resolvent
in terms of a contour integral, \cref{eq:RecursionRInt_z}, which we recall here for convenience :
\begin{align}
\begin{aligned}
    R^\upbeta_g(-z^2,I)=\frac{1}{2\pi i z}\oint_{\qty[-i \infty+\epsilon,i\infty+\epsilon]}\frac{{z'}^2\dd z'}{{z'}^2-z^2}\frac{1}{y(-{z'}^2)}F_g(-z'^2,I),
\end{aligned}
\end{align}
with $\epsilon>0$ and
\begin{align}\label{eq:F_split}
    \begin{aligned}
			F^\upbeta_g(-z^2,I)\coloneqq&\underbrace{\frac{\upbeta-2}{\upbeta}\frac{1}{-2z}\partial_z R^\upbeta_{g-\frac 12}(-z^2,I)}_{\Romannum{1}}\\ 
                        &+\underbrace{R^\upbeta_{g-1}(-z^2,-z^2,I)}_{\Romannum{2}}\\
					&+\underbrace{\sum'_{I\supseteq J, h} R^\upbeta_h(-z^2,J)R^\upbeta_{g-h}(-z^2,I\backslash J)}_{\Romannum{3}}\\
					&+2\underbrace{\sum_{k=1}^{n}\qty[R^\upbeta_0(-z^2,-z^2_k)+\frac 1 \upbeta \frac{1}{\qty(z_k^2-z^2)^2}]R^\upbeta_{g}(-z^2,I\backslash\qty{-z^2_k})}_{\Romannum{4}},
			\end{aligned}
\end{align}
where $\overset{\prime}{\sum}$ is a notation for excluding $R_0(-z^2)$ and $R_0(-z^2,-z^2_k)$ from the sum. Notably, this excludes the cases of $g=0$ where the claimed structure can however be seen already from \cref{eq:relation_genus0}, proven in \cref{sec:Proof_Relation_R}. An important thing to be noted at this point is that the contour integration doesn't yield any additional $\upbeta$ dependence which implies that the induction step is already concluded if the claimed structure, i.e. that for genus $g$ and $n+1$ boundaries, can be observed in $F_g^\upbeta (-z^2,I)$. Additionally, as already remarked above in \cref{sec:Proof_Relation_R}, this shows the applicability of our proof for the more general case of one cut and not necessarily double-scaled matrix models since the dependence on the spectral curve and hence the precise shape of the cut only enters upon contour integration.

We will thus consider each line of \cref{eq:F_split} separately and show that the claimed general form is present. Due to the general form being split into the case of integer and half-integer genus we treat these as two cases for each line. To abbreviate the following discussion we use the shorthand notation of writing a dependence on $-z^2$ just as a dependence on $z$.

\paragraph{\Romannum{1}} For the case of integer $g$, $g-\frac 12$ is half-integer and combining factors one finds that 
\begin{align}
    \Romannum{1}=\frac{\qty(2-\upbeta)^2}{\upbeta^{2g+n}}\sum_{i=1}^g\upbeta^{i-1}\qty[\qty(1-\upbeta)\qty(4-\upbeta)]^{g-i} \frac{1}{2z}\partial_z \mathcal R^i_{g-\frac 12}\qty(z,I),
\end{align}
reproducing the claimed structure, \cref{eq:Resolvent_integer}.

For half-integer $g$, $g-\frac{1}{2}$ is integer. For this case, as for many of the following, it is convenient to use the rewriting 
\begin{align}\label{eq:rel_2_1_4}
    \qty(2-\upbeta)^2=\qty(1-\upbeta)(4-\upbeta)+\upbeta.
\end{align}
By use of this and shifting the summation index, one finds
\begin{align}
\begin{aligned}
    \Romannum{1}=\frac{(2-\upbeta)}{\upbeta^{2g+n}}\frac{1}{2z}\partial_z \Biggl[ &\sum_{i=1}^{g+\frac 12} \mathcal R^{i}_{g-\frac 12}(z,I) \upbeta^{i-1}\qty[\qty(1-\upbeta)\qty(4-\upbeta)]^{g+\frac 12-i} \\
    &+\sum_{i=2}^{g+\frac 12} \mathcal R^{i-1}_{g-\frac 12}(z,I) \upbeta^{i-1}\qty[\qty(1-\upbeta)\qty(4-\upbeta)]^{g+\frac 12-i}\Biggr],
\end{aligned} 
\end{align}
where in the first line we set $\mathcal R^{g+\frac 12}_{g-\frac 12}=\mathcal R^{0}_{g-\frac 12}$ which is possible as $\mathcal R^{g+\frac 12}_{g-\frac 12}$ was not defined before. Taking the derivative inside the bracket and combining the two sums it is apparent that the structure of \cref{eq:Resolvent_halfInteger} is reproduced.

\paragraph{\Romannum{2}} For the case of integer $g$, $g-1$ is integer as well and by combining factors and shifting the index one finds
\begin{align}
    \Romannum{2}=\frac{1}{\upbeta^{2g+n}}\qty[\mathcal R^0_{g-1}\qty(z,z,I)\upbeta^g+\qty(2-\upbeta)^2\sum_{i=2}^g\upbeta^{i-1}\qty[\qty(1-\upbeta)\qty(4-\upbeta)]^{g-i}\mathcal \mathcal R_g^{i-1}\qty(z,z,I)],
\end{align}
reproducing the expected structure. For half-integer $g$, $g-1$ is half-integer and in a similar fashion as for the integer case one finds
\begin{align}
    \Romannum{2}=\frac{\qty(2-\upbeta)}{\upbeta^{2g+n}}\sum_{i=2}^{g+\frac 12} \upbeta^{i-1}\qty[\qty(1-\upbeta)\qty(4-\upbeta)]^{g+\frac 12-i}\mathcal R_g^{i-1}\qty(z,z,I),
\end{align}
being in correspondence with the expected structure for half-integer $g$.

\paragraph{\Romannum{3}} For the case of half-integer genus, there is the possibility of $h$ being an integer, then $g-h$ is half-integer and of $h$ being half-integer which then implies $g-h$ to be integer. For the determination of the structure in terms of $\upbeta$ it suffices, however, to consider the case of half-integer $h$ as for every integer $h$ the case of $h'$ (half-integer) such that $g-h'=h$ has already been considered. Keeping in mind the sum over $J\subseteq I$, for a fixed half-integer $h$ one has to evaluate $R_h^\upbeta(z,J)R^\upbeta_{g-h}(z,I\backslash J)\coloneqq \star$. To abbreviate the following discussion we will drop the arguments of the resolvents, as they are uniquely reconstructible by the lower index. For the evaluation of $\star$ is will be useful to recall the form of the Cauchy product for finite sums, given by
\begin{align}
    \sum_{i=1}^m \sum_{j=1}^k a_i b_j x^{i+j}=\sum_{l=2}^{m+k}x^l\sum_{n=\max\qty(1,l-k)}^{\min\qty(m,l)}a_n b_{l-n}.
\end{align}
Using this and \cref{eq:rel_2_1_4}, one can evaluate the product of the respective general expressions to find
\begin{align}
\begin{aligned}
    \star=&\frac{\qty(2-\upbeta)}{\upbeta^{2g+n}}\sum_{i=1+g-h}^{g+\frac 12}\mathcal R_h^0 \mathcal R_{g-h}^{i-g+h}\upbeta^{i-1}\qty[\qty(1-\upbeta)\qty(4-\upbeta)]^{g+\frac 12 - l}+\\
    &-\frac{\qty(2-\upbeta)}{\upbeta^{2g+n}}\sum_{l=2}^{g+\frac 12} \upbeta^{l-1}\qty[\qty(1-\upbeta)\qty(4-\upbeta)]^{g+\frac 12 - l}\sum_{a=\max\qty(1,l-g+h)}^{\min\qty(g+\frac 12, l)}\mathcal R_h^a \mathcal R_{g-h}^{l-a}+\\
    &+\frac{\qty(2-\upbeta)}{\upbeta^{2g+n}}\sum_{l=1}^{g-\frac 12} \upbeta^{l-1}\qty[\qty(1-\upbeta)\qty(4-\upbeta)]^{g+\frac 12 - l}\sum_{a=\max\qty(1,l+1-g+h)}^{\min\qty(g+\frac 12, l+1)}\mathcal R_h^a \mathcal R_{g-h}^{l+1-a},
\end{aligned}
\end{align}
which is of the form claimed for half-integer $g$, showing the claim for this case. Coming now to the case of integer genus $g$, we have to consider two cases, as now this can be split into an integer $h$ and consequently an integer $g-h$ or a half-integer $h$ which also implies $g-h$ to be half-integer. Considering again the case of half-integer $h$ first, one can evaluate 
\begin{align}
    \star=\frac{\qty(2-\upbeta)^2}{\upbeta^{2g+n}}\sum_{l=1}^{g} \upbeta^{l-1}\qty[\qty(1-\upbeta)\qty(4-\upbeta)]^{g- l}\sum_{a=\max\qty(1,l+\frac 12 -g+h)}^{\min\qty(h+\frac 12, l+1)}\mathcal R_h^a \mathcal R_{g-h}^{l+1-a},
\end{align}
which is of the claimed form. Coming now to the case of integer $h$, the expansion yields
\begin{align}
    \begin{aligned}
        \star=&\frac{1}{\upbeta^{2g+n}}\mathcal R^0_h \mathcal R_{g-h}^0+\\
        &+\qty(2-\upbeta)^2 \sum_{i=1+g-h}^g \mathcal R_h^{i+h-g}\mathcal R_h^0 \upbeta^{i-1}\qty[\qty(1-\upbeta)\qty(4-\upbeta)]^{g- i}+\\
        &+\qty(2-\upbeta)^2 \sum_{i=1+h}^g \mathcal R_h^{j-h}\mathcal R_h^0 \upbeta^{i-1}\qty[\qty(1-\upbeta)\qty(4-\upbeta)]^{g-i}+\\
        &-\qty(2-\upbeta)^2 \sum_{l=2}^{g} \upbeta^{l-1}\qty[\qty(1-\upbeta)\qty(4-\upbeta)]^{g- l}\sum_{a=\max\qty(1,l-g+h)}^{\min\qty(g, l)}\mathcal R_h^a \mathcal R_{g-h}^{l-a}+\\
        &+\qty(2-\upbeta)^2 \sum_{l=1}^{g-1} \upbeta^{l-1}\qty[\qty(1-\upbeta)\qty(4-\upbeta)]^{g- l}\sum_{a=\max\qty(1,l+1-g+h)}^{\min\qty(g, l+1)}\mathcal R_h^a \mathcal R_{g-h}^{l+1-a}.
    \end{aligned}
\end{align}
This result is of the claimed form, showing the statement also for the case of integer $g$, concluding the consideration of contribution \Romannum{3}.

\paragraph{\Romannum{4}} Finally, this contribution is dealt with rather quickly. First, we note that the appearing resolvent $R^\upbeta_g(z,I\backslash\qty{z})$ already is of genus $g$. Consequently, as it is one for $n$ boundaries it only lacks a factor of $\frac{1}{\upbeta}$ to acquire the expected form \cref{eq:Resolvent_halfInteger} or \cref{eq:Resolvent_integer} which is provided by the terms in bracket of contribution \Romannum{4} after noting (cf. \cref{eq:R_0_2_beta})
\begin{align}
    R^\upbeta_0(z_1, z_2)=\frac{1}{\upbeta}R^1_0(z_1,z_2).
\end{align}

With the last argument we have thus shown that for the two cases of integer and half-integer genus $g$ $F_g^\upbeta(z,I)$ indeed is of the form claimed for the respective case. As stated above, this structure is not modified upon computing the actual resolvent from this and thus also $R^\upbeta_g(z,I)$ is of the claimed form. This concludes the induction step. \qed

Apart from the usefulness for the proof, the knowledge of $F^{\upbeta}_g$ in terms of the $\mathcal R^i_g$ enables one to trace the origin of the $\mathcal R^i_{g+1}$ to the combination of $\mathcal R^i_g$ from which it is computed. This, for example, opens up a possibility to simplify the computation of higher genus resolvents by parallelization of the computation of the individual $\mathcal R^i_{g+1}$.

A more direct application, however, is found by observing the $\mathcal{R}^0_g$ terms to originate solely in $\mathcal{R}^0_{g'}$ of lower genus and higher number of arguments or products of these of lower genus and less arguments. Combining this with the basic fact that the $\mathcal{R}^0_g$ appear only for integer genus, one can see directly that the recursion for these terms is just a rescaled version of the recursion for the $\upbeta=2$ matrix model. Now, one observes that the examples of lower genus and number of arguments (in fact it suffices to consider the case of $g=1,n=2$) don't have a dependence on sums of arguments in the denominator, a property which is found for all the terms for the $\upbeta=2$ resolvents, and consequently, using that the recursion is a rescaled $\upbeta=2$ recursion, contributions at higher genus and numbers of arguments likewise don't.

\section{Splitting of the $\upbeta$ dependence}\label{sec:split_upbeta}
In this appendix we explain how to separate the expressions for the contribution to the $n$-boundary resolvent at genus $g\geq \frac 12$ into the invariant basis.

As a first step, the resolvents have to be multiplied by $\upbeta^{2g+n-1}$ in order to remove $\upbeta$ from the denominator, i.e. we define the scaled resolvent $\Tilde{R}^{\upbeta}_g(I)\coloneqq \upbeta^{2g+n-1}R^{\upbeta}_g(I)$. This leaves only a polynomial dependence on $\upbeta$ of maximal order $2g$, which is to be decomposed into the invariant basis. As first step towards this we choose an orthonormal basis for the vector space of polynomials of order $2g$, e.g. the Laguerre polynomials $\mathcal{B}_1=(L_i(\upbeta)|i=0,\dots,2g)$ orthonormal with respect to the scalar product
\begin{equation}
    \langle f,g\rangle:=\int_0^\infty e^{-x} f(x)g(x) \dd x.
\end{equation}
Using this, one can decompose
\begin{align}
    \Tilde{R}^{\upbeta}_g(I)=\sum_{i=0}^{2g}\underbrace{\langle L_i,\Tilde{R}^{\upbeta}_g(I)\rangle}_{\coloneqq a_i(I)} L_i(\upbeta),
\end{align}
where the coefficients $a_i(I)$ carry the dependence on the resolvent's arguments $I=\qty{z_1,\dots,z_n}$. Having done this decomposition, which is an explicit example for the generic decomposition of the resolvent claimed in the main text (\cref{eq:R_decomp_gen}), going over to a desired basis $\mathcal{B}_2=(b_0,\dots,b_{2g})$ only amounts to computing the change of basis for the Laguerre polynomials, i.e. to use 
\begin{align}
    L_i(\upbeta)&=\sum_{j=0}^{2g} \left(M_{\mathcal{B}_2,\mathcal{B}_1}\right)_{ij}b_j(\upbeta),
\end{align}
with
\begin{align}
    M_{\mathcal{B}_2,\mathcal{B}_1}&=M_{\mathcal{B}_1,\mathcal{B}_2}^{-1},\\
    \left(M_{\mathcal{B}_1,\mathcal{B}_2}\right)_{ij}&=\langle L_i,b_j\rangle.
\end{align}
This is always possible for any basis $\mathcal{B}_2$. Motivated by the discussion in the main text we choose the bases
\begin{equation}
    \mathcal{B}_2=\left(\upbeta^0,\dots,\upbeta^g,(2-\upbeta)^2((1-\upbeta)(4-\upbeta))^0\upbeta^{g-1},\dots,(2-\upbeta)^2((1-\upbeta)(4-\upbeta))^{g-1}\upbeta^0\right),
\end{equation}
for integer $g$ and
\begin{equation}
    \mathcal{B}_2=\left(\upbeta^0,\dots,\upbeta^{g-1/2},(2-\upbeta)((1-\upbeta)(4-\upbeta))^0\upbeta^{g-1/2},\dots,(2-\upbeta)((1-\upbeta)(4-\upbeta))^{g-1/2}\upbeta^0\right),
\end{equation}
for half-integer $g$. In fact, we take the invariant ``basis'', which is of course only a basis for the space of invariant polynomials, discussed in the main text and fill up to $2g+1$ basis elements by adding the monomial of the orders not contained in the invariant ``basis''. Putting things together, the decomposition of the resolvent in our desired basis is given by
\begin{align}
    R^{\upbeta}_g(I)=\frac{1}{\upbeta^{2g+n-1}}\sum_{i=0}^{2g}\underbrace{\sum_{i=0}^{2g} a_i(I)\left(M_{\mathcal{B}_2,\mathcal{B}_1}\right)_{ij}}_{\alpha_j(I)}b_j(\upbeta).
\end{align}
The effect of us carefully choosing an invariant basis is now, that the $\alpha_j$ should always vanish in this decomposition for $i\leq g-1$ (integer genus) or $i\leq g-\frac 12$ (half-integer genus). This is what we see for all the computed examples and what we could indeed prove in \cref{sec:Proof_GenStruct}.

\section{Proof of the generality of the surface decomposition}\label{sec:App_Decomposition}

In this appendix we show that any decomposition of a surface of genus $g$ and $n$ geodesic boundaries into the parts prescribed by the arbitrary $\upbeta$ Mirzakhani-like recursion, pictorially represented in \cref{fig:Mirzakhani_Cuttings}, can be ``reshuffled'' to yield either directly the decomposition in \cref{fig:Decomposition} or one deriving from it by adding holes as discussed in the main text. In the following, we refer to such a decomposition as an ordered decomposition. By ``reshuffling'' we mean moving the constituent parts of the surface around in a way that does not change the $\upbeta$ dependence of the whole decomposition.

To do this, we start with an arbitrary decomposition into the allowed building blocks. We will now transform this to one of the ordered cases in two steps. First, we will move all parts containing external boundaries to the left, recreating the first part of \cref{fig:Decomposition}. Second, we will reshuffle the parts building up the remaining, genus carrying, part as to reproduce one of the ordered decompositions. 

For the first part, if $n=1$ we choose the external boundary as the starting point of the genus carrying part and have achieved our goal for the boundary part already since it's empty in this case. If $n\geq2$ there can be at most one 3-holed sphere containing two external boundaries. 

If it is there, choose the boundary to which it is attached as the starting point for the genus carrying part and take out the 3-holed sphere. Then, there are still $n-2$ 3-holed spheres that contain an external boundary and are glued in an orientation as in case a) (Here and in the following, cases x) refer to the ones discussed in \cref{fig:Mirzakhani_Cuttings}) at two of their boundaries. We take out these 3-holed spheres, glueing the structures attached to the two formerly glued boundaries to one another. 

If it is not there, there is one 3-holed sphere containing an external boundary to which something is glued in the way prescribe by a case that is not a). Choose this external boundary as the starting point for the genus carrying part. Similar to the other case, there are now $n-1$ 3-holed spheres that contain an external boundary and are glued in an orientation as in case a) at two of their boundaries. Proceed with them as in the other case.

In both cases, we have taken out a total of $n-1$ 3-holed spheres and have designated a starting boundary for the genus carrying part. We now can assemble the 3-holed spheres in the way done in the first part of \cref{fig:Decomposition} and attach them to the designated starting boundary, by which we have achieved the separation aimed at.

Having done this, we are left with a decomposition of a surface with one geodesic boundary to the left and genus $g$. We will bring it into one of the discussed forms by starting with the 3-holed sphere that contains this boundary and, potentially moving structures attached to it, render the objects attached to it into a form appearing in the ``ordered'' decomposition. We then proceed with the next attached 3-holed sphere in the same way. In order to do this, additionally to the cases a), b), c) and d) discussed in \cref{fig:Mirzakhani_Cuttings} we introduce the additional cases i): like case b) with $Y$ being a crosscap and ii): Like d) with $X$ being the surface of genus $0$ with two boundaries, i.e. the boundaries are glued to one another. 
Having introduced these cases we can state our procedure. First, we note that case a) cannot appear with an unglued boundary since we already removed all external boundaries. It can only appear after a glueing of case d). Consequently, we can define for every appearing 3-holed sphere, apart from these cases a, ``left'' boundary in the obvious way and choose an upper and lower boundary, where we take the convention that if a crosscap is glued to a 3-holed sphere, it is always glued to the lower boundary. The starting point for every step will be a 3-holed sphere $S$ identified by its left boundary, while the starting point for the whole procedure is given by the unique 3-holed sphere containing the remaining external geodesic boundary.

Given $S$, the glueing is as in case
\begin{itemize}
    \item [b)] Then, the part is in the ordered form. Continue by setting $S$ as the first 3-holed sphere of $Y$
    \item[c)]Proceed along the lower boundaries of the glued 3-holed spheres forming $X_1$, until either case i) or ii) are assumed (They are the only possibility for terminating the glueing). 
    \begin{itemize}
        \item[i)] Attach a crosscap to the lower boundary of $S$, attach a 3-holed sphere (that from the end of $X_1$) to the upper boundary. Attach to the lower boundary of this another crosscap and to its upper boundary the rest of $X_1$ (i.e. with the final 3-holed sphere removed). Then attach $X_2$ to the remaining, now unglued boundary of $X_1$. Set $S$ as the first 3-holed sphere of $X_1$.
        \item[ii)] Attach a 3-holed sphere in the a) way to $S$, attach $X_1$ (with the final 3-holed sphere removed) to it. Finally, glue $X_2$ to the unglued boundary of $X_1$ and proceed by setting $S$ as the first 3-holed sphere of $X_1$.
    \end{itemize}
    \item[d)]Then, the part is in the ordered form. Continue by setting $S$ as the 3-holed sphere attached to the type a) 3-holed sphere being glued to the present 3-holed sphere
    \item[i)/ii)]Terminate, the decomposition is in one of the ``ordered'' forms.
\end{itemize}
All of these transformations do not change the $\upbeta$ dependence of the decomposition since either the constituent parts are only moved around or in the case c),ii) the factor of $\frac{1}{\upbeta}$ from the self glueing of the final 3-holed sphere is replaced by that from the a)-type glueing after the prescribed modification.
Furthermore, it is important to note, that in the reactions to the cases that don't terminate the procedure the genus of the (ordered) decomposition left of the new $S$ is increased. Hence, the procedure is guaranteed to terminate since the genus of the decomposed surface is finite. Thus, since we have given all the possible cases how something can be glued to a given $S$ the procedure is complete and yields an ordered decomposition which shows our claim.

\section{Derivation of the SFF from universal RMT for GSE}\label{sec:SFF_derivation}
The integral to evaluate is recalled from the main text to be
\begin{align}
  \kappa_4^s(\tau,\beta)=\int_{0}^{\infty}\dd E e^{-2\beta E}\rho_0(E)- 
\int_{0}^{\infty}\dd E e^{-2\beta E}\rho_0(E)b_4\left(\frac{\tau}{2\pi \rho_0(E)}\right),
\end{align}
with
\begin{align}
    b_4\qty(\frac{\tau}{2\pi \rho_0(E)})=
	\begin{cases}
		1-\frac{\tau}{4\pi\rho_0(E)}+\frac{\tau}{8\pi\rho_0(E)}\log\qty(\abs{1-\frac{\tau}{2\pi\rho_0(E)}}) & \mbox{if } \frac{\tau}{4\pi} \leq \rho_0(E) \\
		0 & \mbox{if } \frac{\tau}{4\pi} \geq \rho_0(E).
	\end{cases}.
\end{align}
This motivates to define $E_\star$ as the solution of $\rho_0(E_\star)=\frac{\tau}{4\pi}$ using which one can write the integral as
\begin{align}
\begin{aligned}\label{eq:kappa_4_split}
    \kappa^{s}_{4}(\tau,\beta)=&\qty[\int_0^{E_\star}\dd{E}e^{-2\beta E} \rho_0(E)+\int_{E_\star}^\infty\dd{E}e^{-2\beta E}\qty(\frac{\tau}{4\pi}-\frac{\tau}{8\pi}\log\qty(\abs{1-\frac{\tau}{2\pi \rho_0(E)}}))]\\
    =&\left[\int_0^{E_\star}\dd{E}e^{-2\beta E} \rho_0(E)+\int_{E_\star}^\infty\dd{E}e^{-2\beta E}\frac{\tau}{4\pi}\right.\\
    &\left.-\int_{E_\star}^\infty\dd{E}e^{-2\beta E}\frac{\tau}{8\pi}\log\qty(\abs{1-\frac{\tau}{2\pi \rho_0(E)}})\right]\\
    &=\int_0^{\infty}\dd{E}e^{-2\beta E} \min\left(\rho_0(E),\frac{\tau}{4\pi}\right)-\int_{E_\star}^\infty\dd{E}e^{-2\beta E}\frac{\tau}{8\pi}\log\qty(\abs{1-\frac{\tau}{2\pi \rho_0(E)}})\\
    &\eqqcolon\kappa_{2}^{s}\qty(\frac\tau2, \beta)-\frac{\tau}{8\pi}\chi\left(\tau,\beta\right),
\end{aligned}
\end{align}
where we can use the known result for the Airy model (\cref{eq:kappa_2_s})
\begin{equation}
    \kappa_{2}^{s}\left(\tau,\beta \right) =\frac{1}{2\sqrt{\pi}}\frac{1}{2^{\frac 5 2}\beta^{3/2}}\Erf\qty(\sqrt{2\beta}\tau).
\end{equation}
What is left to evaluate is
\begin{align}
\begin{aligned}
    \chi\left(\tau,\beta\right)&=\int_{E_\star}^\infty\dd{E}e^{-2\beta E}\log\qty(\abs{1-\frac{\tau}{2\pi \rho_0(E)}})\\
    &=\int_{E_\star}^{4E_\star}e^{-2\beta E}\log\text{\ensuremath{\left(\frac{\tau}{\sqrt{E}}-1\right)}}\text{d}E+\int_{4E_\star}^{\infty}e^{-2\beta E}\log\text{\ensuremath{\left(1-\frac{\tau}{\sqrt{E}}\right)}}\text{d}E,
\end{aligned}
\end{align}
where we now explicitly put $\rho_0$ to be the Airy model density of states, i.e. $\rho_0(E)=\frac{1}{2\pi}\sqrt{E}$. This results in $E_\star=\left(\frac{\tau}{2}\right)^2$. Integrating by parts
yields:
\begin{align}
\begin{aligned}
    \chi\left(\tau,\beta\right)=& \left[\frac{1}{-2\beta}e^{-2\beta E}\log\left(\frac{\tau}{\sqrt{E}}-1\right)\right]_{E=E_\star}^{E=4E_\star}-\int_{E_\star}^{4E_\star}\frac{1}{-2\beta}e^{-2\beta E}\frac{\tau}{2E^{\frac{3}{2}}-2E\tau}\text{d}E\\
 & +\left[\frac{1}{-2\beta}e^{-2\beta E}\log\left(1-\frac{\tau}{\sqrt{E}}\right)\right]_{E=4E_\star}^{\infty}-\int_{4E_\star}^{\infty}\frac{1}{-2\beta}e^{-2\beta E}\frac{\tau}{2E^{\frac{3}{2}}-2E\tau}\text{d}E\\
= & \frac{1}{2\beta}e^{-2\beta E_\star}\underbrace{\log\left(\frac{\tau}{\sqrt{E_\star}}-1\right)}_{0}+\frac{1}{2\beta}\int_{E_\star}^{\infty}e^{-2\beta E}\frac{\tau}{2E^{\frac{3}{2}}-2E\tau}\text{d}E\\
= & \frac{1}{2\beta}\int_{E_\star}^{\infty}e^{-2\beta E}\left(-\frac{1}{2\sqrt{E}\left(\tau-\sqrt{E}\right)}-\frac{1}{2E}\right)\text{d}E.
\end{aligned}
\end{align}
Now we note that the integral $\int_{E_\star}^{\infty}\left(-e^{-2\beta E}\frac{1}{2E}\right)\text{d}E$
is related to the incomplete Gamma function as $-\frac{1}{2}\Gamma\left(0,2\beta E_\star\right)$.
The remaining integral is solved by substituting $x=\sqrt{E},\text{d}x=\frac{1}{2\sqrt{E}}\text{d}E:$
\begin{equation}
-\int_{E_\star}^{\infty}e^{-2\beta E}\frac{1}{2\sqrt{E}\left(\tau-\sqrt{E}\right)}\text{d}E.=-\int_{x_\star}^{\infty}e^{-2\beta x{{}^2}}\frac{1}{\tau-x}\text{d}x\label{eq:shift_operator},
 \end{equation}
(note that $x_\star=\frac{\tau}{2})$.
Now we shift the fraction using the shift operator. Doing this one has to be careful since one has to split the integral in a way such that the different series converge in each domain. Explicitly:
\begin{align}
\begin{aligned}
    -\int_{\frac{\tau}{2}}^{\infty}e^{-2\beta x{{}^2}}\frac{1}{\tau-x}\text{d}x & =-\int_{\frac{\tau}{2}}^{\tau}e^{-2\beta x{{}^2}}\frac{1}{\tau-x}\text{d}x+\int_{\tau}^{\infty}e^{-2\beta x{{}^2}}e^{-\tau\frac{d}{dx}}\frac{1}{x}\text{d}x\label{eq:integral_split}\\
 & =-\int_{\frac{\tau}{2}}^{\tau}e^{-2\beta x{{}^2}}\sum_{n=0}^{\infty}\frac{x^{n}}{\tau^{n+1}}\text{d}x+\int_{\tau}^{\infty}e^{-2\beta x{{}^2}}\sum_{n=0}^{\infty}\frac{\left(-\tau\right)^{n}}{n!}\frac{d^{n}}{dx^{n}}\frac{1}{x}\text{d}x\\
 & =-\int_{\frac{\tau}{2}}^{\tau}e^{-2\beta x{{}^2}}\sum_{n=0}^{\infty}\frac{x^{n}}{\tau^{n+1}}\text{d}x+\int_{\tau}^{\infty}e^{-2\beta x{{}^2}}\sum_{n=0}^{\infty}\frac{\left(-\tau\right)^{n}}{n!}\left(-1\right)^{n}\frac{n!}{x^{n+1}}\text{d}x\\
 & =\sum_{n=0}^{\infty}\left(\int_{\tau}^{\infty}e^{-2\beta x{{}^2}}\frac{\tau^{n}}{x^{n+1}}\text{d}x-\int_{\frac{\tau}{2}}^{\tau}e^{-2\beta x{{}^2}}\frac{x^{n}}{\tau^{n+1}}\text{d}x\right)\\
 & =\sum_{n=0}^{\infty}\frac{1}{2}E_{\frac{n}{2}+1}\left(2\beta\tau^{2}\right)-\frac{1}{4}\left(2^{-n}E_{\frac{1}{2}-\frac{n}{2}}\left(\frac{\beta\tau^{2}}{2}\right)-2E_{\frac{1}{2}-\frac{n}{2}}\left(2\beta\tau^{2}\right)\right),
\end{aligned}
\end{align}
where $E_n(x)$ denotes the exponential integral.

Putting all together, we find
\begin{align}
\begin{aligned}
    \chi(\tau,\beta)& =\frac{1}{2\beta}\int_{E_\star}^{\infty}e^{-2\beta E}\left(-\frac{1}{2\sqrt{E}\left(\tau-\sqrt{E}\right)}-\frac{1}{2E}\right)\text{d}E\\
 & =\frac{1}{2\beta}\left(-\frac{1}{2}\Gamma\left(0,\beta\frac{\tau^{2}}{2}\right)+\sum_{n=0}^{\infty}\left(\frac{1}{2}E_{\frac{n}{2}+1}\left(2\beta\tau^{2}\right)-\frac{1}{4}\left(2^{-n}E_{\frac{1}{2}-\frac{n}{2}}\left(\frac{\beta\tau^{2}}{2}\right)-2E_{\frac{1}{2}-\frac{n}{2}}\left(2\beta\tau^{2}\right)\right)\right)\right)\\
 & =-\frac{1}{4\beta}\Gamma\left(0,\beta\frac{\tau^{2}}{2}\right)+\sum_{n=0}^{\infty}\frac{1}{4\beta}E_{\frac{n}{2}+1}\left(2\beta\tau^{2}\right)-\sum_{n=0}^{\infty}\frac{1}{8\beta}\left(2^{-n}E_{\frac{1}{2}-\frac{n}{2}}\left(\frac{\beta\tau^{2}}{2}\right)-2E_{\frac{1}{2}-\frac{n}{2}}\left(2\beta\tau^{2}\right)\right).
\end{aligned}
\end{align}
We are interested in a series expansion in $\tau$, so we have to expand the occurring special functions. For this, we use \cite{Gradshteyn2000}
\begin{align}
    \Gamma\left(0,\beta\frac{\tau^{2}}{2}\right)&=-\gamma-\log\left(\beta\frac{\tau^{2}}{2}\right)-\sum_{n=1}^{\infty}\frac{\left(-\beta\frac{\tau^{2}}{2}\right)^{n}}{nn!},\\
    E_{n}\left(z\right) & =\frac{\left(-z\right)^{n-1}}{\left(n-1\right)!}\left(-\log z+\psi\left(n\right)\right)-\sum_{\substack{m=0\\
m\neq n-1
}
}\frac{\left(-z\right)^{m}}{\left(m-n+1\right)m!}\qquad n\in\mathbb{N}^{+},
\end{align}
where $\gamma$ denotes the Euler-Mascheroni constant and $\psi$ the Digamma function. The latter can be represented as
\begin{align}
\psi\left(n\right) & =-\gamma+\sum_{m=1}^{n-1}\frac{1}{m}.
\end{align}
Additionally, we use
\begin{align}
E_{n}\left(z\right) & =z^{n-1}\Gamma\left(1-n,z\right),\\
\Gamma\left(a,x\right)&=\Gamma\left(a\right)-\sum_{k=0}^{\infty}\frac{\left(-1\right)^{k}x^{a+k}}{k!\left(a+k\right)}\qquad a\notin-\mathbb{N}_{0},
\end{align}
which can be combined to find
\begin{align}
E_{n}\left(z\right) & =z^{n-1}\Gamma\left(1-n,z\right)\\
 & =z^{n-1}\Gamma\left(1-n\right)-z^{n-1}\sum_{k=0}^{\infty}\frac{\left(-1\right)^{k}z^{1-n+k}}{k!\left(1-n+k\right)}\\
 & =\Gamma\left(1-n\right)z^{n-1}-\sum_{k=0}^{\infty}\frac{\left(-z\right)^{k}}{k!\left(k-n+1\right)}\qquad n\notin\mathbb{N}^{+}.
\end{align}
This can be used to rewrite the latter two occurences of the Exponential integral in $\chi(\tau,\beta)$ as
\begin{align}
E_{\frac{1}{2}-\frac{n}{2}}\left(\frac{\beta\tau^{2}}{2}\right) & =\Gamma\left(\frac{n+1}{2}\right)\left(\frac{\beta\tau^{2}}{2}\right)^{-\frac{n+1}{2}}-\sum_{k=0}^{\infty}\frac{\left(-1\right)^{k}\left(\frac{\beta\tau^{2}}{2}\right)^{k}}{\left(k+\frac{n+1}{2}\right)k!}\\
E_{\frac{1}{2}-\frac{n}{2}}\left(2\beta\tau^{2}\right) & =\Gamma\left(\frac{n+1}{2}\right)\left(2\beta\tau^{2}\right)^{-\frac{n+1}{2}}-\sum_{k=0}^{\infty}\frac{\left(-1\right)^{k}\left(2\beta\tau^{2}\right)^{k}}{\left(k+\frac{n+1}{2}\right)k!}.
\end{align}
The treatment of the first occurence is a more complicated, since we have to distinguish two cases:
\begin{equation}
E_{\frac{n}{2}+1}\left(2\beta\tau^{2}\right)=\begin{cases}
\Gamma\left(-\frac{n}{2}\right)\left(2\beta\tau^{2}\right)^{\frac{n}{2}}-\sum_{k=0}^{\infty}\frac{\left(-1\right)^{k}\left(2\beta\tau^{2}\right)^{k}}{\left(k-\frac{n}{2}\right)k!} & n\notin2\mathbb{N}_{0}\\
\frac{\left(-2\beta\tau^{2}\right)^{\frac{n}{2}}}{\left(\frac{n}{2}\right)!}\left(-\log2\beta\tau^{2}+\psi\left(\frac{n}{2}+1\right)\right)-\sum_{\substack{k=0\\
k\neq\frac{n}{2}
}
}^{\infty}\frac{\left(-2\beta\tau^{2}\right)^{k}}{\left(k-\frac{n}{2}\right)k!} & n\in2\mathbb{N}_{0}.
\end{cases}\label{eq:exp_int_expansion}
\end{equation}
In total we find
\begin{subequations}
\begin{align}
\chi\left(\tau,\beta\right)= & -\frac{1}{4\beta}\left(-\gamma-\log\left(\beta\frac{\tau^{2}}{2}\right)-\sum_{n=1}^{\infty}\frac{\left(-\beta\frac{\tau^{2}}{2}\right)^{n}}{nn!}\right)\\
 & +\sum_{n=0}^{\infty}\frac{1}{4\beta}\left(\frac{\left(-2\beta\tau^{2}\right)^{n}}{n!}\left(-\log2\beta\tau^{2}+\psi\left(n+1\right)\right)-\sum_{\substack{k=0\\
k\neq n
}
}^{\infty}\frac{\left(-2\beta\tau^{2}\right)^{k}}{\left(k-n\right)k!}\right)\\
 & +\sum_{n=0}^{\infty}\frac{1}{4\beta}\left(\Gamma\left(-\frac{2n+1}{2}\right)\left(2\beta\tau^{2}\right)^{\frac{2n+1}{2}}-\sum_{k=0}^{\infty}\frac{\left(-2\beta\tau^{2}\right)^{k}}{\left(k-\frac{2n+1}{2}\right)k!}\right)\\
 & -\sum_{n=0}^{\infty}\frac{1}{8\beta}2^{-n}\left(\Gamma\left(\frac{n+1}{2}\right)\left(\frac{\beta\tau^{2}}{2}\right)^{-\frac{n+1}{2}}-\sum_{k=0}^{\infty}\frac{\left(-\frac{\beta\tau^{2}}{2}\right)^{k}}{\left(k+\frac{n+1}{2}\right)k!}\right)\\
 & +\sum_{n=0}^{\infty}\frac{1}{4\beta}\left(\Gamma\left(\frac{n+1}{2}\right)\left(2\beta\tau^{2}\right)^{-\frac{n+1}{2}}-\sum_{k=0}^{\infty}\frac{\left(-2\beta\tau^{2}\right)^{k}}{\left(k+\frac{n+1}{2}\right)k!}\right).
\end{align}
\end{subequations}
This simplifies considerably as the 1st, 2nd and 4th sum over $k$ can be combined to find:
\begin{align}
\begin{aligned}
    &-\sum_{n=0}^{\infty}\sum_{\substack{k=0\\
k\neq n
}
}^{\infty}\frac{\left(-2\beta\tau^{2}\right)^{k}}{\left(k-n\right)k!}-\sum_{n=0}^{\infty}\sum_{k=0}^{\infty}\frac{\left(-2\beta\tau^{2}\right)^{k}}{\left(k-\frac{2n+1}{2}\right)k!}-\sum_{n=0}^{\infty}\sum_{k=0}^{\infty}\frac{\left(-2\beta\tau^{2}\right)^{k}}{\left(k+\frac{n+1}{2}\right)k!}\\
&=-\sum_{k=0}^{\infty}\left(-2\beta\tau^{2}\right)^{k}\sum_{n=0}^{\infty}\left(\frac{\left(1-\delta_{n,2k}\right)}{\left(k-\frac{n}{2}\right)k!}+\frac{1}{\left(k+\frac{n+1}{2}\right)k!}\right)\\
&=\sum_{k=0}^{\infty}\frac{\left(-2\beta\tau^{2}\right)^{k}}{k!}\qty[\frac{1}{2k+\frac 12}+\qty(2k+\frac 12)\sum_{n=0, n\neq 2k}^{\infty}\frac{1}{\qty(k-\frac n 2)\qty(k+\frac{n+1}{2})}]\\
&=\sum_{k=0}^{\infty}\frac{\left(-2\beta\tau^{2}\right)^{k}}{k!}\qty[\frac{1}{2k+\frac 12}\underbrace{+2 H_{4 k}-2 H_{4 k+1}}_{=-\frac{2}{4k+1}}]\\
&=0,
\end{aligned}
\end{align}
where we denoted the $n$-th harmonic number as $H_n$.
This implies
\begin{subequations}
\begin{align}
\chi\left(\tau,\beta\right)= & -\frac{1}{4\beta}\left(-\gamma-\log\left(\beta\frac{\tau^{2}}{2}\right)-\sum_{n=1}^{\infty}\frac{\left(-\beta\frac{\tau^{2}}{2}\right)^{n}}{nn!}\right)\\
 & +\sum_{n=0}^{\infty}\frac{1}{4\beta}\frac{\left(-2\beta\tau^{2}\right)^{n}}{n!}\left(-\log2\beta\tau^{2}+\psi\left(n+1\right)\right)\\
 & +\sum_{n=0}^{\infty}\frac{1}{4\beta}\Gamma\left(-\frac{2n+1}{2}\right)\left(2\beta\tau^{2}\right)^{\frac{2n+1}{2}}\\
 & -\sum_{n=0}^{\infty}\frac{1}{8\beta}2^{-n}\Gamma\left(\frac{n+1}{2}\right)\left(\frac{\beta\tau^{2}}{2}\right)^{-\frac{n+1}{2}}\label{eq:neg_exponents1}\\
 &+\sum_{k=0}^{\infty}\frac{1}{8\beta}\left(-\frac{\beta\tau^{2}}{2}\right)^{k}\frac{1}{k!}\frac{2 \, _2F_1\left(1,2 k+1;2 k+2;\frac{1}{2}\right)}{2 k+1}\\
 & +\sum_{n=0}^{\infty}\frac{1}{4\beta}\Gamma\left(\frac{n+1}{2}\right)\left(2\beta\tau^{2}\right)^{-\frac{n+1}{2}}\label{eq:neg_exponents2}.
\end{align}
\end{subequations}
This expansion looks like it contains negative powers of $\tau$, however, as can easily be checked, \ref{eq:neg_exponents1} and \ref{eq:neg_exponents2} perfectly cancel each other, so our final result for $\chi(\tau,\beta)$ is

\begin{align}
\begin{aligned}\label{eq:chi_analytic_sum}
    \chi\left(\tau,\beta\right)= & -\frac{1}{4\beta}\left(-\gamma-\log\left(\beta\frac{\tau^{2}}{2}\right)-\sum_{n=1}^{\infty}\frac{\left(-\beta\frac{\tau^{2}}{2}\right)^{n}}{nn!}\right)\\
 & +\sum_{n=0}^{\infty}\frac{1}{4\beta}\frac{\left(-2\beta\tau^{2}\right)^{n}}{n!}\left(-\log2\beta\tau^{2}+\psi\left(n+1\right)\right)\\
 & +\sum_{n=0}^{\infty}\frac{1}{4\beta}\Gamma\left(-\frac{2n+1}{2}\right)\left(2\beta\tau^{2}\right)^{\frac{2n+1}{2}}\\
 &+\sum_{k=0}^{\infty}\frac{1}{8\beta}\left(-\frac{\beta\tau^{2}}{2}\right)^{k}\frac{1}{k!}\frac{2 \, _2F_1\left(1,2 k+1;2 k+2;\frac{1}{2}\right)}{2 k+1}.
\end{aligned}
\end{align}

Combining this with the well-known expansion of $\kappa_2^s(\tau,\beta)$, using \cref{eq:kappa_4_split}, yields the expansion of $\kappa_2^s(\tau,\beta)$ as a power series in $\tau$ and $\beta$ that is presented in the main text.

\begin{figure}[h]
    \centering
    \includegraphics[width=0.6\linewidth]{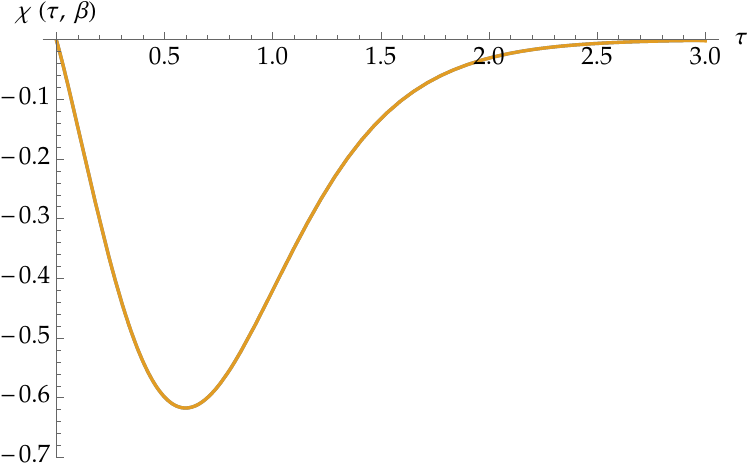}
    \caption{Comparison of a numerical evaluation of the integral definition of $\chi(\tau,\beta)$ at and $\beta=1$ with the analytical result \cref{eq:chi_analytic_sum} summed up to $n=50$.}
    \label{fig:Comparison_GSE_chi}
\end{figure}

For completeness, we give a comparison of this result (summed up to $n=50$) with a numerical evaluation of the integral that defines $\chi(\tau,\beta)$ in \cref{fig:Comparison_GSE_chi}. Here, we choose an exemplary value $\upbeta=1$ and show values of $\tau$ beyond which the function vanishes. As one would expect, the two curves cannot be distinguished.

\section{Determination of the ``cancelling'' functions}\label{sec:det_canc_func}
Here we illustrate our method how to find the ``cancelling'' functions for the logarithmic term and the first terms of second and third type for the $\tau^3\beta^0$ contribution to the $\upbeta=4$ $\tau$-scaled SFF.

What we are looking for generically is a function that asymptotically behaves as
\begin{equation}
a\left(\log\left(bt\tau^{2}\right)+\gamma-3\right)+c\sqrt{t}\tau,
\end{equation}
with $a,b,c\in\mathbb R$ chosen such that subtracting this term cancels the term of the second type and yields the logarithm expected from the universal result.
Furthermore, it should have the defining series expansion
\begin{align}
    d\left(t\tau^{2}\right)^{2}+\order{\qty(t\tau^2)^3},
\end{align}
with $d\in\mathbb{R}$ chosen such that upon adding this expansion the first term of the third type is cancelled.
Specifically, for the case of $\upbeta=4$, these values can be read off to be
\begin{align}
a = 3\qc b =1 \qc c =\sqrt{\frac{\pi}{2}}\qc d =\frac{1}{120}.
\end{align}
We make an ansatz for the logarithmic part, using $\,_{2}F_{2}$
\begin{equation}
A\left(t\tau^{2}\right)^{2}\,_{2}F_{2}\left(2,2;\frac{5}{2},\frac{7}{2};-Bt\tau^{2}\right)\overset{t\rightarrow\infty}{\longrightarrow}\frac{45A}{16B^{2}}\left(\log\left(16Bt\tau^{2}\right)+\gamma-3\right).
\end{equation}
By comparison this fixes $B=\frac{b}{16}$ and $A=\frac{ab^{2}}{720}$. Now we look at the defining expansion for this function which, plugging already the found values for $A$ and $B$, is given to the first orders as
\begin{equation}
A\left(t\tau^{2}\right)^{2}\,_{2}F_{2}\left(2,2;\frac{5}{2},\frac{7}{2};-Bt\tau^{2}\right)=\frac{ab^{2}}{720}t^{2}\tau^{4}-\frac{ab^{2}}{25200}t^{3}\tau^{6}+\dots.
\end{equation}
Next we concern ourselves with taking care of the second type term. Generically, for all type 2 terms we could choose the ansatz
\begin{equation}
C\left(t\tau^{2}\right)^{2}\left(t\tau^{2}\right)^{n}\,_{1}F_{1}\left(\frac{3}{2};\frac{5}{2};-Dt\tau^{2}\right)\overset{t\rightarrow\infty}{\longrightarrow}\left(\sqrt{t}\tau\right)^{2n} \frac{3 \sqrt{\pi } C \sqrt{D}}{4 D^2} \tau t,
\end{equation}
iff $\Re\left(D\right)>0.$ Using $n=0$ to reproduce the present term of
order $\sqrt{t}\tau$ we have
\begin{equation}
C\left(t\tau^{2}\right)^{2}\,_{1}F_{1}\left(\frac{3}{2};\frac{5}{2};-Dt\tau^{2}\right)\overset{t\rightarrow\infty}{\longrightarrow}\frac{C\sqrt{D}}{4D^{2}}3\sqrt{\pi}\tau\sqrt{t}.
\end{equation}
In order to match our expectations, it would thus have to hold that
\begin{align}
\frac{C\sqrt{D}}{4D^{2}}3\sqrt{\pi}=  c
\implies\frac{C^{2}}{16D^{3}}9\pi=  c^{2}
\implies D= & \sqrt[3]{\frac{9\pi C^{2}}{16c^{2}}}.
\end{align}
We have to be careful, squaring in the second step means we apparently find solutions for $c<0$, which, upon closer inspection, do not work out, so we constrain ourselves to $c>0$.
If we consider the defining series for this function we find
\begin{equation}
C\left(t\tau^{2}\right)^{2}\,_{1}F_{1}\left(\frac{3}{2};\frac{5}{2};-Dt\tau^{2}\right)=C\left(t\tau^{2}\right)^{2}-\frac{3}{5}CD\left(t\tau^{2}\right)^{3}+\frac{3}{14}CD^{2}\left(t\tau^{2}\right)^{4}\dots.
\end{equation}
Consequently, we find that for the series expansion to conform to our requirement, it has to hold that
\begin{align}
A+C =d 
\implies C  =d-\frac{ab^{2}}{720}
\implies D =\sqrt[3]{\frac{9\pi\left(d-\frac{ab^{2}}{720}\right)^{2}}{16c^{2}}}.
\end{align}
so we find
\begin{align}
A=\frac{1}{240}\qc
B=\frac{1}{16}\qc
C=\frac{1}{240}\qc
D=\frac{1}{8\sqrt[3]{100}}.
\end{align}
Putting these into our ansatz, we arrive at the expressions for $f_1$ and $f_2$ in the main text.

\newpage
\bibliography{lib}

\end{document}